\documentclass[twocolumn]{aastex631}

\usepackage[T1]{fontenc}

\shorttitle{}
\shortauthors{Hasegawa}
\graphicspath{{./}{figures/}}

\begin{document}

\title{Solid Accretion onto Neptune-Mass Planets I:\\
In-Situ Accretion and Constraints from the metallicity of Uranus and Neptune}

\author[0000-0002-9017-3663]{Yasuhiro Hasegawa}
\affiliation{Jet Propulsion Laboratory, California Institute of Technology, Pasadena, CA 91109, USA}
\email{yasuhiro.hasegawa@jpl.nasa.gov}




\begin{abstract}

The currently available, detailed properties (e.g., isotopic ratios) of solar system planets may provide guides for constructing better approaches of exoplanet characterization.
With this motivation, we explore how the measured values of the deuterium-to-hydrogen (D/H) ratio of Uranus and Neptune can constrain their formation mechanisms.
Under the assumption of in-situ formation, we investigate three solid accretion modes;
a dominant accretion mode switches from pebble accretion to drag-enhanced three-body accretion and to canonical planetesimal accretion, as the solid radius increases.
We consider a wide radius range of solids that are accreted onto (proto)Neptune-mass planets and
compute the resulting accretion rates as a function of both the solid size and the solid surface density.
We find that for small-sized solids, the rate becomes high enough to halt concurrent gas accretion, if all the solids have the same size.
For large-sized solids, the solid surface density needs to be enhanced to accrete enough amounts of solids within the gas disk lifetime.
We apply these accretion modes to the formation of Uranus and Neptune and 
show that if the minimum-mass solar nebula model is adopted,
solids with radius of $\sim 1$ m to $\sim 10$ km should have contributed mainly to their deuterium enrichment;
a tighter constraint can be derived if the full solid size distribution is determined.
This work therefore demonstrates that the D/H ratio can be used as a tracer of solid accretion onto Neptune-mass planets.
Similar efforts can be made for other atomic elements that serve as metallicity indicators.

\end{abstract}



\section{Introduction} \label{sec:intro}

Characterization of exoplanets is one essential means to reveal their origins.
To this goal, the various properties of exoplanets (e.g., mass, radius, and orbital periods) are currently observed and examined.
Atmospheric properties are likely the key quantities that can potentially address a number of the fundamental questions:
the abundance ratio of carbon to oxygen (often referred to as the C/O ratio) may be used as a tracer of the migration history of giant planets 
\citep[e.g.,][]{2011ApJ...743L..16O,2014ApJ...794L..12M};
the abundance of oxygen through the detection of water and the resulting total heavy-element mass can probe the metal enrichment processes of planets 
\citep[e.g.,][]{2014ApJ...793L..27K,2018ApJ...865...32H};
and the detection of the atmosphere around small-sized, rocky planets and the potential biosignatures in the atmosphere
will shed light on the origin of life \citep[e.g.,][]{2018AsBio..18..663S,2021AJ....161..213S}.

Despite the above successes, useful guides are constantly demanded to reliably and efficiently construct a better approach of exoplanet characterization.
In such a context, the importance of the solar system is boosted up considerably;
the solar system bodies are currently the only targets that one can literally visit and reliably access to the detailed properties 
\citep[such as gravitational moments and isotopic ratios, e.g.,][]{2015Sci...350.1815S,2017Sci...356..821B,2019ARA&A..57..113A}.
The availability of these properties already enabled careful vetting of 
which properties can be used to tightly constrain how planets formed and evolved to the current configuration in the solar system \citep[e.g.,][]{2011Natur.473..489D,2017GeoRL..44.4649W}.
Such knowledge is principally applicable to extrasolar environments \citep[e.g.,][]{2020ARA&A..58..727J}.
 
Among the solar system bodies, Uranus and Neptune stand out due to a number of reasons:
These planets are explored for a long time 
\citep[e.g.,][]{1986Icar...67..391B,1995P&SS...43.1517P,1996Icar..124...62P,2011ApJ...726...15H,2013P&SS...77..143N},
however, their formation mechanisms are still elusive \citep[e.g.,][]{2004ARA&A..42..549G,2017AJ....154...98F,2020SSRv..216...38H}.
Hence, more investigations are still required.
The current exoplanet observations confirm the ubiquity of Neptune-sized (and even smaller) planets in the galaxy
\citep[e.g.,][]{2011arXiv1109.2497M,2015ARA&A..53..409W}.
The characterization of small bodies such as asteroids and comets is currently conducted actively
\citep[e.g.,][]{2006Sci...312.1341S,2011ARA&A..49..471M,2015Sci...347A.387A}.
These small bodies are regarded as the leftover of planet-forming materials.
Under this circumstance, one can essentially trace how (proto)Uranus and (proto)Neptune accreted these bodies, 
both by using the detailed properties of the small bodies as the initial condition,
and by examining the resulting (atmospheric) properties of the planets as the final outcome.
Understanding developed for Uranus and Neptune can be applied to a large population of exoplanets.
 
In a series of papers, we undertake such studies, focusing on solid accretion onto Neptune-mass planets and the resulting deuterium-to-hydrogen (D/H) ratio of these planets.
Any atomic elements that can be used as metallicity indicators will constrain formation mechanisms of planets (see Section \ref{sec:disc_trac} for the additional discussion).
However, the D/H ratio measured for solar system bodies is well known to be one important quantity of 
tracing what the solar nebula looked like and how the solar system formed in the nebula 
\citep[e.g.,][]{2014prpl.conf..859C,2018MNRAS.475.2355M,2020SSRv..216...99M,2021SSRv..217....7L};
the existing measurements of the D/H ratio reveal that there is some trend/clustering as a function of the properties (e.g., mass and location) of the solar system bodies, 
which may reflect the formation site/mechanism of the bodies.

In this paper, therefore, we focus on the D/H ratios of Uranus and Neptune that are known to be a factor of $2-3$ higher than the solar nebula value \citep[e.g.,][]{2013A&A...551A.126F}. 
A similar effort was done by \citet{2018MNRAS.475.2355M}, where the canonical planetesimal accretion is investigated.
We expand their study, by considering a wide radius range of solids and the corresponding accretion modes;
different sizes of solids lead to different accretion modes (canonical planetesimal accretion vs drag-enhanced three-body vs pebble accretion).
We will compute below the resulting accretion rates as a function of both the solid radius and the solid surface density,
and show that if the minimum-mass solar nebula model is adopted,
accretion of solids with radius of $\sim 1$ m to $\sim 10$ km likely plays the dominant role in deuterium enrichment of Uranus and Neptune.
A tighter constraint can be obtained by explicitly considering the {\it whole} size distribution and the resulting abundances of solids with {\it all} the radii.

Exploring the link between solid accretion and the resulting atmospheric properties is very timely 
\citep[e.g.,][]{2016ApJ...832...41M,2019ApJ...876L..32H,2020A&A...634A..31V,2020MNRAS.498..680G}.
While it has been explored historically \citep[e.g.,][]{1986Icar...67..391B,1996Icar..124...62P},
a new momentum has recently brought in, thanks to the emergence of the pebble accretion scenario \citep{2010A&A...520A..43O,2012A&A...544A..32L};
as suggested in the literature before \citep[e.g.,][]{2004AJ....128.1348R}, 
small-sized bodies can accelerate the formation of planetary cores and potentially affect the metal enrichment of finally formed planets.

The plan of this paper is as follows.
In Section \ref{sec:d/h}, we summarize the currently known values of the D/H ratio of solar system bodies 
and discuss how these values can be used to constrain the solid distributions (core vs envelope) within Uranus and Neptune at the formation stage.
In Section \ref{sec:gas_solid}, we introduce the disk model used in this work and the gas drag force that affects the dynamics of solids within the gas disk.
We also provide the summary of what solid accretion mode becomes effective for a given radius of solids (see Table \ref{table1}).
The general reader may then proceed directly to Section \ref{sec:app_UN} for the application to Uranus and Neptune.
In Section \ref{sec:in-situ}, we build our formulation to explore how (proto)Uranus and (proto)Neptune accrete gas and solids simultaneously
and determine the dominant accretion mode as a function of solid radius.
The corresponding results are present in Section \ref{sec:res1}.
In Section \ref{sec:app_UN}, we apply the solid accretion model developed in Section \ref{sec:in-situ} to the formation of Uranus and Neptune,
and constrain what radius of solids is plausible for reproducing the current value of the D/H ratio of these planets,
as a function of the solid surface density.
Discussion about the assumptions and idealization adopted in this work is provided in Section \ref{sec:disc}.
Section \ref{sec:conc} is devoted to a concluding remark of this work. 

\section{The Properties of Uranus and Neptune} \label{sec:d/h}

We begin with the consideration of how the measured values of the D/H ratio of Uranus and Neptune can be used to constrain their formation mechanism.
We first summarize the known values of the D/H ratio for solar system bodies and then apply these values to computing the mass budget of Uranus and Neptune at their formation stage.

\subsection{The D/H ratio} \label{sec:d/h_1}

The D/H and other isotopic ratios are considered as one important tracer 
of how planets accrete gas and solids from their natal disks \citep[e.g.,][]{2018MNRAS.475.2355M,2020SSRv..216...99M}.
Accordingly, a number of measurements are available for solar system bodies \citep[e.g.,][references herein]{2015Sci...347A.387A}.
We here summarize them briefly.

The D/H ratio of the Sun is estimated from the measurement of the isotopic composition of helium in the solar wind \citep{1998SSRv...84..239G}.
This estimate is important because it would be comparable to the D/H ratio of the solar nebula ($\mbox{[D/H]}_{\rm SN}$).
\citet{1998SSRv...84..239G} infer that $\mbox{[D/H]}_{\rm SN} = (2.1 \pm 0.5) \times 10^{-5}$.

The D/H ratio of giant planets comes from the HD measurement of their atmospheres;
for Jupiter, its value is $(2.6 \pm 0.7) \times 10^{-5}$ \citep{1998SSRv...84..251M},
and for Uranus and Neptune, their values are $(4.4 \pm 0.4) \times 10^{-5}$ and $(4.1 \pm 0.4) \times 10^{-5}$, respectively \citep{2013A&A...551A.126F}.
The origins of these enhancements by a factor of about 2 (relative to the solar nebula) are the main target of this work.

The D/H ratio is measured for smaller bodies as well.
For them, measurements are done for water;
inner solar system bodies such as the Earth and water-rich meteorites exhibit the D/H ratio of $\sim (1.2-3.0) \times 10^{-4} $  
\citep[e.g.,][]{2012E&PSL.313...56M,2012Sci...337..721A}.
For outer solar system bodies like comets, the D/H ratio ranges from $\sim 2.0 \times 10^{-4}$ to $\sim 6.0 \times 10^{-4}$ 
\citep[e.g.,][]{1995JGR...100.5827B,2012A&A...544L..15B,2015Sci...347A.387A}.

This trend leads to the proposition that a deuterium enrichment profile is an increasing function of the distance from the Sun \citep[e.g.,][]{2014prpl.conf..859C}.

\subsection{Application to (proto)Uranus and (proto)Neptune} \label{sec:d/h_2}

We here apply the above D/H ratios to the formation of Uranus and Neptune.
To make the problem tractable, we adopt the following four assumptions;
the second one is the fundamental assumption of this work,
the first and third ones are reasonable, 
and the forth one would be acceptable.

First, we consider that the solar nebula consists of gas and solids
and assume that gas represents hydrogen and helium, and solids are the other heavy elements including water-ice.
In this work, they are denoted by $XY$ and $Z$, respectively.
Then, the mass ($M_{\rm p}$) of (proto)Uranus/(proto)Neptune is written as
\begin{equation}
\label{eq:M_p}
M_{\rm p}  = M_{XY} + M_Z,
\end{equation}
where $M_{XY}$ is the mass of the gaseous envelope around the planetary core, and $M_{Z}$ is the total mass of heavy elements in the planet.
As discussed in Section \ref{sec:d/h_3}, $M_Z$ should be decomposed into two terms:
\begin{equation}
\label{eq:M_Z}
M_Z \equiv M_{Z,\rm c} + M_{Z,\rm e},
\end{equation}
where $M_{Z,\rm c}$ is the core mass of the planet, 
and $M_{Z,\rm e}$ is the solid mass accreted after core formation and distributed in the envelope.

Second, we assume that the {\it current} value of the D/H ratio of Uranus/Neptune is the direct outcome of gas and solid accretion during the process of forming.
Then, the envelope's {\it bulk} D/H ratio ([D/H]$_{\rm p, e}$) of current Uranus/Neptune can be written as \citep[e.g.,][]{1996P&SS...44.1579L}:
\begin{equation}
\label{eq:D/H_ratio}
\mbox{[D/H]}_{\rm p, e}= \frac{ n_{{\rm H}_2} \mbox{[D/H]}_{{\rm H}_2} +  n_{{\rm H}_2 {\rm O,ice}} \mbox{[D/H]}_{{\rm H}_2 {\rm O,ice}} }{ n_{{\rm H}_2} + n_{{\rm H}_2 {\rm O,ice}} },
\end{equation}
where $n_{{\rm H}_2}$ and $n_{{\rm H}_2 {\rm O,ice}}$ are the number densities of molecular hydrogen and water-ice accreted onto the planet at its formation stage,
and [D/H]$_{{\rm H}_2}$ and [D/H]$_{{\rm H}_2 {\rm O,ice}}$ are their corresponding D/H ratios, respectively.
Note that $n_{{\rm H}_2} $ and $n_{{\rm H}_2 {\rm O,ice}} $ are determined by gas and solid accretion, respectively, 
and have the following relation:
\begin{equation}
\label{eq:ratio_n}
\frac{n_{{\rm H}_2 {\rm O,ice}}}{n_{{\rm H}_2}} = \frac{f_{{\rm H}_2 {\rm O,ice}} M_{Z,e}}{m_{{\rm H}_2 {\rm O}}}  \frac{m_{{\rm H}_2}}{f_{{\rm H}_2} M_{XY}},
\end{equation}
where $f_{{\rm H}_2}$ and $f_{{\rm H}_2 {\rm O,ice}}$ are the mass fractions of molecular hydrogen and water-ice in accreted materials, 
and $m_{{\rm H}_2}$ and $m_{{\rm H}_2 {\rm O}}$ are the mass of molecular hydrogen and water, respectively.
In the solar nebula, $f_{{\rm H}_2} \simeq 0.75$ \citep{2013A&A...551A.126F}.

Third, we assume that $\mbox{[D/H]}_{{\rm H}_2}$ should be comparable to $\mbox{[D/H]}_{\rm SN}$.
This assumption is reasonable because the main constituent of Jupiter is molecular hydrogen and its D/H ratio is similar to $\mbox{[D/H]}_{\rm SN}$ (Section \ref{sec:d/h_1}).

Fourth, we assume that solids accreted onto (proto)Uranus/(proto)Neptune are water-ice rich, 
and the D/H ratio of these solids is well represented by $\mbox{[D/H]}_{{\rm H}_2 {\rm O,ice}}$.
This is an acceptable assumption because the high abundance of water in Neptune's atmosphere can explain the detection of CO \citep{1996P&SS...44.1579L}.
Water is the most abundant volatile in comets  \citep[e.g., ][]{2011ARA&A..49..471M},
and therefore comets may be regarded as (the progenitor of) accreted solids.
Following \citet{2013A&A...551A.126F}, we adopt that $f_{{\rm H}_2 {\rm O,ice}}\simeq 0.72$, 
which corresponds to the ice-to-rock mass ratio of 2.5 \citep[e.g.,][]{2003ApJ...591.1220L}.

Under the above assumptions, equation (\ref{eq:D/H_ratio}) becomes
\begin{equation}
\label{eq:F_pe}
F_{\rm p, e}  \simeq \frac{ 1+ F_{{\rm H}_2 {\rm O,ice}} \frac{ M_{Z, \rm e}}{10 M_{XY}} }{ 1+ \frac{M_{Z, \rm e}}{10M_{XY}} } \simeq 2,
\end{equation}
where the following notations are used for brevity:
\begin{eqnarray}
F_{{\rm H}_2 {\rm O,ice}} = \frac{\mbox{[D/H]}_{{\rm H}_2 {\rm O,ice}}}{\mbox{[D/H]}_{\rm SN}}, \\
F_{\rm p, e}                      = \frac{\mbox{[D/H]}_{\rm p, e}}{\mbox{[D/H]}_{\rm SN}}. \nonumber 
\end{eqnarray}
These represent the deuterium enrichment factor for each object.
Equation (\ref{eq:F_pe}) is re-written as
\begin{equation}
\label{eq:Mz_Mgas}
 \frac{M_{Z, \rm e}}{M_{XY}} = \frac{10}{F_{{\rm H}_2 {\rm O,ice}} -2}.
\end{equation}
Thus, the solid mass accreted after core formation ($M_{Z, \rm e}$) can be computed 
for given values of the envelope mass ($M_{XY}$) and the deuterium enrichment factor of accreted solids ($F_{{\rm H}_2 {\rm O,ice}}$).

In the following, we consider what the mass distribution of solids within Uranus and Neptune at their formation stage looks like, using equation (\ref{eq:Mz_Mgas}).

\subsection{Mass budget} \label{sec:d/h_3}

As discussed in the above section, 
the mass distribution ($M_{XY}$, $M_{Z, \rm c}$, and $M_{Z, \rm e}$) of Uranus/Neptune at their formation stage is  constrained 
by equations (\ref{eq:M_p}), (\ref{eq:M_Z}), and (\ref{eq:Mz_Mgas}) for a given value of $F_{{\rm H}_2 {\rm O,ice}}$.
In order to fully determine the values of $M_{XY}$, $M_{Z, \rm c}$, and $M_{Z, \rm e}$,
one more constraint and explicit specification of  $F_{{\rm H}_2 {\rm O,ice}}$ are necessary.

In this work, we use the results of interior models which infer that $M_Z/M_{\rm p} \simeq 0.76-0.90$ 
\citep[e.g.,][]{2011ApJ...726...15H,2013P&SS...77..143N}.
For $F_{{\rm H}_2 {\rm O,ice}}$,  we parameterize its value ranging from 6 to 30.
The parameterization is motivated both from the existing measurements of the D/H ratio of small solar system bodies (Section \ref{sec:d/h_1})
and from the recent recognition that solar system formation is highly dynamic and protoplanets underwent radial migration during the process of forming 
\citep[e.g.,][]{1999Natur.402..635T,2005Natur.435..459T,2011Natur.475..206W}.
In other words, the current locations of planets do not necessarily correspond to their formation sites.
Accordingly, it is plausible to consider a wide range of $F_{{\rm H}_2 {\rm O,ice}}$.

We are now in a position to compute the values of $M_{XY}$, $M_{Z, \rm c}$, and $M_{Z, \rm e}$.
We first confirm that $M_Z$ should be decomposed (i.e., equation (\ref{eq:M_Z})).
This is simply because if $M_Z=M_{Z, \rm e}$, then 
\begin{equation}
\frac{M_Z}{M_{\rm p}} = \frac{10}{F_{{\rm H}_2 {\rm O,ice}} +8}.
\end{equation}
For the range of $6 \la F_{{\rm H}_2 {\rm O,ice}} \la 30$,
the resulting values ($0.26 \la M_Z/M_{\rm p} \la 0.71$) become smaller than the predictions made by interior models.

\begin{figure}
\begin{center}
\includegraphics[width=8.5cm]{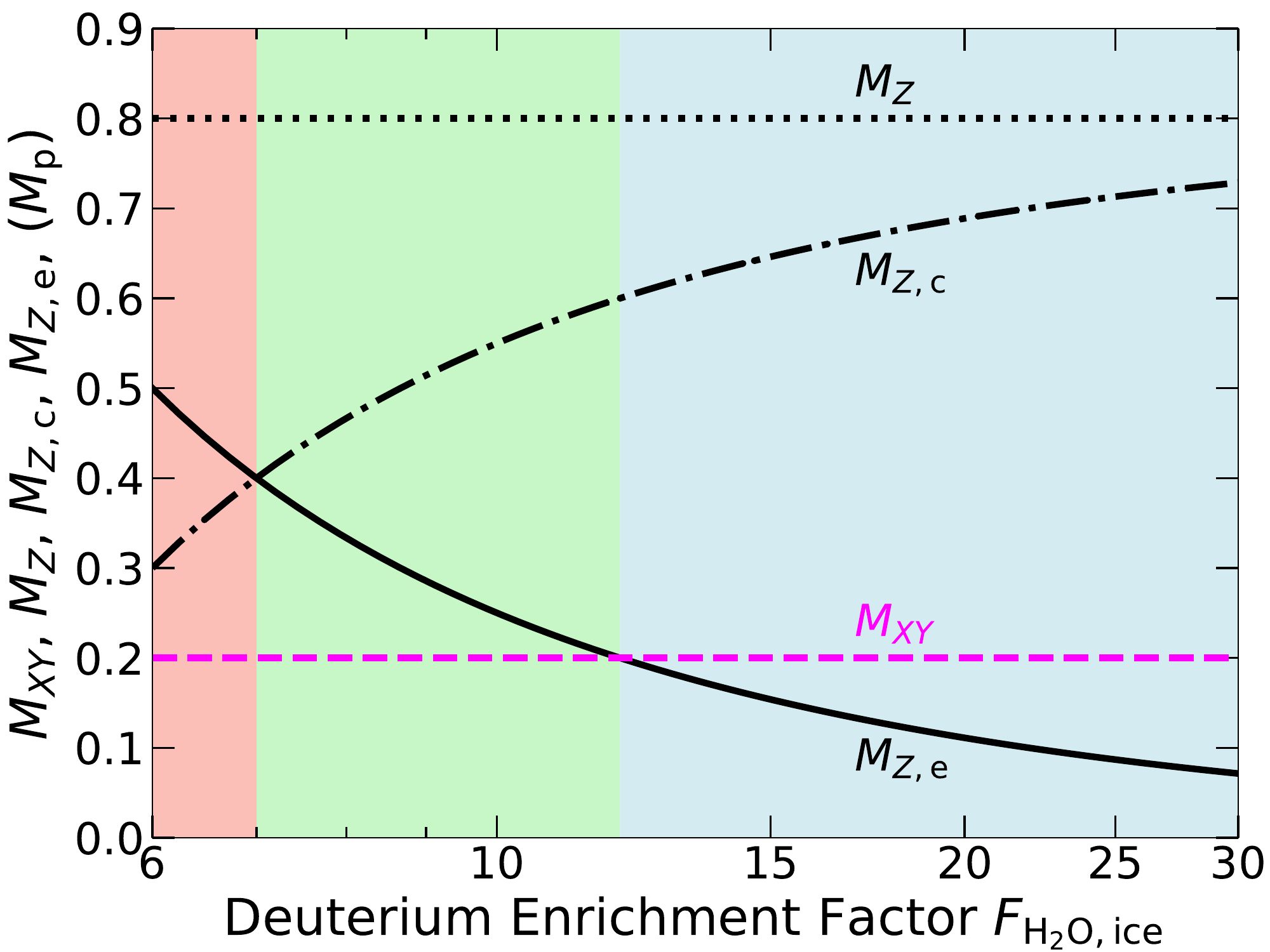}
\caption{The mass distribution of gas and solids within Uranus and Neptune at their formation stage, as a function of $F_{{\rm H}_2 {\rm O,ice}}$.
The envelope mass ($M_{XY}$) is denoted by the magenta dashed line,
the total solid mass ($M_Z$) is by the black dotted line,
the solid mass within the planetary core ($M_{Z, \rm c}$) is by the black dash-dotted line,
and the solid mass accreted after core formation and distributed in the envelope ($M_{Z, \rm e}$) is by the black solid lines.
The variation of $M_{Z, \rm c}$ and $M_{Z, \rm e}$ divides the parameter space into three regions (see the three shaded regions):
Case 1: $M_{XY} < M_{Z, \rm c} < M_{Z, \rm e}$ (the red shaded region), 
Case 2: $M_{XY} <  M_{Z, \rm e} < M_{Z, \rm c}$ (the green shaded region), 
and Case 3: $ M_{Z, \rm e} < M_{XY}  < M_{Z, \rm c}$ (the blue shaded region).
The additional description for each case is provided in the main text.}
\label{fig1}
\end{center}
\end{figure}

We then compute the mass distribution of solids within (proto)Uranus/(proto)Neptune, using the assumption that $M_Z/M_{\rm p} \simeq 4/5$.
Figure \ref{fig1} shows the resulting values of $M_{Z, \rm e}$ and $M_{Z, \rm c}$ as a function of $F_{{\rm H}_2 {\rm O,ice}}$.
As expected, the value of $M_{Z, \rm e}$ decreases with increasing the value of $F_{{\rm H}_2 {\rm O,ice}}$.
There are three possible cases to reproduce the current values of the D/H ratio of Uranus and Neptune:
Case 1: $M_{XY} < M_{Z, \rm c} < M_{Z, \rm e}$ (the red shaded region),  
Case 2: $M_{XY} <  M_{Z, \rm e} < M_{Z, \rm c}$ (the green shaded region), 
and Case 3: $ M_{Z, \rm e} < M_{XY}  < M_{Z, \rm c}$ (the blue shaded region).
For Case 1, the deuterium enrichment factor of accreted solids is low.
As a result, a large amount of solids need to be accreted to reproduce the current value of the D/H ratio.
This in turn dictates reduction of the core mass since the total solid mass is fixed.
For Case 2, the deuterium enrichment factor of accreted solids becomes higher,
and the contribution of accreted solids to the total solid mass becomes lower than the core mass.
For Case 3, accreted solids are further enriched in deuterium, and hence their mass contribution becomes even less.

In this work, we consider Case 3 mainly; 
previous studies suggest that formation sites of Uranus and Neptune may be $\ga 10$ au \citep[e.g.,][]{2005Natur.435..459T},
at which the deuterium enrichment factor of solids may be $\ga 10$ \citep[e.g.,][]{2011ApJ...734L..30K}.

Note that the adopted values of the ice-to-rock ratio and $M_Z/M_{\rm p}$ lead to the rock mass of $0.22 M_{\rm p}$;
equivalently, $\sim 3.2 M_{\oplus}$ for Uranus and  $\sim 3.8 M_{\oplus}$ for Neptune, 
where $M_{\oplus}$ is the Earth mass.
It is obvious that the rock mass becomes larger if a smaller value of the ice-to-rock ratio is used.

In the following sections, 
we determine what size of solid particles need to be accreted onto (proto)Uranus and (proto)Neptune 
as a function of the solid surface density of the solar nebula,
to reproduce the mass distributions of solids constrained by their D/H ratios (see Figure \ref{fig1}).

\section{The Properties of Gas and Solids in the Solar Nebula} \label{sec:gas_solid}

Before considering solid accretion onto (proto)Uranus and (proto)Neptune,
we here describe the properties of gas and solids in the solar nebula.
Given the dynamical picture of solar system formation,
the birth places of Uranus and Neptune may be much closer to the Sun than the current locations.
In this section, the position of $r=10$ au is picked as a reference value to compute the nebular properties.
A parameter study is conducted in Section \ref{sec:app_UN_param}.

\subsection{Disk model}

We adopt the so-called minimum mass solar nebula (MMSN) model \citep[e.g.,][]{1981PThPS..70...35H}
that is widely used in the literature to explore the formation of both the solar and extrasolar systems.

The MMSN model describes the gas ($\Sigma_{\rm d, g}$) and solid ($\Sigma_{\rm d, s}$) surface density profiles as
\begin{eqnarray}
\label{eq:sigma_g}
\Sigma_{\rm d, g} & \simeq & 5.4 \times 10 f_{\rm d,g} \left( \frac{r}{10 \mbox{ au}}\right)^{-3/2} \mbox{ g cm}^{-2}, \\
\label{eq:sigma_d}
\Sigma_{\rm d, s}    & \simeq  &  7.7 \times 10^{-1} f_{\rm d,s} \left( \frac{r}{10 \mbox{ au}}\right)^{-3/2} \mbox{ g cm}^{-2}, 
\end{eqnarray}
where $f_{\rm d,g}$ and  $f_{\rm d,s} $ are the gas and solid enhancement/reduction factors, respectively.
Hereafter, the suffix "d" represents the nebular disk.
In this paper, we set that $f_{\rm d,g}=1$ and use $f_{\rm d,s}$ as a parameter.
Note that in order to account for the enhancement of solids due to the formation of water-ice,
$\Sigma_{\rm d,s}$ is increased by a factor of $\sim 4.2$ at $r \geq 2.7$ au in the MMSN model.
The transition location is known as the snow (or ice) line.
The recent measurements of the solar elemental abundances imply that 
the value of this factor is likely overestimated \citep{2009ARA&A..47..481A}.
With the adopted value of the ice-to-rock ratio (Section \ref{sec:d/h_2}), the enhancement becomes 3.5.
We therefore increase $\Sigma_{\rm d, s}$ by that value in equation (\ref{eq:sigma_d});
additional solid enhancement/reduction will be taken into account, by changing the value of $f_{\rm d,s}$.

The nebular (gas and dust) temperature at the midplane region is written, under the optically thin assumption, as
\begin{equation}
T_{\rm d} \simeq 8.9 \times 10 \left( \frac{r}{10 \mbox{ au}}\right)^{-1/2} \mbox{ K}.
\end{equation}

Accordingly, other nebular quantities at the midplane region such as the gas volume density ($\rho_{\rm d, g}$), the sound speed ($c_{\rm s}$), 
the aspect ratio ($h_{\rm d}$), and the quantity charactering the pressure gradient of the nebular gas ($\eta_{\rm d}$)
are given as
\begin{eqnarray}
\rho_{\rm d, g} & =  &\frac{\Sigma_{\rm d,g}}{\sqrt{2 \pi} H_{\rm g}} \simeq 2.4 \times 10^{-12} f_{\rm d,g}  \left(  \frac{r}{10 \mbox{ au}} \right)^{-11/4} \mbox{ g cm}^{-3}, \\
c_{\rm s} & = & \sqrt{\frac{k_{\rm B} T_{\rm d}}{m_{\rm g}}} \simeq 5.6 \times 10^{-1} \left(  \frac{r}{10 \mbox{ au}} \right)^{-1/4} \mbox{ km s}^{-1}, \\
h_{\rm d} & =  & \frac{c_{\rm s}}{ v_{\rm Kep}}  \simeq 5.9 \times 10^{-2} \left(  \frac{r}{10 \mbox{ au}} \right)^{1/4}, \\
\eta_{\rm d} & = & - \frac{1}{2}  \left( \frac{c_{\rm s}}{ v_{\rm Kep}} \right)^2 \frac{d \ln P_{\rm d}}{d \ln r}  \simeq 5.7 \times 10^{-3} \left(  \frac{r}{10 \mbox{ au}} \right)^{1/2},
\end{eqnarray}
where $H_{\rm g} = c_{\rm s} / \Omega$ is the gas pressure scale height, $\Omega= \sqrt{GM_{\odot}/r^3}$ is the orbital angular frequency around the Sun,
$k_{\rm B}$ is the Bolzmann's constant, $m_{\rm g}=2.34$ is the mean molecular weight of the nebular gas in the atomic mass unit, 
$v_{\rm Kep}= r \Omega$ is the Keplerian velocity, and $P_{\rm d}= \rho_{\rm d, g} c_{\rm s}^2$ is the gas pressure.

\subsection{Drag force}

The dynamics of solids is fundamental for determining the solid accretion rate onto (proto)planets.
When the motion of the nebular gas is modeled effectively by the laminar flow (equivalently, turbulent motion is very weak), 
aerodynamical drag arising from the nebular gas plays an important role for solid dynamics.
The resulting drag force can be classified into several regimes.
In this work, we adopt such an assumption and consider three regimes that are well explored in the literature \citep[e.g.,][]{2010A&A...520A..43O}.

The corresponding stopping (or friction) time ($t_{\rm s}$) can be given as 
\citep{1976PThPh..56.1756A,1977MNRAS.180...57W}
\begin{equation}
t_{\rm s} =  \left\{ 
                           \begin{array} {l}
                                                 { \displaystyle \frac{\rho_{\rm s} s}{\rho_{\rm d, g} v_{\rm th} }   }                    (s< \frac{9}{4}l_{\rm mfp}) \\
                                                 { \displaystyle \frac{8\rho_{\rm s} s}{3 C_{\rm d}\rho_{\rm d, g} v_{\rm rel}^{\rm d-s} }}  (s> \frac{9}{4}l_{\rm mfp}), 
                           \end{array} 
                     \right.
\end{equation}
where $\rho_{\rm s}=2$ g cm$^{-3}$ and $s$ are the bulk density and the radius of solid particles, respectively,
$l_{\rm mfp} = m_{\rm g}/(\sigma_{\rm mol} \rho_{\rm d,g})$ is the mean free path of gas particles,
$\sigma_{\rm mol}= 2 \times 10^{-15}$ cm$^2$ is the collisional cross-section between molecular gas particles \citep{1986Icar...67..375N}, 
$v_{\rm th} = \sqrt{8/\pi}c_{\rm s}$ is the mean thermal velocity of the nebular gas,
$v_{\rm rel}^{\rm d-s}$ is the relative velocity between the nebular gas and the solid particles,
and $C_{\rm d}$ is the drag coefficient.
For the Stokes regime,
we adopt that $C_{\rm d} = 24 / \mbox{Re}_{\rm p}$, following \citet{1977MNRAS.180...57W}, 
where $\mbox{Re}_{\rm p} = 2 s v_{\rm rel}^{\rm d-s}/ \nu_{\rm mol}$ is the particle Reynolds number,
and $\nu_{\rm mol}=l_{\rm mfp} v_{\rm th}/2$ is the molecular viscosity of the nebular gas.
For the Quadratic regime, 
we use that $C_{\rm d} = 2$, following \citet{1976PThPh..56.1756A}.

Consequently, $t_{\rm s}$ is written as 
\begin{equation}
\label{eq:t_s}
t_{\rm s} =  \left\{ 
                           \begin{array} {l}
                                                 { \displaystyle \frac{\rho_{\rm s} s}{\rho_{\rm d, g} v_{\rm th} }   }                          (s< \frac{9}{4}l_{\rm mfp}; \mbox{ the Epstein regime}) \\
                                                 { \displaystyle \frac{4\rho_{\rm s} s^2}{ 9 \rho_{\rm d, g} v_{\rm th} l_{\rm mfp} } }   \\
                                                 {\displaystyle   (\frac{9}{4}l_{\rm mfp} < s < 3 l_{\rm mfp} \frac{ v_{\rm th}}{v_{\rm rel}^{\rm d-s}} };   \mbox{ the Stokes regime}) \\
                                                 { \displaystyle \frac{4\rho_{\rm s} s}{3 \rho_{\rm d, g} v_{\rm rel}^{\rm d-s} } }           
                                                 { \displaystyle (3 l_{\rm mfp} \frac{ v_{\rm th}}{v_{\rm rel}^{\rm d-s}}< s  }; \mbox{ the Quadratic regime}). \\
                           \end{array} 
                     \right.
\end{equation}

In the following calculations, we use either $t_{\rm s}$ or the dimensionless Stokes number (St$=t_{\rm s} \Omega$).

\subsection{Characteristic radii of solid particles} \label{sec:chara_radius}

The formulae of $t_s$ (equation (\ref{eq:t_s})) allow one to determine two characteristic radii of solid particles.
In addition, several characteristic radii can be defined in our problem.
We here summarize them in order to elucidate the problem under consideration (see Table \ref{table1}).

\begin{table*}
\begin{minipage}{17cm}
\centering
\caption{The characteristic radii of solid particles that involve with deuterium enrichment of Uranus and Neptune}
\label{table1}
{\scriptsize
\begin{tabular}{c||c|c|c|c|c|c|c} 
\hline
Regime          & \multicolumn{3}{c|}{Epstein}                                                                                                         & Stokes                                 & \multicolumn{3}{c}{Quadratic}                  \\ \hline \hline
Name             & Pebble                                   &  Boulder                                     & Small fragment                        & Medium fragment               & Large fragment & Comet  & Planetesimal      \\ \hline
Size                & $s \la 34$ cm                        & 34 cm $\la s \la 4.1$ m              & 4.1m $\la s \la 18$ m               & 18 m $\la s \la 46$ m             & 46 m $\la s \la 1$ km  &  1 km $\la s \la 110$ km & 110km $\la s $            \\
                      & equation (\ref{eq:s_pa_Ep}) & equation (\ref{eq:s_nogap_Ep}) & equation (\ref{eq:s_max_Ep})  & equation (\ref{eq:s_max_St})      & & &   equation (\ref{eq:s_Qu_max})    \\ \hline
Eccentricity     &   \multicolumn{2}{c|}{shear dominated}                                     & \multicolumn{5}{c}{dispersion dominated}                                                                                     \\ \hline
Accretion        & $\sim$ Hill radius                  &  \multicolumn{6}{c}{Larger than the core radius}                                                                          \\
radius             &                                              &  \multicolumn{5}{c}{}                                                                                                                  &                               \\   \hline
Accretion        & Pebble                                  &  \multicolumn{5}{c|}{Drag-enhanced three-body}                                                                      & Two-body               \\
mode              & accretion                               &  \multicolumn{5}{c|}{}                                                                                                                  &                               \\ \hline
Section treated & Section \ref{sec:acc_peb}  &  Section \ref{sec:acc_bou}       &  Section \ref{sec:acc_sma_frag} &  Section \ref{sec:acc_mid_frag}   &  \multicolumn{2}{c|}{Section \ref{sec:acc_lar_com}}    &  Section \ref{sec:acc_pla}   \\ \hline
Key equation &  equation (\ref{eq:Mdot_Z_pa})  &  equation (\ref{eq:Mdot_Z_3b}) &  equation (\ref{eq:Mdot_Z_3b_Ep}) &   equation (\ref{eq:Mdot_Z_3b_St})   &  \multicolumn{2}{c|}{equation (\ref{eq:Mdot_Z_3b_Qu})}    &   equation (\ref{eq:Mdot_Z_disp})   \\ 
\hline                                  
\end{tabular}
}
\end{minipage}
\end{table*}

The first characteristic radius ($s_{\rm max}^{\rm Ep}$) is defined at the transition, where the drag force switches from the Epstein to the Stokes regime.
This radius is computed readily when the nebular model is chosen,
which is given as
\begin{equation}
\label{eq:s_max_Ep}
s_{\rm max}^{\rm Ep} = \frac{9}{4} l_{\rm mfp} \simeq 1.8 \times 10^{3}  f_{\rm d,g}^{-1} \left(  \frac{r}{10 \mbox{ au}} \right)^{11/4} \mbox{ cm}.
\end{equation}
Thus, the solid particles with the size of $s\la$ 18m are in the Epstein regime in our nebular model.

The second one is the transition radius ($s_{\rm max}^{\rm St}$) that changes from the Stokes to the Quadratic regime.
This radius is a function of $v_{\rm rel}^{\rm d-s}$, and hence depends more on the problem setup.
In this work, we investigate solid accretion onto planetary cores under
the assumption that the formation of the cores is already complete (see Section \ref{sec:in-situ_1}).
For this case, we find that the velocity of solid particles around the cores is determined mainly by the gravitational force arising from the cores (Section \ref{sec:solid_ecc}),
and thereby $v_{\rm rel}^{\rm d-s} \approx v_{\rm rel}^{\rm c-s}$, 
where $v_{\rm rel}^{\rm c-s}$ is the relative velocity between the cores and the solid particles.
As shown in Section \ref{sec:solid_ecc}, the corresponding, equilibrium eccentricity ($e^{\rm Qu}_{\rm eq} \simeq v_{\rm rel}^{\rm c-s}/v_{\rm Kep}$) is computed self-consistently,
by equating its damping timescale ($4\rho_{\rm s} s / (3 \rho_{\rm d, g} v_{\rm rel}^{\rm c-s}$) in equation (\ref{eq:t_s})) 
with its excitation timescale (equation (\ref{eq:t_ext})).
Consequently, $s_{\rm max}^{\rm St}$ is written as (see equation (\ref{eq:e_Qu_eq}))
\begin{eqnarray}
\label{eq:s_max_St}
s_{\rm max}^{\rm St}  & = & \left[  3 \sqrt{\frac{8}{\pi}}   \left( \frac{9}{2\sqrt{3}} \right)^{-1/5}  \frac{m_{\rm g} h_{\rm d}}{\sigma_{\rm mol} \rho_{\rm d, g} h_{\rm H}} 
                                     \left( \frac{\rho_{\rm s}}{\rho_{\rm d, g} r} \right)^{-1/5} \right]^{5/6}  \\ \nonumber
                   & \simeq & 4.6 \times 10^{3} f_{\rm d, g}^{-2/3} \left( \frac{M_{Z, \rm c}}{10 M_{\oplus}} \right)^{-5/18} \left(  \frac{r}{10 \mbox{ au}} \right)^{53/24} \mbox{ cm},
\end{eqnarray}
where $h_{\rm H}= (M_{Z, \rm c}/(3M_{\odot}))^{1/3}$ is the reduced Hill radius.
Accordingly, the solid particles with the size range of 18m $\la s \la$ 46m are in the Stokes regime in our setup.

We now consider other characteristic radii of solid particles.
These radii can be defined, based on their dynamical properties.
One important property is the eccentricity ($e$) of solid particles.
To proceed, we introduce the Hill eccentricity ($e_{\rm H} = e/h_{\rm H}$).
When $e_{\rm H} < 1$, the particles are in the so-called shear dominated regime, 
and $v_{\rm rel}^{\rm c-s}$ is determined by the Keplerian shear.
On the other hand, when $e_{\rm H} > 1$, the particles are in the so-called dispersion dominated regime,
and $v_{\rm rel}^{\rm c-s}$ is controlled by the eccentricity of the particles.
This distinction becomes important in this work,
because it can be used to specify which formula of solid accretion rates will be adopted (equation (\ref{eq:Mdot_Z_disp}) vs equation (\ref{eq:Mdot_Z_shear})).
As shown in Section \ref{sec:solid_rate}, the critical eccentricity ($e_{\rm H} =1$) is readily computed (see equation (\ref{eq:s_nogap_Ep2})).
The corresponding radius ($s_{\rm shear}^{\rm Ep}$) in the Epstein regime is given as 
\begin{eqnarray}
\label{eq:s_nogap_Ep}
s_{\rm shear}^{\rm Ep} & =         & \frac{8 \sqrt{3}}{27} \frac{\rho_{\rm d,g} v_{\rm th}}{\rho_{\rm s} h_{\rm H} \Omega} \\ \nonumber
                                       & \simeq & 4.1 \times 10^{2} f_{\rm d, g} \left( \frac{M_{Z, \rm c}}{10 M_{\oplus}} \right)^{-1/3} \left(  \frac{r}{10 \mbox{ au}} \right)^{-3/2} \mbox{ cm}.
\end{eqnarray}
Consequently, the dynamics of the particles with the size of $\ga 4.1$ m is in the dispersion dominated regime,
and the gravitational interaction from the planetary cores plays an important role in determining the eccentricity of these planets.
This justifies the assumption adopted in deriving $s_{\rm max}^{\rm St}$ (see equation  (\ref{eq:s_max_St})).

Another dynamical property that can be used to define characteristic radii is how effective the drag force is to affect the orbit of solid particles.
If the drag force is so strong that the orbit of solid particles is not approximated well by a Keplerian orbit,
the so-called pebble accretion comes into play.
\citet{2012ApJ...747..115O} show that
the Keplerian orbit approximation holds well for the particles whose the Stokes number is larger than 2.
We find that in our problem setup, 
the corresponding radius ($s_{\rm pe}^{\rm Ep}$) resides in the Epstein regime and is written as (see equation (\ref{eq:t_s}))
\begin{equation}
\label{eq:s_pa_Ep}
s_{\rm pe}^{\rm Ep}  = 2 \frac{\rho_{\rm d,g} v_{\rm th}}{\rho_{\rm s} \Omega}  \simeq  3.4 \times 10 f_{\rm d, g} \left(  \frac{r}{10 \mbox{ au}} \right)^{5/4} \mbox{ cm}.
\end{equation}
Thus, pebble accretion becomes effective for the particles with the size of $s\la$ 34 cm in our setup.

Even if the orbit of solid particles is approximated well by a Keplerian orbit,
the gas drag force can enhance their accretion rate onto (proto)planets \citep{2010A&A...520A..43O}.
We label the transition radius as $s^{\rm Qu}_{\rm 3b, max}$.
As shown in Section \ref{sec:acc_pla}, the corresponding calculation is straightforward, but relatively cumbersome and dependent on some model parameters.
We therefore list up only the value as
\begin{equation}
\label{eq:s_Qu_max}
s^{\rm Qu}_{\rm 3b, max} \la 1.1 \times 10^2 \mbox{ km},
\end{equation}
which is obtained from the self-consistent calculation for the case that $M_{Z, \rm c} = 10 M_{\oplus}$.

In the followings, we use the nomenclature summarized in Table \ref{table1} and explore solid accretion in each regime individually.

\section{Accretion of Neptune-mass planets} \label{sec:in-situ}

We here explore accretion onto Neptune-mass planets.
We first discuss our problem setup with the adopted assumptions.
We then summarize our formulation to compute the gas and solid accretion rates onto Neptune-mass planets.
Our results are present in Section \ref{sec:res1}.

\subsection{Problem setup} \label{sec:in-situ_1}

We here describe our problem setup and summarize assumptions used in this work.

We adopt the core accretion picture;
core accretion is the leading model of giant planet formation, which is widely used in the literature \citep[e.g.,][]{1996Icar..124...62P,2014prpl.conf..643H}.
In this scenario, the formation of planets is divided mainly into two stages: core formation and gas accretion onto cores.
It is well recognized historically that the additional accretion of solids during the gas accretion stage is 
the key process for better understanding the metal enrichment of giant planets
both in the solar and extrasolar systems \citep[e.g.,][]{1996Icar..124...62P,2016ApJ...832...41M,2018ApJ...865...32H}.
The importance of such accretion is also supported by the current D/H ratio of Uranus and Neptune (Section \ref{sec:d/h_3}).
With this picture, we adopt the following four assumptions to make the problem tractable in our simplified calculations.

First, we assume that a planetary core already forms and is surrounded by a swarm of solids in the gas-rich solar nebula;
equivalently, we explore gas and solid accretion onto the core after core formation is complete.
This is the fundamental assumption of this work.

Second, we assume that a contracting envelope around the planetary core is radiative.
While this assumption may not be realistic,
it leads to self-consistent results within our framework.
The additional discussion is provided in Appendix \ref{app_env}.

Third, we assume that the gas and dust opacities of the planetary envelope do not change by gas and solid accretion.
For gas opacity, it would be acceptable because we mainly consider Case 3 in Figure \ref{fig1} (that is, $M_{XY} > M_{Z, \rm e}$).
For this case, the mean molecular weight of the envelope gas is increased by less than a factor of 2,
even if all the accreted solids evaporate in the planetary envelope.
We have confirmed that such an increase changes the envelope gas density only by a factor of few (Appendix \ref{app_env}),
which would be acceptable in our simplified calculations.
As described in Section \ref{sec:in-situ_2}, the gas opacity does not affect the gas accretion timescale in our formulation.
For dust opacity, it is well recognized in the literature that it changes significantly in planetary envelopes 
\citep{2010Icar..209..616M,2014A&A...566A.141M,2014ApJ...789L..18O}.
However, the actual value is uncertain, and hence we parameterize it.
We find that in our formulation,
it regulates the gas accretion timescale, but not the envelope structure (Section \ref{sec:solid_radius}).

Fourth, we assume that concurrent accretion of gas and solids occurs in-situ.
In this and next sections, the position of a protoplanet is set at $r_{\rm p}=10$ au.
We conduct a parameter study about $r_{\rm p}$ in Section \ref{sec:app_UN_param}.

\subsection{Gas accretion} \label{sec:in-situ_2}

We start off from the examination of gas accretion.
Suppose that a planetary core with the mass of $M_{Z,\rm c}$ forms at a certain location ($r= r_{\rm p}$).
Then, the core undergoes gas accretion when it becomes more massive than the so-called critical core mass 
\citep[$M_{Z, \rm c}^{\rm crit}$, e.g., ][]{1980PThPh..64..544M,1982P&SS...30..755S,1986Icar...67..391B}:
\begin{equation}
\label{eq:M_c_crit}
M_{Z, \rm c}^{\rm crit} \simeq 10 M_{\oplus}  \left( \frac{ \kappa_{\rm grain} }{ \kappa_{\rm grain}^{\rm ISM} } \right)^a \left( \frac{ \dot{M}_{Z, \rm c} }{1 0^{-6} M_{\oplus} \mbox{yr}^{-1} } \right)^b,
\end{equation}
where we adopt that $a \simeq b \simeq 0.25$, following \citet{2000ApJ...537.1013I}, 
$ \kappa_{\rm grain}$ is the dust opacity of the envelope around the core,
and $\dot{M}_{Z, \rm c}$ is the solid accretion rate onto the core.
In the above equation, $ \kappa_{\rm grain}$ is normalized by $\kappa_{\rm grain}^{\rm ISM} = 1$ g cm$^{-2}$ that corresponds to the ISM value \citep{1996Icar..124...62P}.
Following the third assumption in this section, the dust opacity in the planetary envelope is parameterized, using a $f-$factor:
\begin{equation}
f_{\rm grain} = \frac{ \kappa_{\rm grain} }{ \kappa_{\rm grain}^{\rm ISM} }.
\end{equation}
We hereafter refer to $f_{\rm grain}$ as the dust opacity factor.
As shown in Section \ref{sec:app_UN_gas}, $f_{\rm grain}$ should have the range from $\sim 0.1$ to $\sim 10$ 
to reproduce the estimated value of $M_{XY}$ for Uranus and Neptune within the gas disk lifetime (Figure \ref{fig9}).
Therefore, we here consider that $f_{\rm grain} = 0.1, 1,$ and 10 as examples.

The onset of gas accretion is determined by the condition that $M_{Z, \rm c} \ga M_{Z, \rm c}^{\rm crit}$;
if the solid accretion rate onto the core is high enough that  $M_{Z, \rm c} \simeq M_{Z, \rm c}^{\rm crit}$,
then the envelope can maintain hydrostatic equilibrium due to the heat generated by the gravitational energy released from infalling solids.
We refer to such a critical solid accretion rate as 
\begin{equation}
\label{eq:dotM_c_crit}
\dot{M}_{Z, \rm c}^{\rm crit} \equiv \dot{M}_{Z, \rm c} \mbox{ when } M_{Z, \rm c} \simeq M_{Z, \rm c}^{\rm crit}.
\end{equation}
On the other hand,
if $M_{Z, \rm c} \ga M_{Z, \rm c}^{\rm crit}$, the pressure gradient of the envelope cannot support its gas against the core's gravity,
even if the heat from infalling solids is taken into account.
Consequently, the hydrostatic configuration no longer exists, 
and the envelope starts contracting.
This contraction leads to the gas flow from the surrounding disk to the Bondi radius of the core within which
the inflow gas is bound to the core.
Accordingly, the gas accretion rate onto the core is essentially comparable to the contraction rate.
During contraction, the heat generated by the contraction itself can support the envelope gas,
so that envelope contraction proceeds in a quasi-hydrostatic fashion.

The resulting gas accretion timescale is given as, 
also known as the Kelvin-Helmholtz timescale \citep[e.g.,][]{2000ApJ...537.1013I}:
\begin{equation}
\label{eq:tau_KH}
\tau_{\rm KH} = 10^{c} f_{\rm grain} \left( \frac{M_{Z, \rm c}}{ 10 M_{\oplus} } \right)^{-d} \mbox{yr},
\end{equation}
where we set that $c=7$ and $d=4$, following \citet{1997Icar..126..282T}.

This phase of gas accretion continues until the accretion rate exceeds the rate of gas supply from the surrounding protoplanetary disk.
This transition is characterized by the condition that $\tau_{\rm KH}$ becomes comparable to the timescale of $\tau_{\rm hydro} (\equiv M_{Z, \rm c}/ \dot{M}_{XY, \rm hydro}$),
where \citep{2016ApJ...823...48T}
\begin{eqnarray}
\label{eq:Mpdot_hydro}
\dot{M}_{XY, \rm hydro} & =          &  0.29 h_{\rm d}^{-2}  \left( \frac{M_{Z, \rm c}}{M_*} \right)^{4/3} \Sigma_{\rm d,g} r^2 \Omega  \\ \nonumber
                                   & \simeq &  3.1 \times 10^{-3} f_{\rm d,g}  \left( \frac{M_{Z, \rm c}}{10 M_{\oplus}} \right)^{4/3}  \left(  \frac{r}{10 \mbox{ au}} \right)^{-3/2}
                                       \frac{M_{\oplus}}{\mbox{yr}}.
\end{eqnarray}
The above equation is derived from two-dimensional hydrodynamical simulations \citep{2002ApJ...580..506T}.
When $\tau_{\rm KH} < \tau_{\rm hydro}$,
gas accretion onto the core is regulated by disk evolution (i.e., $\dot{M}_{XY, \rm hydro}$).

\begin{figure*}
\begin{minipage}{17cm}
\begin{center}
\includegraphics[width=8.3cm]{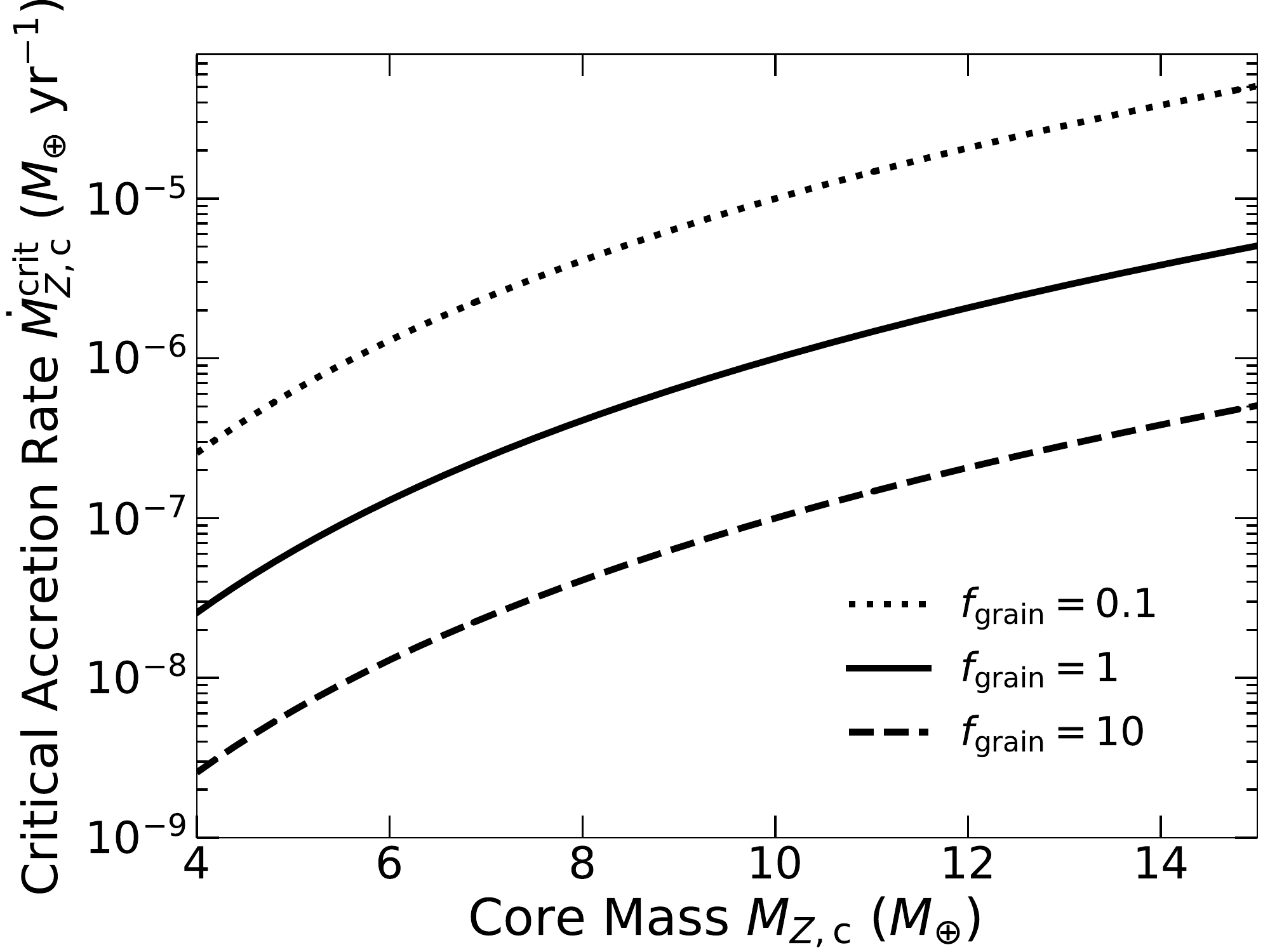}
\includegraphics[width=8.3cm]{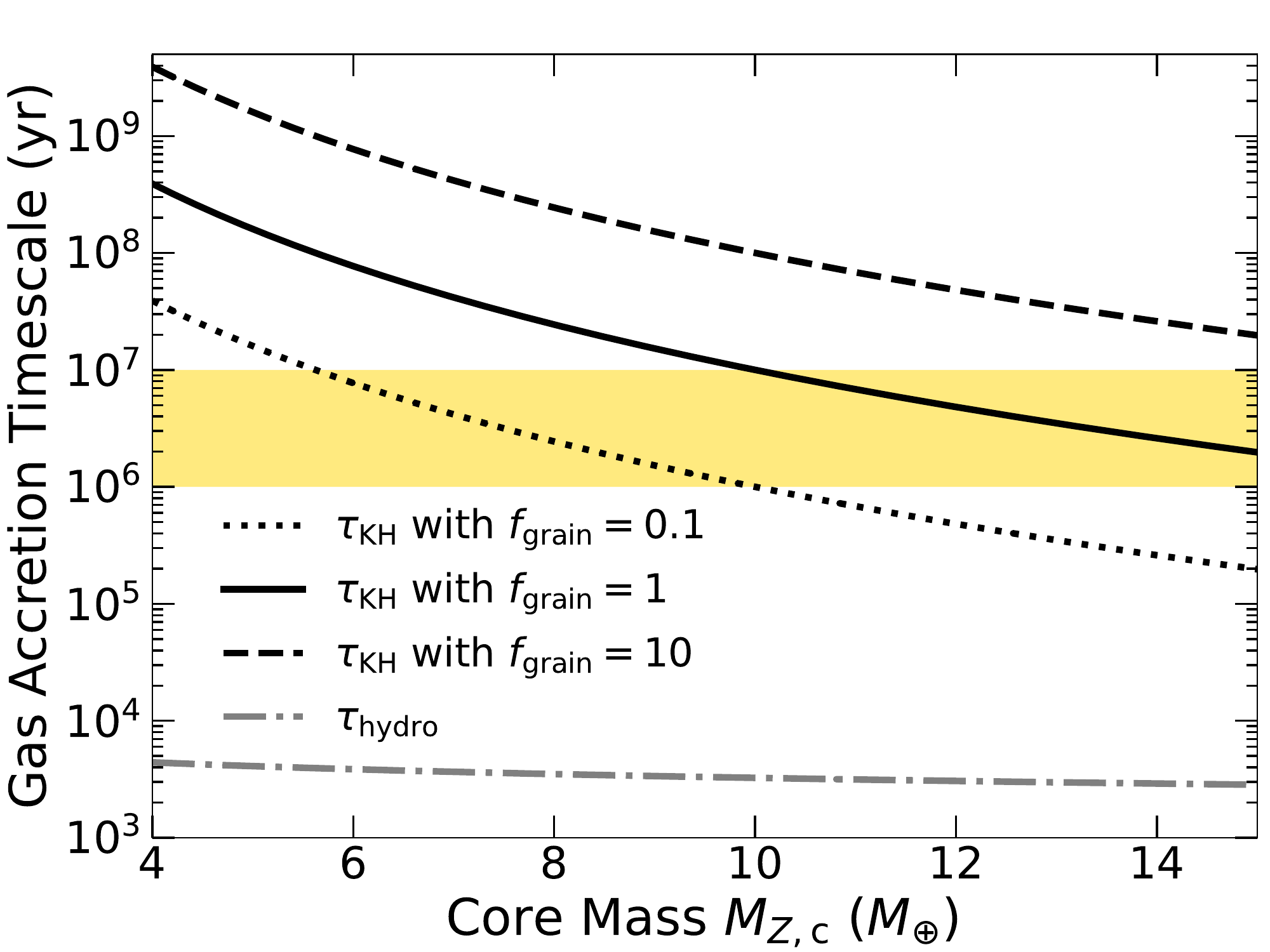}
\caption{The resulting values of $\dot{M}_{Z, \rm c}^{\rm crit}$, $\tau_{\rm KH}$, and $\tau_{\rm hydro}$ as functions of $M_{Z, \rm c}$ and $f_{\rm grain}$.
On the left panel, $\dot{M}_{Z, \rm c}^{\rm crit}$ is computed (see equation (\ref{eq:dotM_c_crit})).
The dotted, the solid, and the dashed black lines denote the cases that $f_{\rm grain}=0.1$, 1, and 10, respectively.
Gas accretion begins when the solid accretion rate onto planetary cores becomes lower than these lines.
On the right panel, the values of $\tau_{\rm KH}$ with the cases that $f_{\rm grain}=0.1$, 1, and 10 are represented by the dotted, the solid, and the dashed black lines, respectively.
For $\tau_{\rm hydro}$, its value is computed at $r_{\rm p} = 10$ au, using the MMSN model with $f_{\rm d,g}=1$ (equation (\ref{eq:sigma_g}), see the dash-dotted, gray line).
The yellow shaded region denotes the characteristic lifetimes of the solar nebula.
Given that $\tau_{\rm KH} > \tau_{\rm hydro}$ for all the cases,
the gas accretion rate onto the cores of Uranus and Neptune is regulated by $\tau_{\rm KH}$.}
\label{fig2}
\end{center}
\end{minipage}
\end{figure*}

We now specify which gas accretion recipe becomes important when computing the solid accretion rate.
Figure \ref{fig2} shows how the values of $\dot{M}_{Z, \rm c}$, $\tau_{\rm KH}$, and $\tau_{\rm hydro}$ behave for given values of $f_{\rm grain}$ and $M_{Z, \rm c}$.
The value of $M_{Z, \rm c}$ varies from $4 M_{\oplus}$ to $15 M_{\oplus}$,
which are possible for Uranus and Neptune, based on Figure \ref{fig1}.
Our calculations show that 
$\dot{M}_{Z, \rm c}^{\rm crit}$ is an increasing function of $M_{Z, \rm c}$, but a decreasing function of $f_{\rm grain}$ (see the left panel);
this is simply because envelopes contract readily when planetary cores are massive and the envelopes' metallicity is low for a given value of $\dot{M}_{Z, \rm c}$.
Gas accretion starts if solid accretion on planetary cores is lower than $\dot{M}_{Z, \rm c}^{\rm crit}$ in the plot.
In Section \ref{sec:app_UN}, we use this condition to constrain plausible values of the solid surface density enhancement factor ($f_{\rm d,s}$)
to reproduce the D/H ratio of Uranus/Neptune.
On the right panel, the gas accretion timescales are depicted;
we find that $\tau_{\rm KH}$ is always longer than $\tau_{\rm hydro}$ for the possible range of the core mass.

It is thus sufficient to consider gas accretion determined by $\tau_{\rm KH}$ in the following calculations.

\subsection{Solid accretion: Overview} \label{sec:solid_view}

We are in the position to explore solid accretion.
We here provide the overview of how solid accretion rates are computed in our formulation.

Computations of solid accretion rates begin with determination of an accretion mode for a given size of solids.
As discussed in Section \ref{sec:chara_radius},
the presence of gas drag affects the orbit of solid particles, 
which leads to two characteristic solid radii: $s^{\rm Ep}_{\rm pe}$ and $s^{\rm Qu}_{\rm 3b, max}$ (Table \ref{table1});
when solids are larger than $s^{\rm Qu}_{\rm 3b, max}$, the effect of gas drag becomes negligible and the canonical planetesimal accretion picture can be adopted.
In this work, the accretion mode is referred to as "two-body" because the corresponding eccentricity is modeled by the two-body scattering formula (see the next section).
For solids with the size range of $s^{\rm Ep}_{\rm pe} < s < s^{\rm Qu}_{\rm 3b, max}$, the drag force is strong enough to enhance solid accretion, compared with the canonical picture.
This mode is called "drag-enhanced three-body", following \citet{2010A&A...520A..43O}.
When solids are smaller than $s^{\rm Ep}_{\rm pe}$, the Keplerian orbit approximation breaks down, and the accretion mode switches to the so-called pebble accretion.
Thus, there are three accretion modes in our formulation.

Once an accretion mode is specified, then the resulting accretion rate is computed.
In principle, the accretion rate becomes a function of both eccentricities of solids and the effective accretion radius of the core.
We therefore determine these two quantities, which are discussed below.

\subsection{Eccentricity of solids} \label{sec:solid_ecc}

We describe how the eccentricity of solids is computed in this work.

We first point out that the eccentricity of solids becomes one fundamental parameter 
when solids are in the dispersion-dominated regime (Section \ref{sec:chara_radius}).
Thus, computations of eccentricity become important for the two-body and drag-enhanced three-body modes (Table \ref{table1}).

We then compute equilibrium eccentricity.
In our problem setup, the eccentricity excitation of solids by the core can be modeled by the two-body scattering formula \citep[][]{2008ApJ...684.1416S},
namely, the scattering cross-section is given by the the Bondi radius of the core with respect to $v_{\rm rel}^{\rm c-s}$.
The resulting eccentricity excitation timescale ($t_{\rm ext}$) is written as \citep[e.g.,][]{2008ApJ...684.1416S}
\begin{equation}
\label{eq:t_ext}
t_{\rm ext} \simeq \frac{8 \sqrt{3}}{27} \frac{e_{\rm H}^4}{h_{\rm H} \Omega}.
\end{equation}
The corresponding, equilibrium eccentricity ($e_{\rm eq}$) is computed, by equating $t_{\rm ext}$ with $t_{\rm s}$ (see equation (\ref{eq:t_s})).
For instance, the equilibrium eccentricity ($e_{\rm eq}^{\rm Qu}$) in the Quadratic regime is given as
\begin{eqnarray}
\label{eq:e_Qu_eq}
\frac{e^{\rm Qu}_{\rm eq}}{h_{\rm H}}  & =         &    \left( \frac{9}{2 \sqrt{3}} \frac{\rho_{\rm s} s}{\rho_{\rm d,g} r}\right)^{1/5}  \\ \nonumber
                                                            & \simeq &  1.1 \times 10 f_{\rm d,g}^{-1/5} \left(  \frac{s}{100 \mbox{ km}} \right)^{1/5} \left(  \frac{r}{10 \mbox{ au}} \right)^{7/20}.
\end{eqnarray}

When solids are in the shear-dominated regime, the Keplerian shear controls the relative velocity between solids and the core.
For the drag-enhanced three-body mode, we use $e_{\rm H}=3.2$, following \citet{2010A&A...520A..43O};
this value is estimated in the gas free limit.
The effect of gas drag is incorporated when computing the accretion rate (equation (\ref{eq:Mdot_Z_3b})).
For the pebble accretion mode, we adopt the approach of \citet{2012ApJ...747..115O} (see Section \ref{sec:solid_rate} for the detail).

\subsection{Accretion radius of solids} \label{sec:solid_radius}

The presence of the gaseous envelope around the core plays an important role in determining the solid accretion rate;
during a close encounter with the core, solids can interact with its (physically extended) envelope.
This interaction can reduce their speed due to aerodynamical drag arising from the envelope gas,
and eventually enhance the solid accretion rate considerably \citep{2003A&A...410..711I}.
In our formulation, this enhancement becomes effective for solids in the two-body accretion mode.

The critical envelope density ($\rho_{XY}^{\rm crit}$) at which incoming solids are captured is given as \citep{2003A&A...410..711I}
\begin{equation}
\label{eq:rho_XY_crit}
\rho_{XY}^{\rm crit} \equiv \frac{2}{3} \left( \frac{(v_{\rm rel}^{\rm c-s})^2}{2 G M_{Z, \rm c}} + \frac{1}{R_{\rm H}} \right) \rho_{\rm s} s,
\end{equation}
where $R_{\rm H} = r h_{\rm H} $ is the Hill radius.
Equivalently, the accretion radius of the protoplanet ($R_{\rm p, a}$) is computed from that $\rho_{XY} (R_{\rm p, a}) = \rho_{XY}^{\rm crit}$.

In order to take into account the effect of the envelope, we model the properties of the envelope, following \citet{2012ApJ...747..115O}.
We adopt the second and third assumptions, as discussed in Section \ref{sec:in-situ_1} (also see Appendix \ref{app_env}).

The envelope density profile ($\rho_{XY} (R)$) is then written as \citep{2012ApJ...747..115O}
\begin{equation}
\label{eq:rho_XY}
R \simeq  \frac{R_{\rm c, B}}{\gamma} \left\{ 
                           \begin{array} {l}
                                                  \left\{  1+ \frac{1 }{\gamma} \left[ 2 W_{\rm neb} \left( \frac{\rho_{XY} (R)}{\rho_{\rm d, g}} -1 \right)  \right. \right. \\
                                                                      \left. \left.          +   \log \left(  \frac{\rho_{XY} (R)}{\rho_{\rm d, g}} \right) \right] \right\}^{-1}  \\
                                                 ( \rho_{\rm d, g} \leq  \rho_{XY} (R) \leq \rho_{XY, \rm tr} (R_{\rm tr})  ) \\ \\
                                                 \left\{  \frac{R_{\rm c, B}}{\gamma R_{\rm tr}} 
                                                                     + \frac{4 (4 W_{\rm neb})^{1/3}}{\gamma}  \right. \\
                                                          \times \left.  \left[ \left(  \frac{\rho_{XY} (R)}{\rho_{\rm d, g}}\right)^{1/3} -    \left( \frac{\rho_{XY, \rm tr} (R_{\rm tr})}{\rho_{\rm d, g}} \right)^{1/3} \right] \right\}^{-1} \\
                                                 ( \rho_{XY, \rm tr} (R_{\rm tr}) \leq  \rho_{XY} (R) ), 
                           \end{array} 
                     \right.
\end{equation}
where $R$ is the distance measured from the core center, $R_{\rm c, B}= GM_{Z, \rm c}/c_s^2$ is the Bondi radius with respect to the sound speed,
$\gamma=1.4$ is the adiabatic index of the disk gas,
$\rho_{XY, \rm tr} / \rho_{\rm d, g} =  1 / (5 W_{\rm neb})$, $R_{\rm tr}$ is the transition radius computed from the above equation at $\rho_{XY} (R= R_{\rm tr}) = \rho_{XY, \rm tr}$.
This transition radius is effectively comparable to the radiative-convective boundary (Appendix \ref{app_env}).
The dimensionless parameter ($W_{\rm neb}$) is defined as
\begin{equation}
\label{eq:W_neb}
W_{\rm neb} = \frac{3 \kappa_{\rm grain} L}{ 64 \pi \sigma_{\rm SB}} \frac{\rho_{\rm d, g} c_{\rm s}^2}{G M_{Z, \rm c} T_{\rm d}^4},
\end{equation}
where the luminosity of the accreting protoplanet ($L$) is given as
\begin{equation}
\label{eq:Lum}
L = \frac{G M_{Z, \rm c} \dot{M}_{\rm p}}{R_{\rm p}},
\end{equation}
$\sigma_{\rm SB}$ is the Stefan Boltzmann constant,
$\dot{M}_{\rm p}$ is the accretion rate onto the protoplanet that consists of the core and envelope, 
$R_{\rm p} = f_{\rm L} R_{\rm c, B}$ is the effective radius of the protoplanet, and $f_{\rm L}$ is an adjusting factor.

\begin{figure}
\begin{center}
\includegraphics[width=8.3cm]{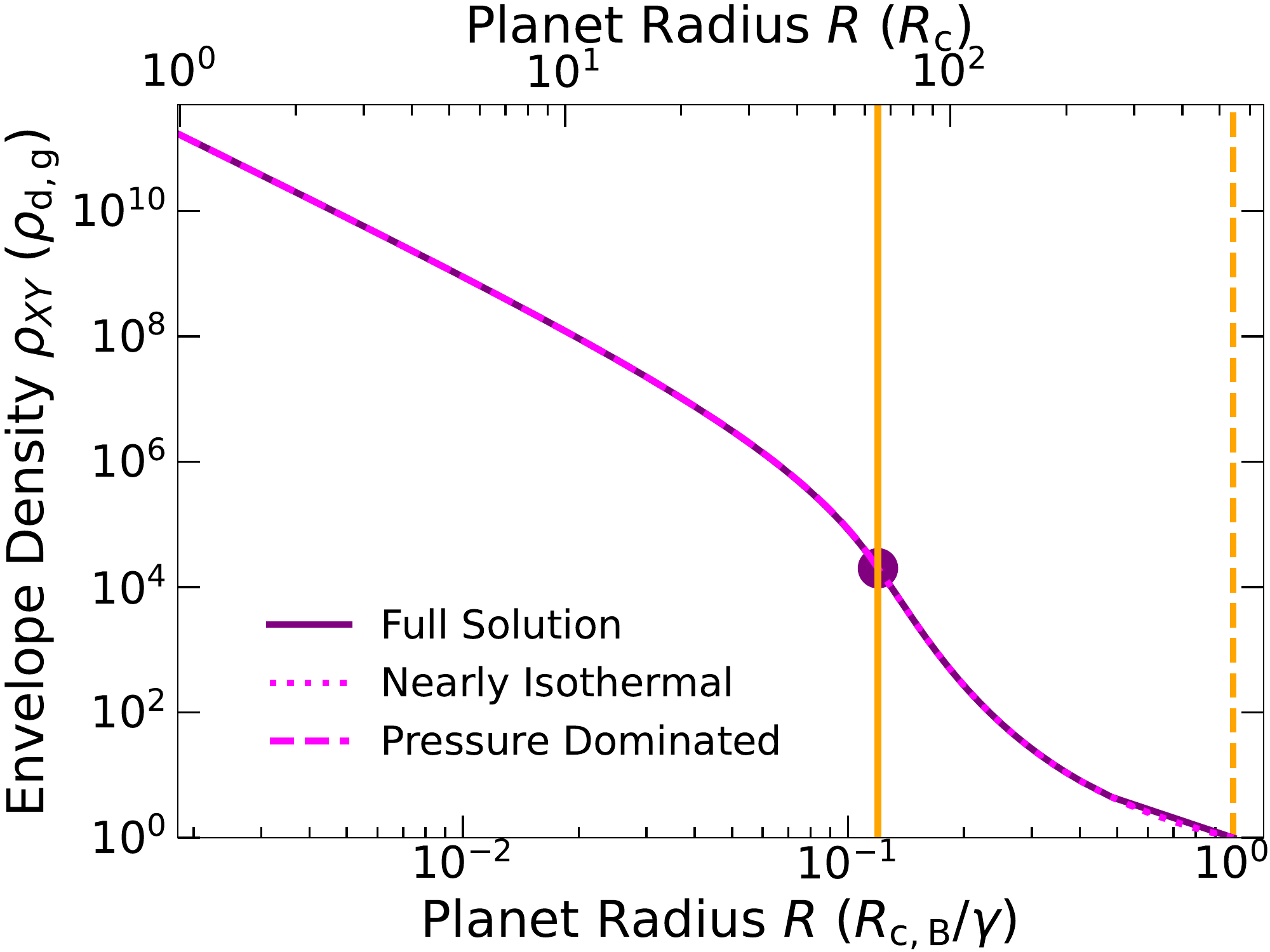}
\caption{The envelope density profile for the case that that $W_{\rm neb} = 10^{-5}$.
The full, analytical solution is denoted by the solid purple line (see Appendix \ref{app_env}), 
and the approximate solutions for the nearly isothermal and pressure dominated regimes are by the dotted and dashed magenta lines, respectively (see equation (\ref{eq:rho_XY})).
For reference, the dashed and solid vertical orange lines represent the transition locations from the solar nebula to the planetary envelope 
and from the nearly isothermal regime to the pressure dominated one, respectively.
The purple dot labels the latter transition, that is, $(R, \rho_{XY})=(R_{\rm tr},\rho_{{XY}, \rm tr})$.} 
\label{fig3}
\end{center}
\end{figure}

As an example, Figure \ref{fig3} shows the resulting density profile for the case that $W_{\rm neb} = 10^{-5}$.
One finds that as $R$ decreases, the density increases relatively rapidly in the nearly isothermal regime (that is, $ \rho_{\rm d, g} \leq  \rho_{XY} (R) \leq \rho_{XY, \rm tr} (R_{\rm tr})$).
The profile goes to $R^3$ in the pressure-dominated regime (that is, $\rho_{XY, \rm tr} (R_{\rm tr}) \leq  \rho_{XY} (R)$).
The latter is the well-known profile for radiative zones \citep[e.g.,][]{1982P&SS...30..755S,2003A&A...410..711I}.

Under the first assumption in Section \ref{sec:in-situ_1}, and considering Case 3 of Figure \ref{fig1} mainly in this work,
$\dot{M}_{\rm p}$ is re-written as
\begin{equation}
\label{eq:dotM_p}
\dot{M}_{\rm p} =  \dot{M}_{{\rm p}, XY} +  \dot{M}_{{\rm p}, Z} \simeq \dot{M}_{{\rm p}, XY} \simeq \frac{M_{Z, \rm c}}{ \tau_{\rm KH}}.
\end{equation}
This equation can be viewed as the mathematical expression that the envelope is heated by the contraction itself.
Previous studies show that considerable contraction occurs in a radiative zone \citep[e.g.,][]{2000ApJ...537.1013I}.
Such contraction is achieved by cooling.
Also, contraction heats up the envelope and establishes quasi-hydrostatic evolution, as discussed in Section \ref{sec:in-situ_2}.
Under the quasi-hydrostatic evolution, heating and cooling should equilibrate effectively.
Our simplified treatment of gas accretion cannot model such equilibrium.
Since the luminosity is proportional to the gas accretion rate in our formulation (equations (\ref{eq:Lum}) and (\ref{eq:dotM_p})),
the (quasi-)equilibrium situation might be mimicked, by adjusting the value of $f_{\rm L}$ (see Section \ref{sec:disc_cav} for the additional discussion).
We find that when $f_{\rm L} =1/3$, the resulting luminosity becomes comparable to the results of \citet{2000ApJ...537.1013I};
hence we use that value in this work. 

Consequently, the dependence of $f_{\rm grain}$ disappears in $W_{\rm neb}$:
\begin{eqnarray}
\label{eq:W_neb2}
W_{\rm neb} 
                   & \simeq  &1.4 \times 10^{-5} f_{\rm d,g} \left( \frac{f_{\rm L}}{1/3} \right)^{-1} \left( \frac{10^{-c}}{10^{-7}} \right)  \\ \nonumber
                    & \times &    \left( \frac{\kappa_{\rm grain}^{\rm ISM}}{1 \mbox{ cm}^{2} \mbox{ g}^{-1}} \right)  \left( \frac{M_{Z, \rm c}}{10 M_{\oplus}} \right)^{d} \left( \frac{r}{10 \mbox{ au}} \right)^{-7/4}.
\end{eqnarray}
Note that $\kappa_{\rm grain}^{\rm ISM}$ appears in the above equation 
because the gas accretion timescale is normalized for the case that $\kappa_{\rm grain} = \kappa_{\rm grain}^{\rm ISM}$ (see equation (\ref{eq:tau_KH})).
This in turn indicates that $W_{\rm neb}$ is independent of the grain opacity in our formulation.
On the contrary, $L$ is a function of the grain opacity as
\begin{eqnarray}
L  & \simeq &  4.6 \times 10^{-9}  f_{\rm grain}^{-1}  \left( \frac{f_{\rm L}}{1/3} \right)^{-1} \left( \frac{10^{-c}}{10^{-7}} \right)  \\ \nonumber
    & \times & \left( \frac{M_{Z, \rm c}}{10 M_{\oplus}} \right)^{(1+d)} \left( \frac{r}{10 \mbox{ au}} \right)^{-1/2} L_{\odot},
\end{eqnarray}
where $L_{\odot}$ is the solar luminosity.

In summary, the density profile is independent of the grain opacity since it is a function of $W_{\rm neb}$ (but not $L$),
while the gas accretion timescale is dependent on the grain opacity since $L$ is a function of $f_{\rm grain}$ in our simplified model.

\subsection{Solid accretion rate} \label{sec:solid_rate}

Armed with the above formulation, we now describe the solid accretion rate.
We consider three modes as discussed in Section \ref{sec:solid_view} (also see Table \ref{table1}).

We begin with the two-body mode.
In this mode, the solid accretion rate is determined by three competing processes \citep[e.g.,][]{2007ApJ...666..447Z,2008ApJ...684.1416S}:
scattering of solids by the core, eccentricity damping of the solids by the disk gas, 
and the expansion of the feeding zone of the core by gas accretion.
When the expansion of the feeding zone is faster than the eccentricity damping of scattered solids,
that is, the core accretes the disk gas very efficiently,
then the core can accrete these scattered solids,
since they are still in the feeding zone.
This in turn indicates that while the dynamics of solids tends to open up a gap in their disk,
efficient gas accretion closes the gap.
On the other hand, when the expansion of the feeding zone is slower than the eccentricity damping of scattered solids,
then the scattered solids can be located outside the feeding zone of the core.
This eventually leads to gap formation in the solid disk.
As demonstrated by \citet{2018ApJ...865...32H}, 
the presence of gaps in solid disks around planets plays the major role in reproducing the heavy-element content trend of giant planets both in the solar and extrasolar systems.
Thus, the competition of the three processes can result in the formation of gaps in solid disks,
and hence this accretion mode can divide into two sub-modes due to the presence of gaps or no gaps.

The presence of gaps (or no gaps) effectively corresponds to the dispersion dominated regime (or the shear dominated regime). 
This is because gap formation is caused by the damping of relatively high eccentricities \citep[i.e., $e_{\rm H} > 1$,][]{2008ApJ...684.1416S}.
Consequently, the solid accretion rate ($\dot{M}_{{\rm p}, Z}^{\rm disp}$) onto the protoplanet for the gap-opening case is written as \citep{2008ApJ...684.1416S}
\begin{eqnarray}
\label{eq:Mdot_Z_disp}
\dot{M}_{{\rm p}, Z}^{\rm disp} & \simeq & 0.36 \times 10^{c'}   f_{\rm d, s}  f_{R}^{1/2} \eta^{d'}_{\rm gap}  \\ \nonumber
                                                 & \times  &   \left( \frac{\rho_{\rm c}}{3 \mbox{ g cm}^{-3} } \right)^{1/2} \left( \frac{R_{\rm c}}{R_{\oplus}} \right)^2 M_{\oplus} \mbox{ yr}^{-1},
\end{eqnarray}
where $c' \simeq -6$, $d' \simeq 1.4$, $\rho_{\rm c}$ and $R_{\rm c}$ are the bulk density and the radius of the core, respectively,
$f_{R} = R_{\rm p, a}/R_{\rm c}$ is the enhancement factor for the accretion radius due to the presence of the envelope, $R_{\oplus}$ is the radius of the Earth,  and
\begin{equation}
\label{eq:eta}
\eta_{\rm gap}  =  \frac{8}{\tau_{\rm KH}} \left( \frac{e_{\rm H}^2}{t_{\rm s}} \right)^{-1},
\end{equation}
where the gap opening condition is given by $\eta_{\rm gap}<1$.
Note that $\dot{M}_{{\rm p}, Z}^{\rm disp}$ is a function of $\eta_{\rm gap}$ and hence $f_{\rm grain}$.
This is because solid accretion is determined by 
the competition between eccentricity damping of the solids by the disk gas and the expansion of the feeding zone of the core by gas accretion,
as discussed above.
It should also be pointed out that the power of $f_{R}$ reflects the simulation results, 
where solid accretion occurs in the two dimensional fashion \citep{2008ApJ...684.1416S}.

As described above, gap formation may not be realized when the eccentricity of solids is in the shear dominated regime.
Through the complete examination, we have found that this arises in the Epstein regime (Table \ref{table1}).
The corresponding radius of the solid particles ($s^{\rm Ep}_{\rm shear}$) is computed as ($t_{\rm ext} = t_{\rm s} $with $e_{\rm H}=1$)
\begin{equation}
\label{eq:s_nogap_Ep2}
s_{\rm shear}^{\rm Ep} =  \frac{8 \sqrt{3}}{27} \frac{\rho_{\rm d,g} v_{\rm th}}{\rho_{\rm s} h_{\rm H} \Omega}. 
\end{equation}
The resulting solid accretion rate ($\dot{M}_{{\rm p}, Z}^{\rm shear} $) is given as \citep{2008ApJ...684.1416S}
\begin{eqnarray}
\label{eq:Mdot_Z_shear}
\dot{M}_{{\rm p}, Z}^{\rm shear}  & \simeq & 0.36 \times 10^{c''}   f_{\rm d, s}  f_{R}^{1/2} \xi^{d''} \\ \nonumber
                                                     & \times & \left( \frac{\rho_{\rm c}}{3 \mbox{ g cm}^{-3} } \right)^{1/2} \left( \frac{R_{\rm c}}{R_{\oplus}} \right)^2 M_{\oplus} \mbox{ yr}^{-1},
\end{eqnarray}
where $c'' \simeq -6$, $d'' \simeq 0.8$, and
\begin{equation}
\xi = \frac{74 \pi}{h_{\rm H} \Omega  \tau_{\rm KH} }.
\end{equation}

We then consider the drag-enhanced three-body mode.
This accretion mode is realized 
when accreted solids have a relatively small Stokes number and/or their eccentricity is relatively small \citep{2010A&A...520A..43O}.
The corresponding accretion rate ($\dot{M}_{{\rm p}, Z}^{\rm 3b}$) is given as \citep{2012ApJ...747..115O}:
\begin{equation}
\label{eq:Mdot_Z_3b}
\dot{M}_{{\rm p}, Z}^{\rm 3b}  =  2 \Sigma_{\rm d, s} r^2 h_{\rm H}^2 \Omega \frac{1}{\rm St} \exp \left[ - \left( \frac{0.7 e_{\rm H}}{\rm St} \right)^5 \right] e_{\rm H}, 
\end{equation}
where it is assumed that the accretion takes place in the two dimensional plane due to low (or negligible) turbulence.

We finally discuss the pebble accretion mode.
This accretion mode has recently received considerable attention,
as it can accelerate core formation significantly even in the outer part of disks \citep[e.g.,][]{2012A&A...544A..32L}.
We here adopt the formulation developed in \citet{2010A&A...520A..43O,2012ApJ...747..115O}.

The solid accretion rate via pebble accretion ($\dot{M}_{{\rm p}, Z}^{\rm pe}$) is given as 
\begin{equation}
\label{eq:Mdot_Z_pe}
\dot{M}_{{\rm p}, Z}^{\rm pe} = 2 \Sigma_{\rm d,s} b_{\rm set} \Delta v_{\rm set},
\end{equation}
where $b_{\rm set}$ is the effective pebble accretion radius of the protoplanet,
$\Delta v_{\rm set}$ is the approaching velocity of pebbles toward the protoplanet,
and it is again assumed that accretion takes place in the two-dimensional fashion.

The explicit descriptions of $b_{\rm set}$ and $\Delta v_{\rm set}$ are given as, respectively
\citep{2012ApJ...747..115O}
\begin{equation}
\label{eq:b_set}
b_{\rm set} = b_{{\rm set}, 0} \exp \left[ - \left( \frac{\rm St}{2} \right)^{0.65} \right],
\end{equation}
and 
\begin{equation}
\label{eq:Delta_v}
\Delta v_{\rm set} = \eta_{\rm d} r \Omega + \frac{3}{2} b_{\rm set} \Omega,
\end{equation}
where \citep{2010A&A...520A..43O}
\begin{equation}
\label{eq:b_set0}
3 b_{{\rm set}, 0}^3 + 2 \eta_{\rm d} r b_{{\rm set}, 0}^2 - 24 \mbox{St} R_{\rm H}^3  = 0.
\end{equation}
As shown by \citet{2012A&A...544A..32L}, pebble accretion can divide into two regimes,
depending on which approaching velocity ($\eta_{\rm d} r \Omega$ vs $3 b_{\rm set} \Omega /2$) is dominant  \citep[also see][]{2016A&A...589A..15S}.
We have found that $\eta_{\rm d} r \Omega \ll 3 b_{\rm set} \Omega /2$ in our configuration,
which is referred to as the Hill accretion regime in \citet{2012A&A...544A..32L}.
In this regime, the second term in equation (\ref{eq:b_set0}) becomes negligible,
and $b_{{\rm set}, 0}$ can be written as
\begin{equation}
b_{{\rm set}, 0} \simeq 2 \mbox{St}^{1/3} R_{\rm H}.
\end{equation}
Since the exponential function in equation (\ref{eq:b_set}) becomes an order of unity for small St,
the condition that $\eta_{\rm d} r \Omega \ll 3 b_{\rm set} \Omega /2$ can be re-written as
\begin{equation}
M_{Z, \rm c} \gg \frac{\eta_{\rm d}^3 M_{\odot}}{9 \mbox{St}} \simeq 3.4 \times 10^{-3} \left( \frac{\rm St}{2} \right)^{-1} \left( \frac{r}{10 \mbox{ au}} \right)^{3/2} M_{\oplus}.
\end{equation}
This equation confirms that it is sufficient to consider the Hill accretion regime in this work.

Consequently, $\dot{M}_{{\rm p}, Z}^{\rm pe} $ is given as
\begin{equation}
\label{eq:Mdot_Z_pa}
\dot{M}_{{\rm p}, Z}^{\rm pe}  \simeq 12 \Sigma_{\rm d,s} \mbox{St}^{2/3} R_{\rm H}^2 \Omega  \exp \left[ - 2 \left( \frac{\rm St}{2} \right)^{0.65} \right].
\end{equation}

In the following sections, we use the above formulation and explore solid accretion in each regime (see Table \ref{table1}).

\section{Resulting solid accretion rates} \label{sec:res1}

The solid accretion rate is computed when the following five quantities are specified:
the solid surface density ($f_{\rm d,s}$), the core mass ($M_{Z, \rm c}$), its position ($r_{\rm p}$), its envelope dust opacity factor ($f_{\rm grain}$), and the radius of accreted solids ($s$).
We here consider certain sets of these parameters to explore how equilibrium eccentricity of accreted solids is computed and what is the resulting accretion rate;
an additional parameter study is undertaken in Section \ref{sec:app_UN}.

Table \ref{table2} summarizes the sets of parameters adopted in this section.
Among them, the values of $M_{Z, \rm c}$ and $f_{\rm grain}$ are motivated by 
constraints obtained from the current values of the D/H ratio of Uranus and Neptune (Sections \ref{sec:d/h} and \ref{sec:app_UN}).

\begin{table*}
\begin{minipage}{17cm}
\centering
\caption{The summary of the parameter study}
\label{table2}
{\scriptsize
\begin{tabular}{c|c|c|c|c|c} 
\hline
Section                                      &  Accreted solid                       &  Core mass ($M_{Z, \rm c}$)          & Position of a planet ($r_{\rm p}$) & the solid surface density ($f_{\rm d,s}$)  & the dust opacity factor ($f_{\rm grain}$) \\ \hline \hline
Section \ref{sec:acc_pla}           & planetesimals                        & 10 $M_{\oplus}$, 12 $M_{\oplus}$  & 10 au                                           &  1                                                            & 1                                                          \\ \hline
Section \ref{sec:acc_lar_com}   & comets and large fragments  & 10 $M_{\oplus}$                             & 10 au                                           &  1                                                            & 1                                                          \\ \hline
Section \ref{sec:acc_mid_frag}  & medium fragments                 & 10 $M_{\oplus}$                             & 10 au                                           &  $10^{-2}$, 1                                          & 1                                                          \\ \hline
Section \ref{sec:acc_sma_frag} & small fragments                     & 10 $M_{\oplus}$                             & 10 au                                           &  $10^{-2}$, 1                                          & 1                                                          \\ \hline
Section \ref{sec:acc_bou}          & boulders                                 & 10 $M_{\oplus}$                             & 10 au                                           &  $10^{-2}$, 1                                          & 1                                                          \\ \hline
Section \ref{sec:acc_peb}          & pebbles                                 & 10 $M_{\oplus}$                             & 10 au                                           &  $10^{-3}$, $10^{-2}$                             & 1                                                          \\ \hline
\end{tabular}
}
\end{minipage}
\end{table*}

\subsection{Accretion of planetesimals} \label{sec:acc_pla}

We here focus on planetesimal accretion. 
We determine when the accretion radius exceeds the core radius (that is, $f_{\rm R} > 1$)
and which accretion mode ($\dot{M}_{{\rm p}, Z}^{\rm 3b}$ vs $\dot{M}_{{\rm p}, Z}^{\rm disp}$) is dominant.
We choose $f_{\rm d,s}=1$, and $M_{Z, \rm c}=10M_{\oplus}$ and $12M_{\oplus}$ (Table \ref{table2}).

The first condition ($f_{\rm R} > 1$) is specified analytically as follows.
The envelope density in the vicinity of the core can be approximated as (equation (\ref{eq:rho_XY}))
\begin{eqnarray}
\label{eq:rho_XY_Rc}
\frac{\rho_{XY} (R_{\rm c})}{\rho_{\rm d,g}} & \simeq & \frac{R_{\rm d, B}^3}{256 W_{\rm neb} R_{\rm c}^3} \\ \nonumber
                                                                    &  = &  \frac{ \pi \sigma_{\rm SB} f_{\rm L}}{12  f_{\rm grain} \kappa^{\rm ISM}_{\rm grain} } 
                                                                              \frac{ G (GM_{Z, \rm c})^3 T_{\rm d}^4 \tau_{\rm KH}}{ \rho_{\rm d,g} R_{\rm c}^3 c_{\rm s}^{10}} \\ \nonumber
                                                             & \simeq & 1.2 \times 10^{11} f_{\rm d, g}^{-1} \left( \frac{ f_{\rm L}}{1/3} \right) \left( \frac{10^c}{10^7}\right) \\ \nonumber
                                                             & \times &  \left( \frac{M_{Z, \rm c}}{10 M_{\oplus}} \right)^{2-d}  \left( \frac{r}{10 \mbox{ au} } \right)^{13/4}.                                                                                                     
\end{eqnarray}
When the dispersion dominated regime (i.e., $e_{\rm H}>1$) is considered, 
$\rho_{XY}^{\rm crit} $ can be simplified as (equation (\ref{eq:rho_XY_crit}))
\begin{eqnarray}
\label{eq:rho_XY_crit1}
\frac{\rho_{XY}^{\rm crit} }{\rho_{\rm d,g}} & \simeq & \frac{(v_{\rm rel}^{\rm c-s})^2}{3 GM_{Z,\rm c}} \frac{ \rho_{\rm s} s}{\rho_{\rm d,g}} 
                                                                = \frac{(e/h_{\rm H})^2}{9 h_{\rm H}} \frac{ \rho_{\rm s} s}{\rho_{\rm d,g}r} \\ \nonumber
                                                                & \simeq & 2.9 \times 10^{7} f_{\rm d,g}^{-1}\left( \frac{e/h_{\rm H}}{10} \right)^2 \left( \frac{M_{Z, \rm c}}{10 M_{\oplus}} \right)^{-1/3} \\ \nonumber
                                                                 & \times &   \left( \frac{s}{100 \mbox{ km} } \right) \left( \frac{r}{10 \mbox{ au} } \right)^{7/4}.
\end{eqnarray}
Then, the maximum eccentricity ($e^{\rm Qu}_{\rm max}$) of solids that can be captured by the protoplanet becomes,
by equating equation (\ref{eq:rho_XY_Rc}) with equation (\ref{eq:rho_XY_crit1}),
\begin{eqnarray}
\frac{e^{\rm Qu}_{\rm max}}{h_{\rm H}} & \simeq & \left(
                                             \frac{3 \pi \sigma_{\rm SB} f_{\rm L} h_{\rm H}}{4 f_{\rm grain} \kappa^{\rm ISM}_{\rm grain} } \frac{r}{ \rho_{\rm s} s}
                                               \frac{ G (GM_{Z, \rm c})^3 T_{\rm d}^4 \tau_{\rm KH}}{ R_{\rm c}^3 c_{\rm s}^{10}}   
                                               \right)^{1/2} \\ \nonumber
                               & \simeq & 6.4 \times 10^2  \left( \frac{ f_{\rm L}}{1/3} \right)^{1/2} \left( \frac{10^c}{10^7} \right)^{1/2}    \left( \frac{s}{100 \mbox{ km} } \right)^{-1/2} \\ \nonumber
                                  & \times &   \left( \frac{M_{Z, \rm c}}{10 M_{\oplus}} \right)^{(7/3-d)/2}  \left( \frac{r}{10 \mbox{ au} } \right)^{3/4}. 
\end{eqnarray}
Note that this large eccentricity is the value theoretically computed in the extreme limit;
in reality, the planetesimals with the eccentricity of $e^{\rm Qu}_{\rm eq} \la 10$ mainly contribute to solid accretion.

Consequently, the size of solids ($s^{\rm Qu}_{XY, \rm max}$) that satisfy the condition $f_{\rm R} > 1$ is given as, using equation (\ref{eq:e_Qu_eq}),
\begin{eqnarray}
s^{\rm Qu}_{XY, \rm max} & \la &       \left( \frac{9}{2 \sqrt{3}} \frac{\rho_{\rm s} }{\rho_{\rm d,g} r}\right)^{-2/7} \\ \nonumber
                                          & \times &           \left(   
                                                                    \frac{3 \pi \sigma_{\rm SB} f_{\rm L} h_{\rm H}}{4 f_{\rm grain} \kappa^{\rm ISM}_{\rm grain}} \frac{r}{ \rho_{\rm s} }
                                                                    \frac{ G (GM_{Z, \rm c})^3 T_{\rm d}^4 \tau_{\rm KH}}{ R_{\rm c}^3 c_{\rm s}^{10}}       
                                                     \right)^{5/7} \\ \nonumber
                                        & \simeq & 3.4 \times 10^4   f_{\rm d,g}^{2/7} \left( \frac{ f_{\rm L}}{1/3} \right)^{5/7} \left( \frac{10^c}{10^7} \right)^{5/7}      \\ \nonumber
                                        & \times & \left( \frac{M_{Z, \rm c}}{10 M_{\oplus}} \right)^{5(7/3-d)/7} \left( \frac{r}{10 \mbox{ au} } \right)^{4/7}  
                                        \mbox{ km}.      
\end{eqnarray}
This equation indicates that the condition that $f_{\rm R} > 1$ is always met in our configuration; 
the accretion radius is larger than the core radius for all the solids under consideration.

\begin{figure}
\begin{center}
\includegraphics[width=8.3cm]{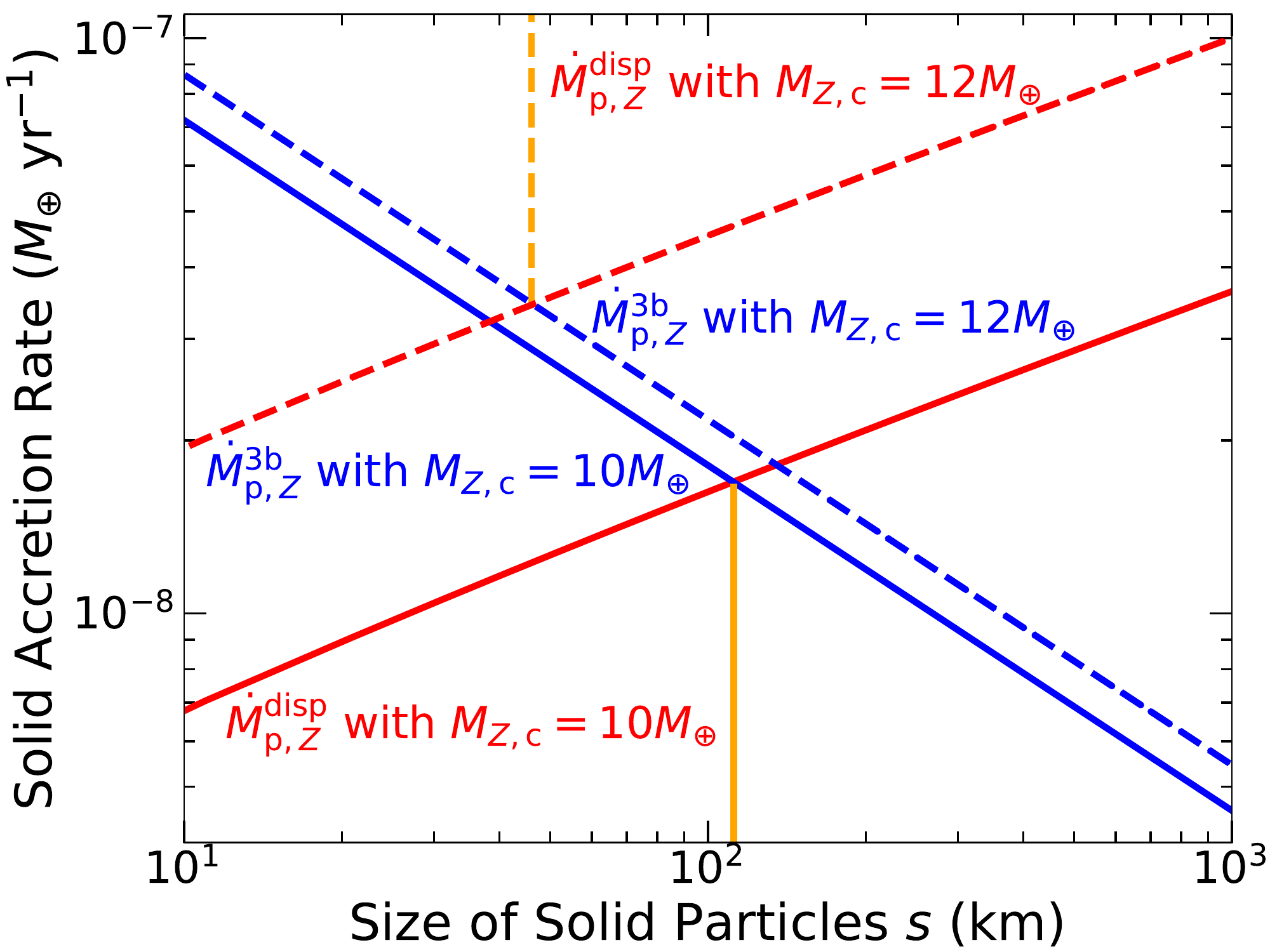}
\caption{The accretion rate of planetesimals as a function of their radius for the case that $f_{\rm d,s}=1$ (Table \ref{table2}).
The values of $\dot{M}_{{\rm p}, Z}^{\rm disp}$ are denoted by the red lines, and those of $\dot{M}_{{\rm p}, Z}^{\rm 3b}$ by the blue lines.
Two values of the core mass are considered: $M_{Z, \rm c}= 10 M_{\oplus}$ and $12M_{\rm \oplus}$ (see the solid and dashed lines, respectively).
The dominant accretion mode switches from $\dot{M}_{{\rm p}, Z}^{\rm 3b}$ to $\dot{M}_{{\rm p}, Z}^{\rm disp}$
at $r_{\rm s} \simeq 1.1 \times 10^2$ km for the case that $M_{Z, \rm c}= 10 M_{\oplus}$ and at $r_{\rm s} \simeq 46$ km for the case that $M_{Z, \rm c}= 12 M_{\oplus}$ (see the vertical orange lines).}
\label{fig4}
\end{center}
\end{figure}

We now explore the condition that $\dot{M}_{{\rm p}, Z}^{\rm 3b} > \dot{M}_{{\rm p}, Z}^{\rm disp}$.
As done above, one can principally compute analytically the characteristic radius of solids that satisfy the condition.
We, however, find that the corresponding calculation is rather cumbersome due to the factor of $f_{\rm R}$ (see equation (\ref{eq:Mdot_Z_disp})).
Therefore, we determine the radius graphically.
Figure \ref{fig4} shows the resulting values of $\dot{M}_{{\rm p}, Z}^{\rm disp}$ and $\dot{M}_{{\rm p}, Z}^{\rm 3b}$ as a function of $s$ and $M_{Z, \rm c}$.
We find that $\dot{M}_{{\rm p}, Z}^{\rm disp}$ is an increasing function of both the radius of accreted solids and the core mass.
This behavior can be understood when the dependence of $\eta_{\rm gap}$ on these two quantities is examined;
in the Quadratic, dispersion dominated regime, $\eta_{\rm gap}$ becomes, using equations (\ref{eq:t_s}), (\ref{eq:e_Qu_eq}), and (\ref{eq:eta}),
\begin{eqnarray}
\label{eq:eta_gap}
\eta_{\rm gap} 
       & \simeq &  \frac{32}{3  h_{\rm H} \Omega \tau_{\rm KH}} \left( \frac{9}{2 \sqrt{3}} \right)^{-3/5} \left( \frac{\rho_{\rm s} s}{\rho_{\rm d,g} r} \right)^{2/5} \\ \nonumber
       & \simeq & 5.5 \times 10^{-2}    f_{\rm d, g}^{-2/5}   \left(  \frac{f_{\rm grain}}{1} \right)^{-1}  \left(  \frac{10^c}{10^7} \right)^{-1}                 \\ \nonumber
       & \times & \left(  \frac{M_{Z, \rm c}}{10 M_{\oplus}} \right)^{d-1/3}     \left(  \frac{s}{100 \mbox{ km}} \right)^{2/5} \left(  \frac{r}{10 \mbox{ au}} \right)^{11/5}.
\end{eqnarray}
Equivalently, gap formation occurs readily for small-sized solids, which consequently leads to the reduction in $\dot{M}_{{\rm p}, Z}^{\rm disp}$.
For $\dot{M}_{{\rm p}, Z}^{\rm 3b}$, it increases with decreasing the solid radius and increasing the core mass.
This is simply because, when St is large enough (see equations  (\ref{eq:t_s}), (\ref{eq:e_Qu_eq}), and (\ref{eq:Mdot_Z_3b})),
\begin{eqnarray}
\label{eq:Mdot_Z_3b_Qu}
\dot{M}_{{\rm p}, Z}^{\rm 3b}  & \simeq &  \frac{3}{2}  \left( \frac{9}{2 \sqrt{3}} \right)^{2/5} \Sigma_{\rm d, s} r^2 h_{\rm H}^3 \Omega  \left( \frac{\rho_{\rm s} s}{\rho_{\rm d,g} r} \right)^{-3/5}  \\ \nonumber
                                              &\simeq & 1.8 \times 10^{-8} f_{\rm d,g}^{3/5} f_{\rm d,s}  \left( \frac{M_{Z, \rm c}}{10 M_{\oplus}} \right) \\ \nonumber
                                              & \times &  \left( \frac{s}{100 \mbox{ km}} \right)^{-3/5}       \left(  \frac{r}{10 \mbox{ au}} \right)^{-41/20}
                                                               M_{\oplus} \mbox{ yr}^{-1}.
\end{eqnarray}

Thus, the condition that $\dot{M}_{{\rm p}, Z}^{\rm 3b} > \dot{M}_{{\rm p}, Z}^{\rm disp}$ is realized
when the core with the mass of $10 M_{\oplus}$ accretes the solids with the size of $\la 1.1 \times 10^2$ km
and the core with the mass of $12 M_{\oplus}$ accretes the solids with the size of $\la $ 46 km.

\subsection{Accretion of comets and large fragments} \label{sec:acc_lar_com}

We briefly discuss the accretion of comets and large fragments (Table \ref{fig1}).
We explicitly distinguish between them,
because solid particles with the size of $\la 1$ km tend to be fragile against their collisions \citep[e.g.,][]{1999Icar..142....5B},
and hence their size distributions and abundances would be different from each other.
However, they are both in the dispersion dominated regime,
their accretion mode is the drag-enhanced three-body,
and the accretion radius is extended beyond the core radius due to the presence of the envelope.
The corresponding accretion rate is thus given by equation (\ref{eq:Mdot_Z_3b_Qu}).
We confirm that the resulting values still satisfy the condition that $M_{Z, \rm c} \ga M_{Z, \rm c}^{\rm crit}$ (equation (\ref{eq:M_c_crit})) 
for the case that $f_{\rm d, s}=1$ and $M_{Z, \rm c}=10M_{\oplus}$ (Table \ref{table2});
gas accretion proceeds simultaneously.

\subsection{Accretion of medium fragments} \label{sec:acc_mid_frag}

We now explore the accretion of medium fragments.
The main difference with the previous section is that accreted solids are in the Stokes regime (see Table \ref{table1}).
Then, the equilibrium eccentricity ($e^{\rm St}_{\rm eq}$) is written as (see equations (\ref{eq:t_s}) and (\ref{eq:t_ext}))
\begin{eqnarray}
\frac{e^{\rm St}_{\rm eq}}{h_{\rm H}} & =          &  \left( \frac{3}{2\sqrt{3}} h_{\rm H} \Omega \frac{ \rho_{\rm s} s^2 }{ \rho_{\rm d, g} v_{\rm th} l_{\rm mfp} }  \right)^{1/4}  \\ \nonumber
                                                         & \simeq & 1. 9  \left(  \frac{M_{Z, \rm c}}{10 M_{\oplus}} \right)^{1/12} \left(  \frac{s}{30 \mbox{ m}} \right)^{1/2} 
                                                                              \left(  \frac{r}{10 \mbox{ au}} \right)^{-5/16}.
\end{eqnarray}

\begin{figure}
\begin{center}
\includegraphics[width=8.3cm]{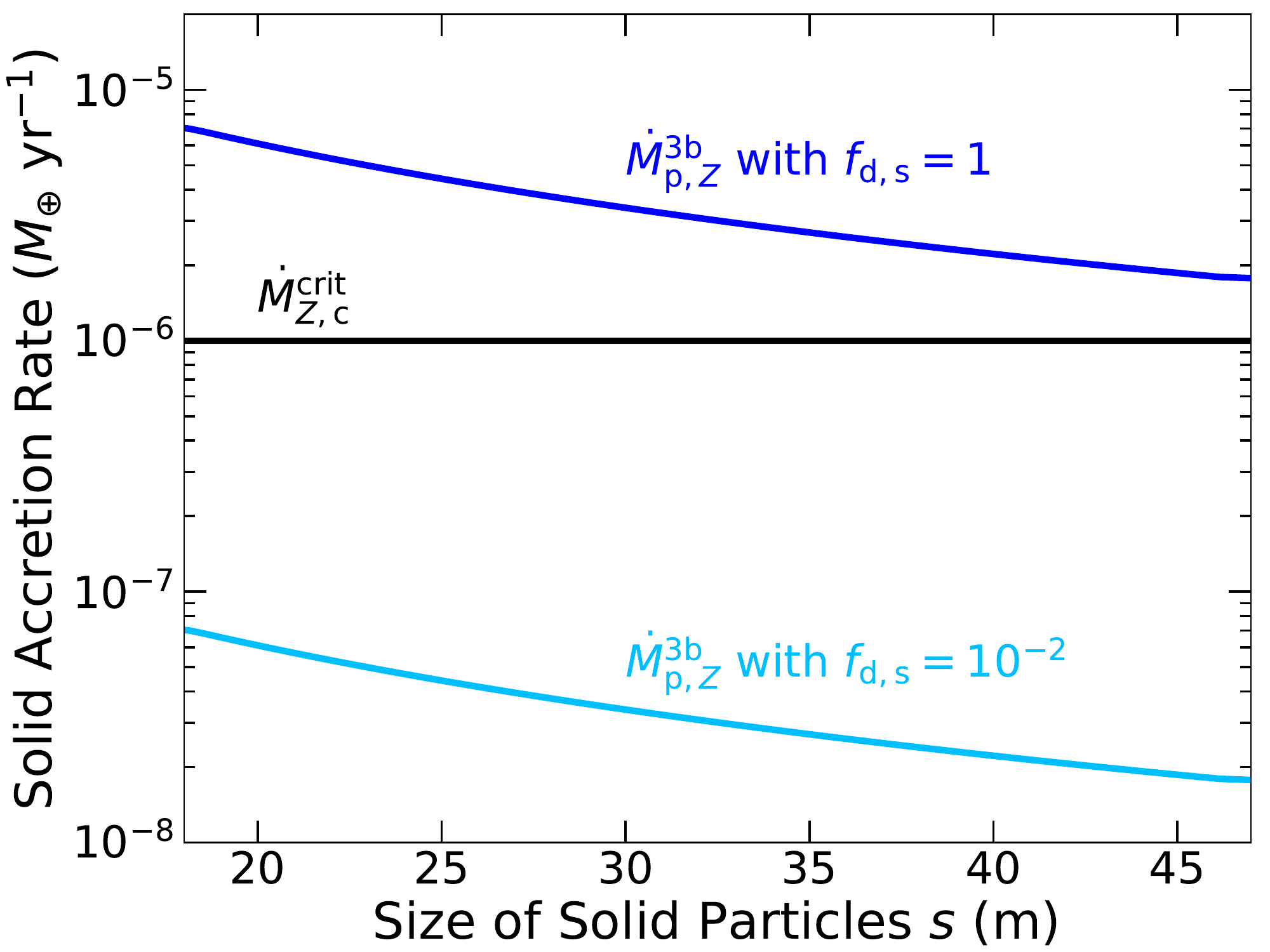}
\caption{The accretion rate of medium fragments as a function of their radius for the case that $M_{Z, \rm c}= 10 M_{\oplus}$ (Table \ref{table2}).
The values of $\dot{M}_{{\rm p}, Z}^{\rm 3b}$ with $f_{\rm d,s}=10^{-2}$ and 1 are denoted by the light blue and blue lines, respectively. 
For comparison purpose, the critical solid accretion rate ($\dot{M}_{Z, \rm c}^{\rm crit}$) is shown by the black line (see Figure \ref{fig2}).
The value of $\dot{M}_{{\rm p}, Z}^{\rm 3b}$ with $f_{\rm d,s}=1$ exceeds that of $\dot{M}_{Z, \rm c}^{\rm crit}$ for all the solid radii.
This indicates that concurrent gas accretion is halted if all the solids constituting $\Sigma_{\rm d, s}$ are in the Stokes regime.}
\label{fig5}
\end{center}
\end{figure}

The resulting accretion rate is determined by $\dot{M}_{{\rm p}, Z}^{\rm 3b}$ (see equation (\ref{eq:Mdot_Z_3b})):
\begin{eqnarray}
\label{eq:Mdot_Z_3b_St}
\dot{M}_{{\rm p}, Z}^{\rm 3b}  & \simeq &  \frac{27}{4 \sqrt{3}} \left( \frac{2\sqrt{3}}{3} \right)^{3/4} \Sigma_{\rm d, s} r^2 h_{\rm H}^{9/4} \Omega^{1/4}  
                                                                    \left( \frac{ \rho_{\rm s} s^2 }{ \rho_{\rm d, g} v_{\rm th} l_{\rm mfp} }  \right)^{-3/4} \\ \nonumber
                                              &\simeq & 3.4 \times 10^{-6} f_{\rm d,s}  \left( \frac{M_{Z, \rm c}}{10 M_{\oplus}} \right)^{3/4} \\ \nonumber
                                              & \times &  \left( \frac{s}{30 \mbox{ m}} \right)^{-3/2}       \left(  \frac{r}{10 \mbox{ au}} \right)^{-1/16}
                                                               M_{\oplus} \mbox{ yr}^{-1}.
\end{eqnarray}
Figure \ref{fig5} shows its behavior; 
for this case, we adopt that $M_{Z, \rm c}=10 M_{\oplus}$ and examine the dependence of $f_{\rm d,s}$ ($f_{\rm d,s}=10^{-2}$ and 1, see Table \ref{table2}).
While its dependence is obvious, it becomes important;
as the solid size becomes smaller, 
$\dot{M}_{{\rm p}, Z}^{\rm 3b}$ monotonically increases and eventually exceeds the critical value 
beyond which gas accretion is halted due to such efficient solid accretion ($\dot{M}_{Z, \rm c}^{\rm crit}$, see equation (\ref{eq:dotM_c_crit})).
Our results show that it occurs for the case that $f_{\rm d,s}=1$.
This, however, does not necessarily mean that concurrent gas accretion is not realized,
because the mass budget of $\Sigma_{\rm d, s}$ consists of all the solids with various sizes.
Instead, this suggests that the size distribution and abundance of solids are the key 
to reliably specifying which accretion mode in what regime becomes important for Neptune-mass planets
and accurately computing the total mass of solids accreted onto these planets.

\subsection{Accretion of small fragments} \label{sec:acc_sma_frag}

We here discuss the accretion of small fragments (Table \ref{table1}).
As with the case for the previous section,
the main differences are the regime of the drag force and the corresponding, equilibrium eccentricity.
For this case, the eccentricity is given as (see equations (\ref{eq:t_s}) and (\ref{eq:t_ext}))
\begin{eqnarray}
\frac{e^{\rm Ep}_{\rm eq}}{h_{\rm H}} & =          &  \left( \frac{27}{8\sqrt{3}} h_{\rm H} \Omega \frac{ \rho_{\rm s} s}{ \rho_{\rm d, g} v_{\rm th} }  \right)^{1/4}  \\ \nonumber
                                                         & \simeq & 1. 3  f_{\rm d,g}^{-1/4} \left(  \frac{M_{Z, \rm c}}{10 M_{\oplus}} \right)^{1/12} \left(  \frac{s}{10 \mbox{ m}} \right)^{1/4} 
                                                                              \left(  \frac{r}{10 \mbox{ au}} \right)^{3/8}.
\end{eqnarray}

\begin{figure}
\begin{center}
\includegraphics[width=8.3cm]{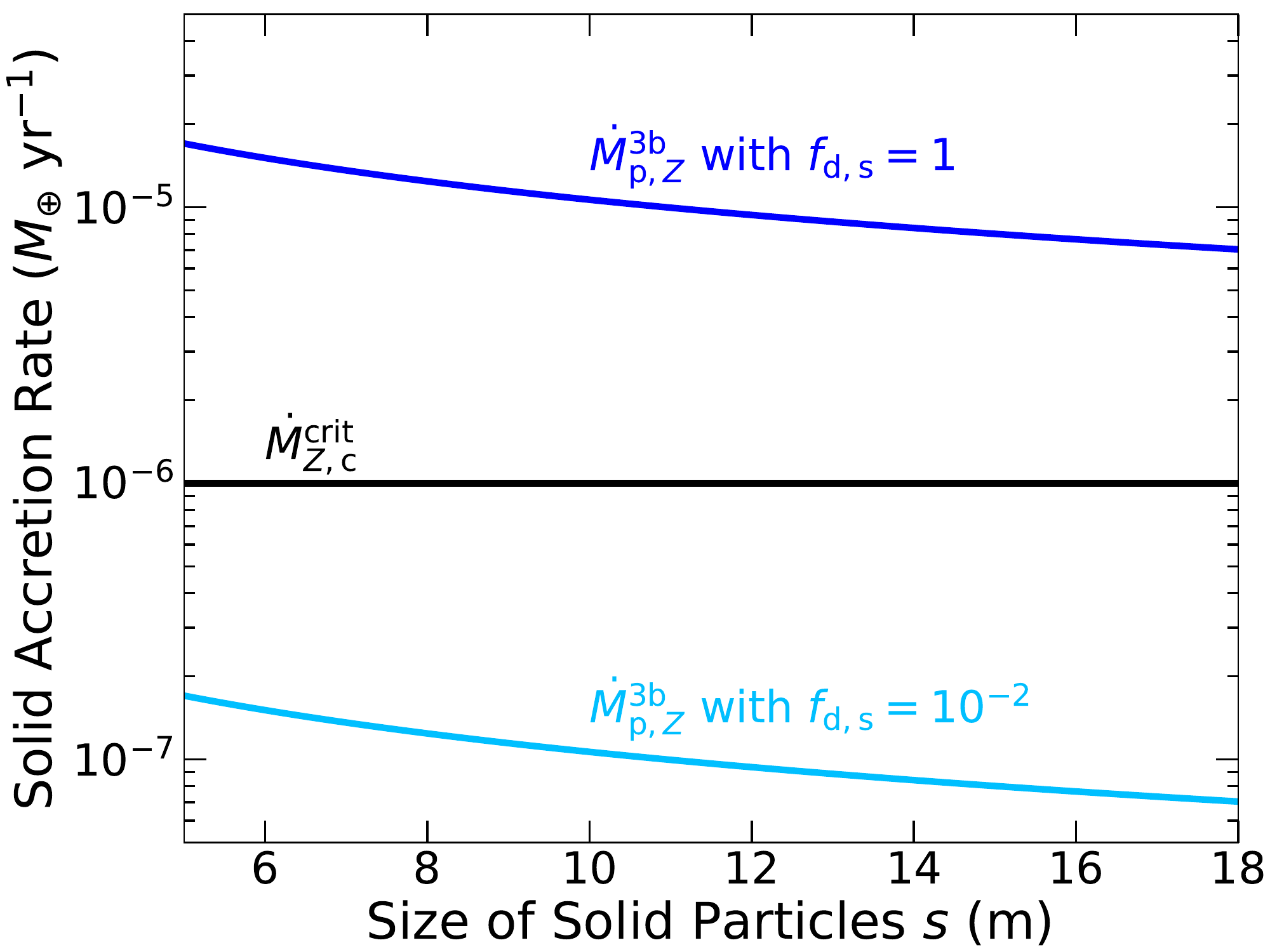}
\caption{The accretion rate of small fragments as a function of their radius for the case that $M_{Z, \rm c}= 10 M_{\oplus}$  (as in Figure \ref{fig5}, also see Table \ref{table2}).
Solid accretion can stop concurrent gas accretion for the case that  $f_{\rm d,s}=1$.}
\label{fig6}
\end{center}
\end{figure}

Figure \ref{fig6} shows the resulting accretion rate ($\dot{M}_{{\rm p}, Z}^{\rm 3b}$) for the case that $M_{Z, \rm c}= 10 M_{\oplus}$ (Table \ref{table2}).
As seen in the previous section,
we find that $\dot{M}_{{\rm p}, Z}^{\rm 3b}$ exceeds $\dot{M}_{Z, \rm c}^{\rm crit}$ if $f_{\rm d,s}=1$ is adopted.

Mathematically, $\dot{M}_{{\rm p}, Z}^{\rm 3b}$ is written as
\begin{eqnarray}
\label{eq:Mdot_Z_3b_Ep}
\dot{M}_{{\rm p}, Z}^{\rm 3b}  & \simeq & 2 \left(  \frac{27}{4 \sqrt{3}} \right)^{1/4} \Sigma_{\rm d, s} r^2 h_{\rm H}^{9/4} \Omega^{1/4}  
                                                                    \left( \frac{ \rho_{\rm s} s}{ \rho_{\rm d, g} v_{\rm th} }  \right)^{-3/4} \\ \nonumber
                                              &\simeq & 1.1 \times 10^{-5} f_{\rm d,g}^{3/4} f_{\rm d,s}  \left( \frac{M_{Z, \rm c}}{10 M_{\oplus}} \right)^{3/4} \\ \nonumber
                                              & \times &  \left( \frac{s}{10 \mbox{ m}} \right)^{-3/4}       \left(  \frac{r}{10 \mbox{ au}} \right)^{-17/8}
                                                               M_{\oplus} \mbox{ yr}^{-1}.
\end{eqnarray}

\subsection{Accretion of boulders} \label{sec:acc_bou}

We now explore the accretion of boulders (Table \ref{table1}).
For this case, we find that the computed equilibrium eccentricity of accreted solids becomes  $e_{\rm H}<1$,
and hence they reside in the shear dominated regime.
Therefore, we adopt that $e_{\rm H} = 3.2$ (Section \ref{sec:solid_ecc}), 
and $\dot{M}_{{\rm p}, Z}^{\rm disp}$ is replaced with $\dot{M}_{{\rm p}, Z}^{\rm shear}$ for the two-body mode (see equation (\ref{eq:Mdot_Z_shear})).
We, however, confirm that $\dot{M}_{{\rm p}, Z}^{\rm 3b} > \dot{M}_{{\rm p}, Z}^{\rm shear}$ for the case that $M_{Z, \rm c}= 10 M_{\oplus}$.
Accordingly, we focus on the drag-enhanced three-body mode.

\begin{figure}
\begin{center}
\includegraphics[width=8.3cm]{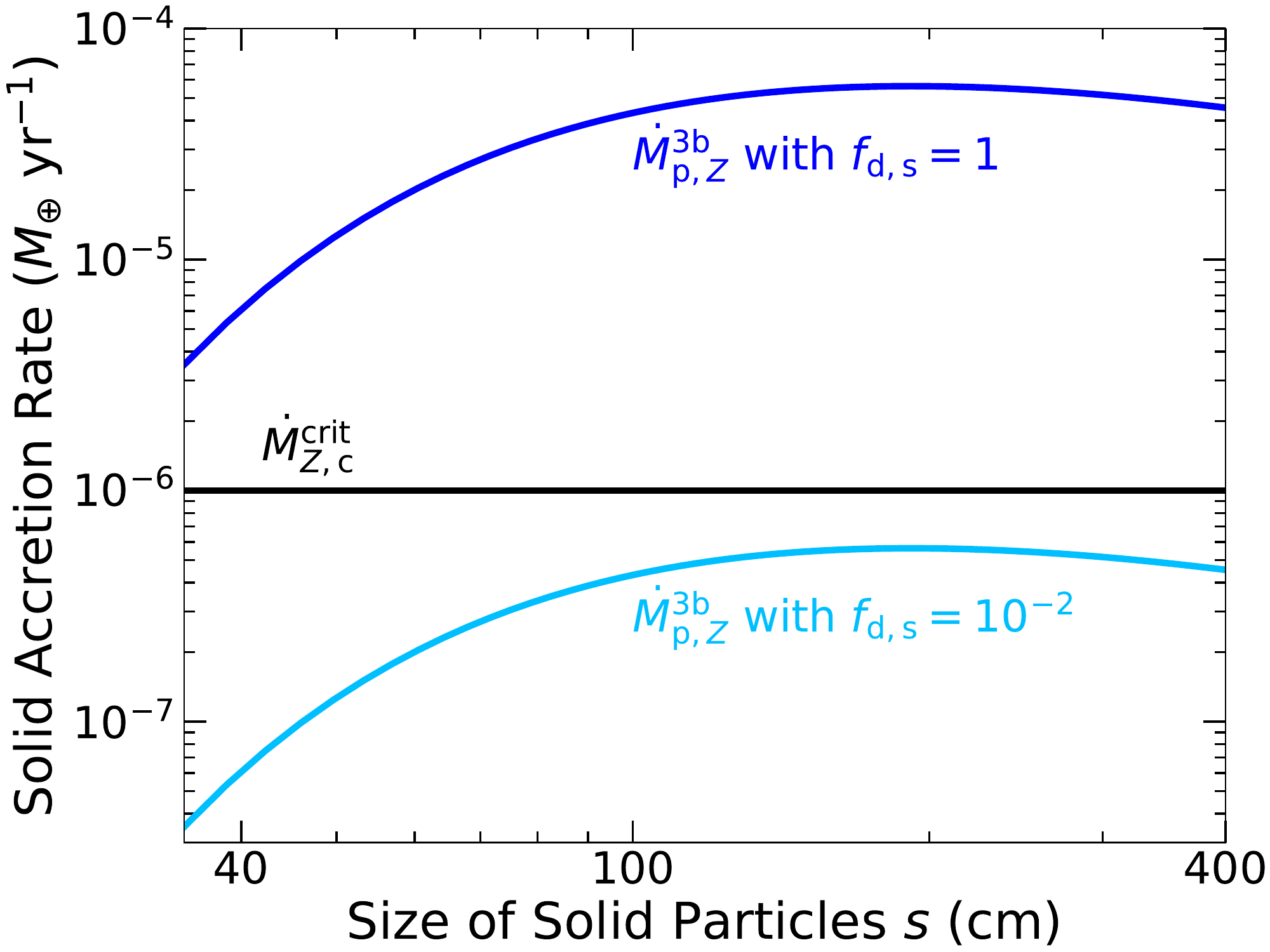}
\caption{The accretion rate of boulders as a function of their radius for the case that $M_{Z, \rm c}= 10 M_{\oplus}$ (as in Figure \ref{fig5}, also see Table \ref{table2}).
As with the case for the accretion of small and medium fragments,
$\dot{M}_{{\rm p}, Z}^{\rm 3b} > \dot{M}_{Z, \rm c}^{\rm crit}$ with $f_{\rm d,s}=1$.
The accretion rate drops rapidly when $s$ becomes smaller than about 100 cm,
which results from the factor of the exponential function (see equation (\ref{eq:Mdot_Z_3b})).}
\label{fig7}
\end{center}
\end{figure}

Figure \ref{fig7} shows the resulting accretion rate (Table \ref{table2}).
Our results show that the corresponding accretion rate is higher than the critical value for the case that $f_{\rm d,s}=1$;
this can be avoided if $f_{\rm d,s}$ is reduced to $\sim 10^{-2}$ or lower.
The rapid decrease in $\dot{M}_{{\rm p}, Z}^{\rm 3b}$ originates from the factor of the exponential function (see equation (\ref{eq:Mdot_Z_3b})).

\subsection{Accretion of pebbles} \label{sec:acc_peb}

\begin{figure}
\begin{center}
\includegraphics[width=8.3cm]{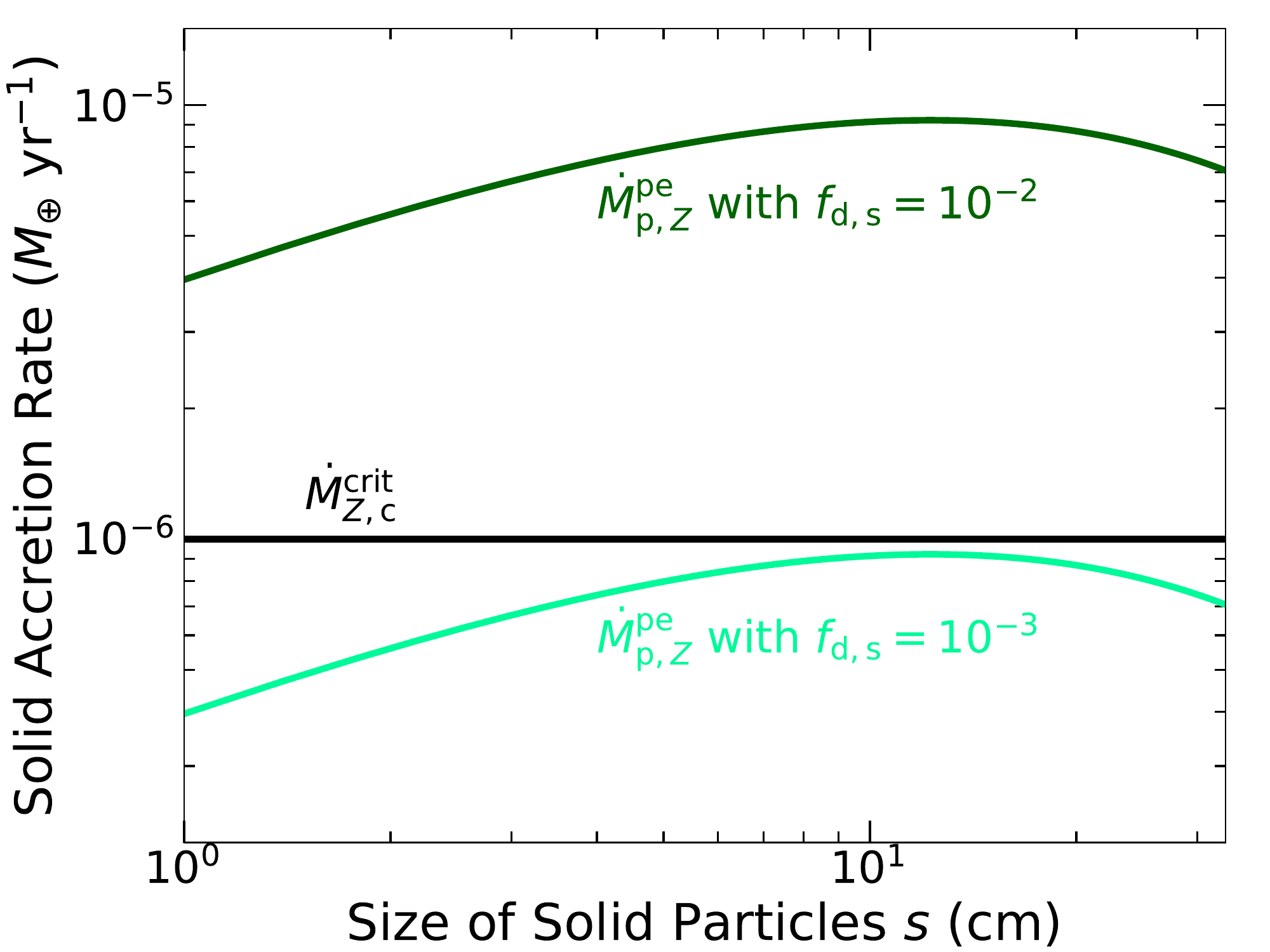}
\caption{The accretion rate of pebbles as a function of their radius for the case that $M_{Z, \rm c}= 10 M_{\oplus}$ (Table \ref{table2}).
The values of $\dot{M}_{{\rm p}, Z}^{\rm pe}$ with $f_{\rm d,s}=10^{-3}$ and $10^{-2}$ are denoted by the light green and green lines, respectively. 
As in Figure \ref{fig5}, $\dot{M}_{Z, \rm c}^{\rm crit}$ is shown by the black line for comparison purpose.
Efficient pebble accretion can stop concurrent gas accretion for the case that $f_{\rm d, s}$ is a few times larger than $10^{-3}$.}
\label{fig8}
\end{center}
\end{figure}

We finally investigate the accretion of pebbles (Table \ref{table2}).
Figure \ref{fig8} shows its resulting behavior.
As expected, pebble accretion is so efficient that concurrent gas accretion can be halted when $f_{\rm d,s}$ is a few time larger than $10^{-3}$.

\section{Application to the D/H ratios of Uranus and Neptune} \label{sec:app_UN}

In this section, we apply the gas and solid accretion discussed in Sections \ref{sec:in-situ} and \ref{sec:res1} to the formation of Uranus and Neptune.
We determine under what conditions, their D/H ratios can be reproduced.
To proceed, we use the mass distributions derived from the D/H ratio (Figure \ref{fig1});
equivalently, $M_{Z, \rm c}$ becomes a function of $F_{{\rm H}_2 {\rm O,ice}}$.
We will show below that the gas accretion timescale constrains possible values of $f_{\rm grain}$.
Accordingly, the required parameters to compute solid accretion rates change from five ($f_{\rm d,s}$, $M_{Z, \rm c}$, $r_{\rm p}$, $f_{\rm grain}$, $s$) 
to four ($f_{\rm d,s}$, $F_{{\rm H}_2 {\rm O,ice}}$, $r_{\rm p}$, $s$).
Therefore, we here compute plausible values of $f_{\rm d,s}$ as a function of $s$, $r_{\rm p}$, and $F_{{\rm H}_2 {\rm O,ice}}$.

\subsection{Gas accretion} \label{sec:app_UN_gas}

We first explore gas accretion.
We focus on its timescale ($\Delta t_{XY}$) that is required to reproduce the envelope mass ($M_{XY}$) of Uranus and Neptune (Figure \ref{fig1}).
Mathematically, $\Delta t_{XY}$ can be computed as (see equation (\ref{eq:tau_KH}))
\begin{eqnarray}
\label{eq:delta_t_XY}
\Delta t_{XY} & \simeq & \frac{M_{XY}}{\dot{M}_{XY}}  \simeq \frac{M_{XY}}{M_{Z, \rm c}} \tau_{\rm KH} \\ 
                      & \simeq &   2 \times 10^6 f_{\rm grain} \left( \frac{10^c}{10^7} \right) \left( \frac{M_{XY}}{2M_{\oplus}} \right) \left( \frac{M_{Z, \rm c}}{10M_{\oplus}} \right)^{-(d+1)} \mbox{ yr}, \nonumber
\end{eqnarray}
under the assumption that the mass growth rate during the gas accretion stage is determined by $\dot{M}_{XY} \simeq M_{Z, \rm c} / \tau_{\rm KH}$ (equation (\ref{eq:dotM_p})).

\begin{figure*}
\begin{minipage}{17cm}
\begin{center}
\includegraphics[width=8.3cm]{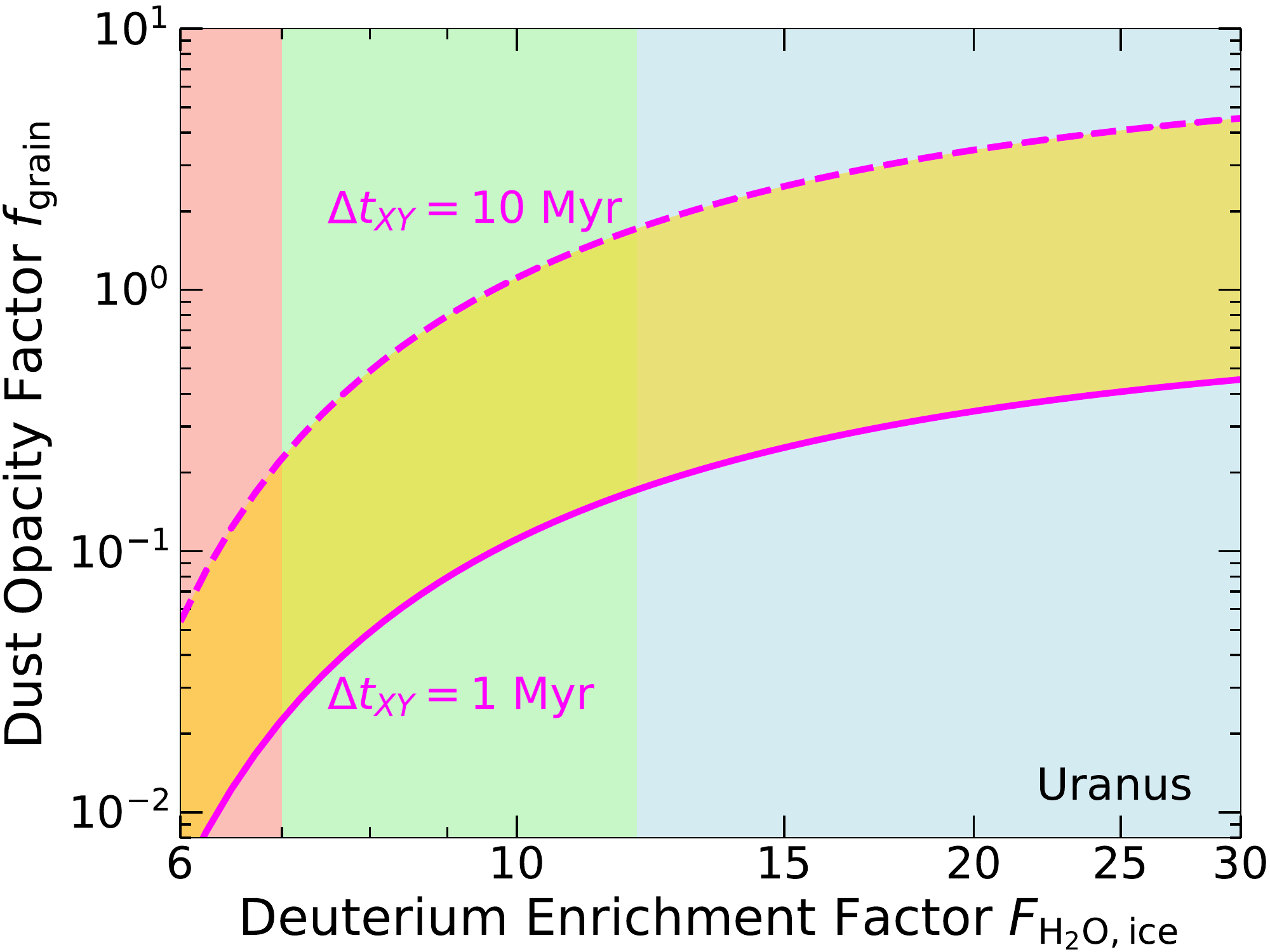}
\includegraphics[width=8.3cm]{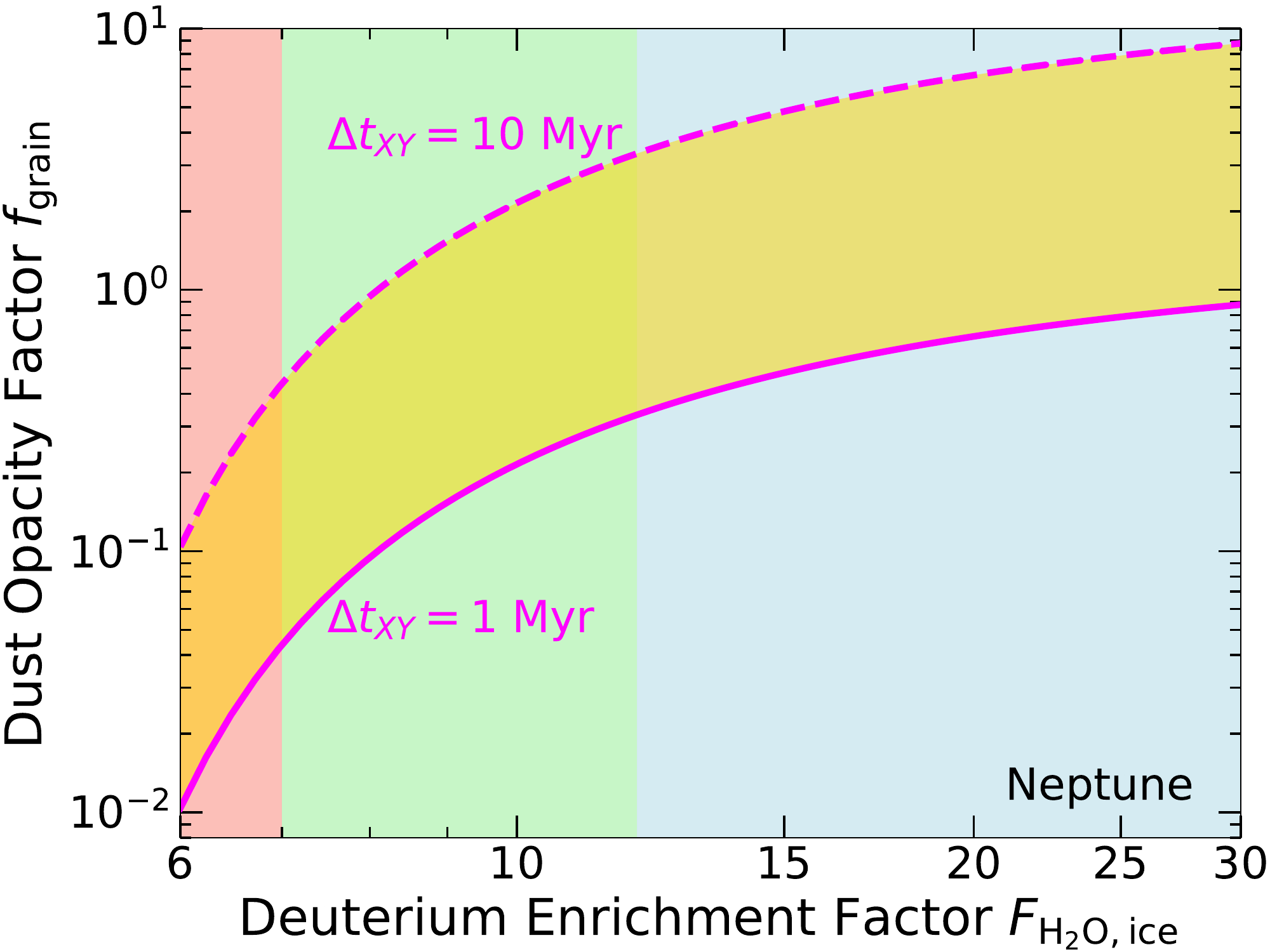}
\caption{The dust opacity factor ($f_{\rm grain}$) required to reproduce the envelope mass ($M_{XY}$) of Uranus and Neptune
as a function of the deuterium enrichment factor ($F_{{\rm H}_2 {\rm O,ice}}$) of accreted solids on the left and right panels, respectively.
The solid and dashed magenta lines denote the cases that the gas accretion timescale ($\Delta t_{XY}$) corresponds to 1 and 10 Myr, respectively.
The red, green, and blue shaded regions are defined in Figure \ref{fig1}.
The range of possible values of $f_{\rm grain}$ is represented by the yellow shaded region.
Both the cores of Uranus and Neptune can accrete the gas mass of $M_{XY}$ within the disk lifetime for certain values of $f_{\rm grain}$.}
\label{fig9}
\end{center}
\end{minipage}
\end{figure*}

The above equation indicates that combined with the results of Figure \ref{fig1},
$f_{\rm grain}$ can be uniquely determined for given values of $\Delta t_{XY}$ and $F_{{\rm H}_2 {\rm O,ice}}$.
Since gas accretion should be completed within the gas disk lifetime, that is $\Delta t_{XY} \simeq 10^6-10^7$ yr,
a possible range of $f_{\rm grain}$ becomes a function of $F_{{\rm H}_2 {\rm O,ice}}$.
Figure \ref{fig9} shows the results.
For completeness, $F_{{\rm H}_2 {\rm O,ice}}$ varies from 6 to 30.
We find that $f_{\rm grain}$ increases with increasing $F_{{\rm H}_2 {\rm O,ice}}$.
This is simply because of the $M_{Z, \rm c}$ dependence on $f_{\rm grain}$.
Our results therefore suggest that when (proto)Uranus/(proto)Neptune obtains the current D/H ratio via concurrent gas and solid accretion,
their envelope dust opacity factor should be in the range of 0.1 to 10 at the formation stage.

In the following sections, we will use these values to constrain solid accretion.
Note that the above range of $f_{\rm grain}$ should be viewed as an upper bound;
a lower limit should be derived from either core formation timescales or dust evolution calculations in planetary envelopes (Section \ref{sec:disc_dust}).

\subsection{Solid accretion}

We now investigate solid accretion.
We first compute a critical value of $f_{\rm d,s}$ ($f_{\rm d,s} ^{\rm crit}$). 
Since the critical solid accretion rate ($\dot{M}_{Z, \rm c}^{\rm crit}$) is given by equation (\ref{eq:dotM_c_crit}),
$f_{\rm d,s} ^{\rm crit}$ can be defined as (see equations (\ref{eq:tau_KH}) and (\ref{eq:delta_t_XY}))
\begin{eqnarray}
\label{eq:f_d_s_crit}
f_{\rm d,s} ^{\rm crit} & \equiv & \frac{M_{Z, \rm e} / \Delta t_{XY} }{\dot{M}_{Z, \rm c}^{\rm crit}}  \\ \nonumber
                      & \simeq &    \left( \frac{10^c}{10^7} \right)^{-1}  \left( \frac{ M_{Z, \rm e} }{ 2M_{\oplus}} \right)  \left( \frac{M_{XY}}{2M_{\oplus}} \right)^{-1}        
                       \left( \frac{M_{Z, \rm c}}{10M_{\oplus}} \right)^{1-1/b+d} , 
\end{eqnarray}
where $M_{Z, \rm e}$ is the solid mass in the envelope at the formation stage (see equation (\ref{eq:M_Z})).
Note that the dependence on $f_{\rm grain}$ disappears in the configuration adopted in this work (i.e., $a=b$, see equation (\ref{eq:M_c_crit})).
Therefore, $f_{\rm d,s} ^{\rm crit}$ is uniquely determined for a given value of  $F_{{\rm H}_2 {\rm O,ice}}$.

\begin{figure*}
\begin{minipage}{17cm}
\begin{center}
\includegraphics[width=8.3cm]{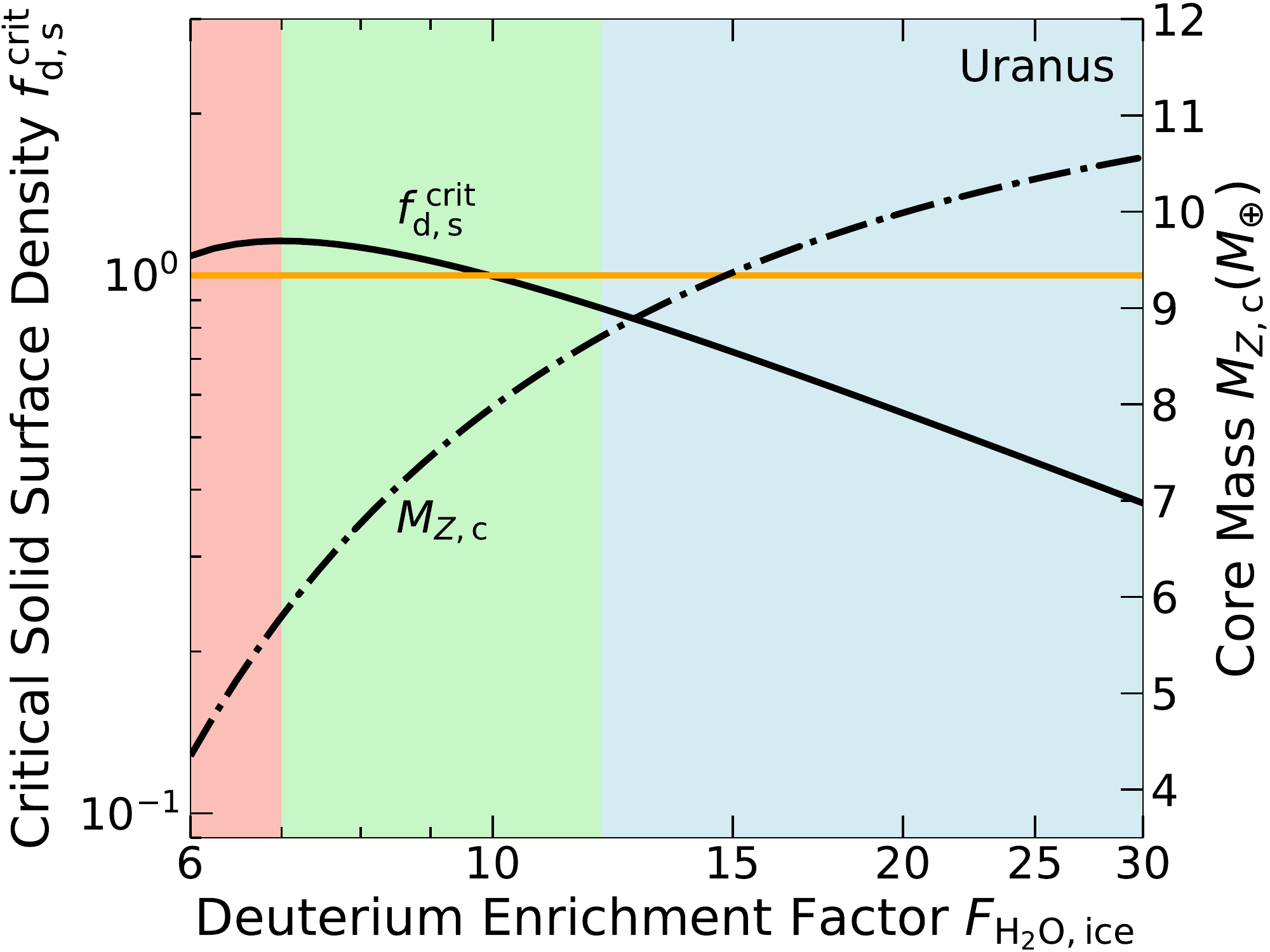}
\includegraphics[width=8.3cm]{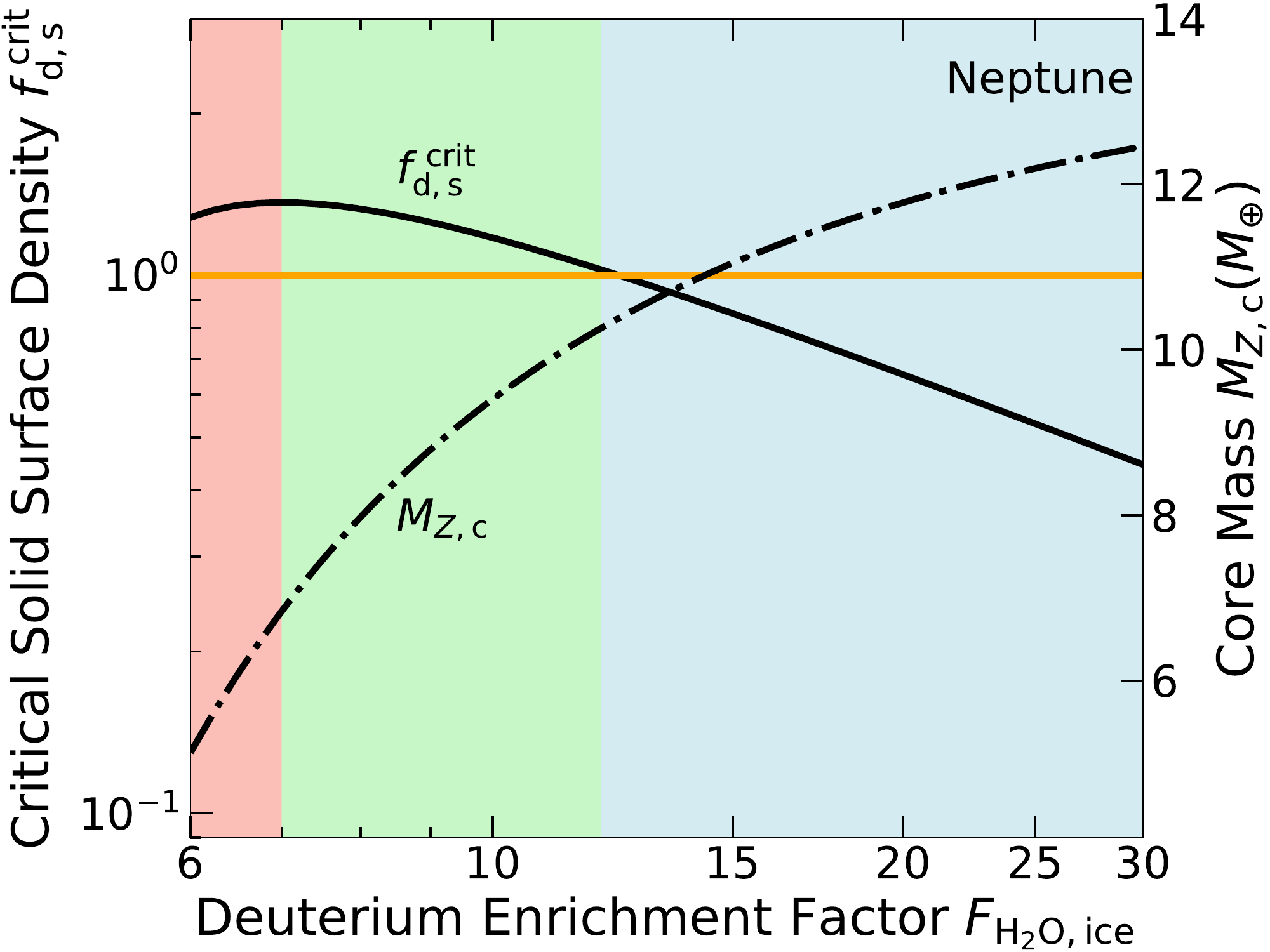}
\caption{ 
The critical solid surface density enhancement/reduction factor ($f_{\rm d,s} ^{\rm crit}$) as a function of $F_{{\rm H}_2 {\rm O,ice}}$ 
for Uranus and Neptune on the left and right panels, respectively (see the black solid line).
Their corresponding core mass is denoted by the black dash-dotted line (the right axis).
For comparison purpose, the orange solid line is plotted, which corresponds to the MMSN model (i.e., $f_{\rm d,s}=1$).
The value of $f_{\rm d,s} ^{\rm crit}$ determines whether a certain solid accretion mode can achieve concurrent accretion.}
\label{fig10}
\end{center}
\end{minipage}
\end{figure*}

Figure \ref{fig10} shows the resulting behavior of $f_{\rm d,s} ^{\rm crit}$ (the left axis).
For reference, the core mass is also plotted (the right axis).
The value of $f_{\rm d,s} ^{\rm crit}$ can be used to judge whether concurrent solid accretion is realized;
when a certain accretion mode leads to the condition that $f_{\rm d,s} \la f_{\rm d,s} ^{\rm crit}$,
the resulting accretion rate is so high that concurrent accretion cannot take place
unless the solid surface density is reduced somehow.
On the other hand, when $f_{\rm d,s} \ga f_{\rm d,s} ^{\rm crit}$,
the corresponding accretion rate is slow enough to realize concurrent accretion.
If $f_{\rm d,s} > 1$, then some density enhancement (relative to the MMSN) is needed to complete solid accretion with the disk lifetime.

We then explore solid accretion onto (proto)Uranus and (proto)Neptune.
As done in equation (\ref{eq:f_d_s_crit}), we compute the value of $f_{\rm d, s}$ that is needed to reproduce the value of $M_{Z, \rm e}$ (see Figure \ref{fig1}):
\begin{equation}
\label{eq:f_d_s}
f_{\rm d,s} =  \frac{M_{Z, \rm e} / \Delta t_{XY} }{\dot{M}_{{\rm p}, Z}/f_{\rm d, s} },  
\end{equation}
where the most dominant solid accretion mode is adopted for $\dot{M}_{{\rm p}, Z}$ in each regime 
($\dot{M}_{{\rm p}, Z}^{\rm disp}$ vs $\dot{M}_{{\rm p}, Z}^{\rm 3b}$ vs $\dot{M}_{{\rm p}, Z}^{\rm pe}$, see Table \ref{table1}, also see Section \ref{sec:in-situ}).

Equation (\ref{eq:f_d_s}) differs from equation (\ref{eq:f_d_s_crit}) in terms of the following two points:
the first one is that $f_{\rm d,s}$ is now a function of $f_{\rm grain}$ through $ \Delta t_{XY}$ (see equation (\ref{eq:delta_t_XY})).
In this work, we use the possible values of $f_{\rm grain}$ constrained by gas accretion (Figure \ref{fig9}),
that is, $ \Delta t_{XY} = 10^6$ yr or $10^7$ yr.
The second point is that $\dot{M}_{{\rm p}, Z}$ is a function of $r_{\rm p}$, $s$, and $f_{\rm grain}$ when $\dot{M}_{{\rm p}, Z} = \dot{M}_{{\rm p}, Z}^{\rm disp}$ (see equation (\ref{eq:Mdot_Z_disp})),
and a function of $r_{\rm p}$ and $s$ when $\dot{M}_{{\rm p}, Z} = \dot{M}_{{\rm p}, Z}^{\rm 3b}$ or $\dot{M}_{{\rm p}, Z}^{\rm pe}$ (see equation (\ref{eq:Mdot_Z_3b}) or (\ref{eq:Mdot_Z_pe}));
hence, $f_{\rm d,s}$ also becomes a function of $r_{\rm p}$ and $s$, and the additional dependence of $f_{\rm grain}$ comes in for the former case.
In this section, we set that $r_{\rm p}=10$ au and compute $f_{\rm d,s}$ as a function of $F_{{\rm H}_2 {\rm O,ice}}$ for a given value of $s$;
a parameter study about $r_{\rm p}$ will be undertaken in Section \ref{sec:app_UN_param}.

\begin{figure*}
\begin{minipage}{17cm}
\begin{center}
\includegraphics[width=8.3cm]{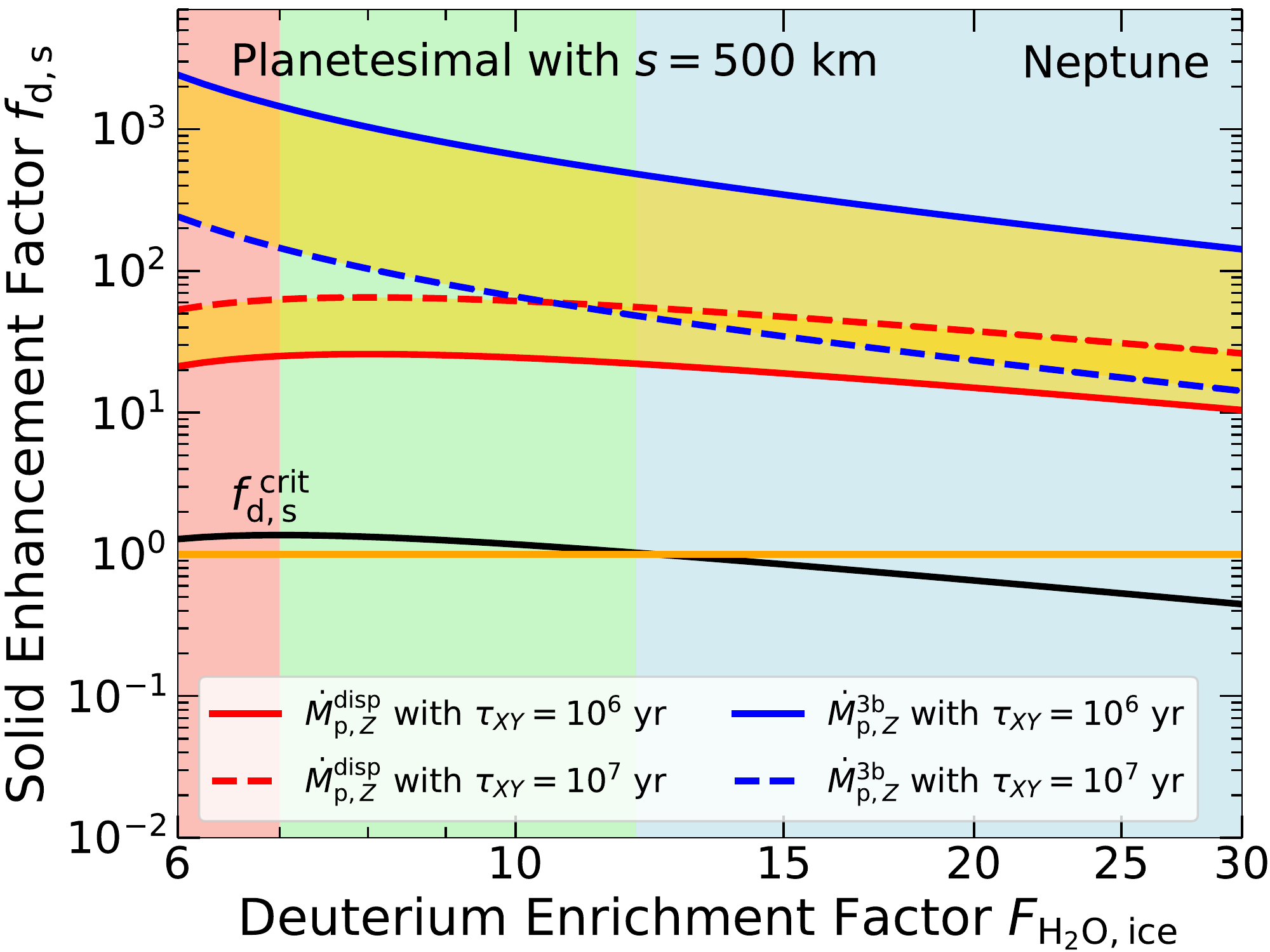}
\includegraphics[width=8.3cm]{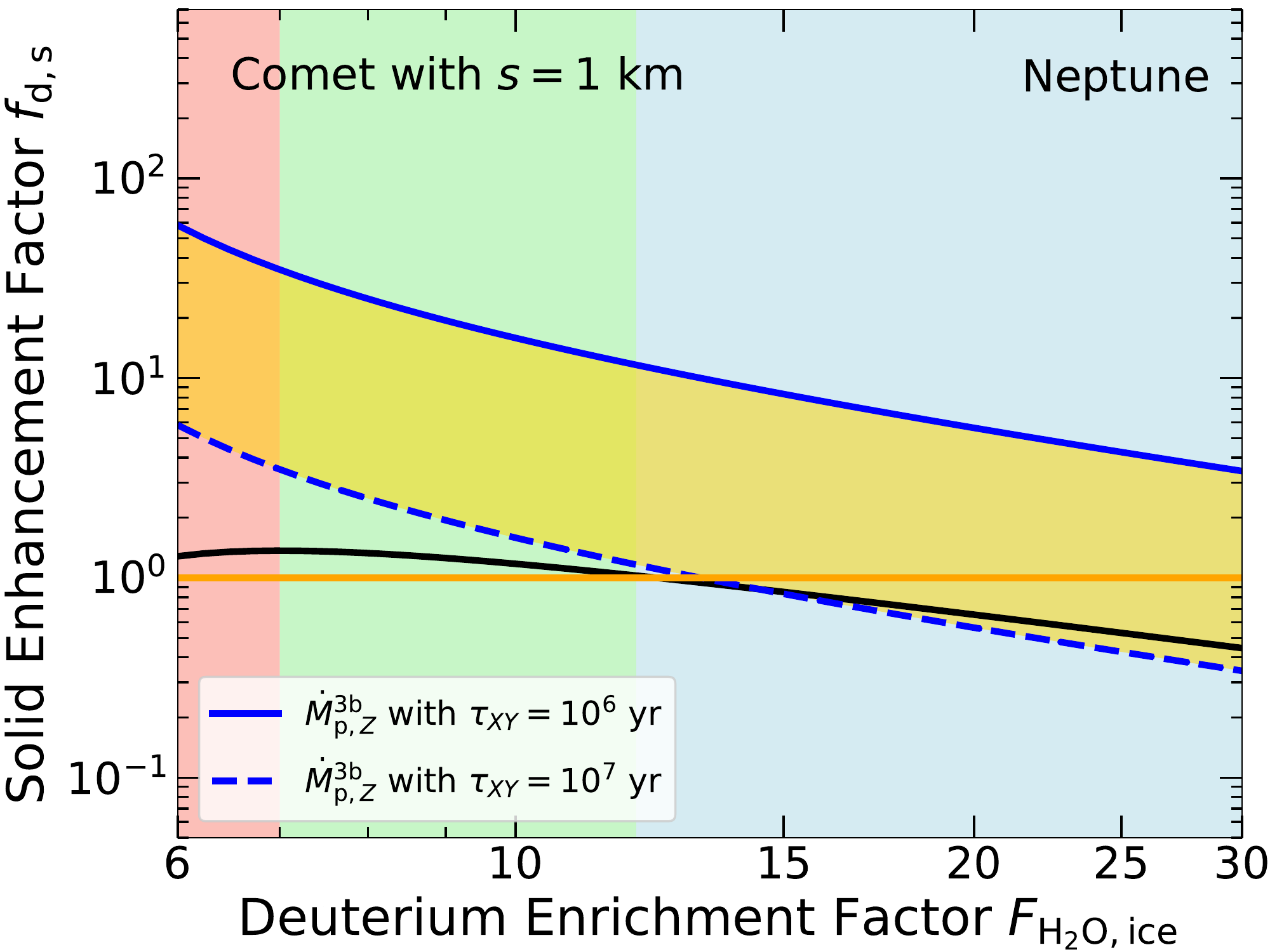}
\includegraphics[width=8.3cm]{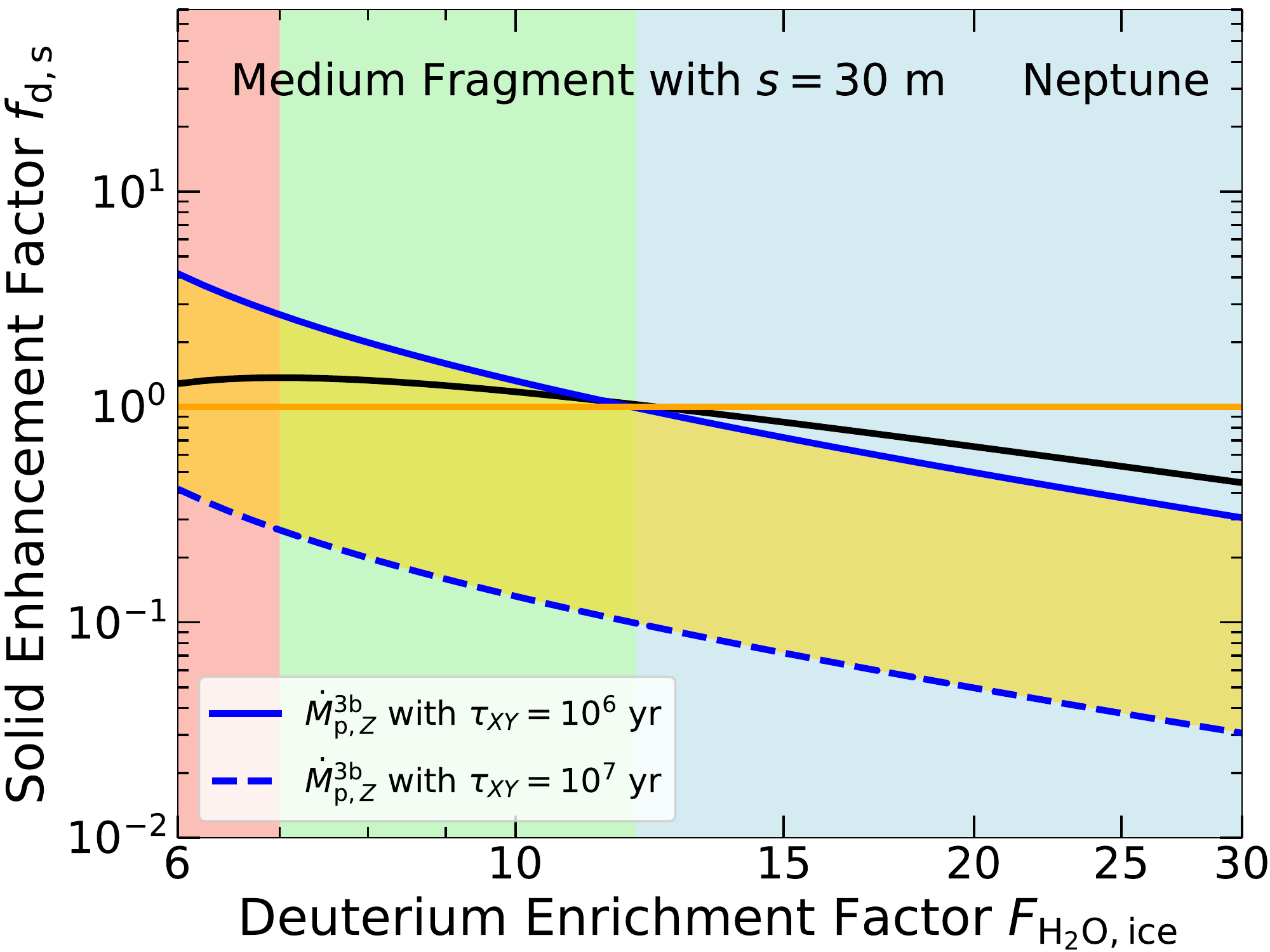}
\includegraphics[width=8.3cm]{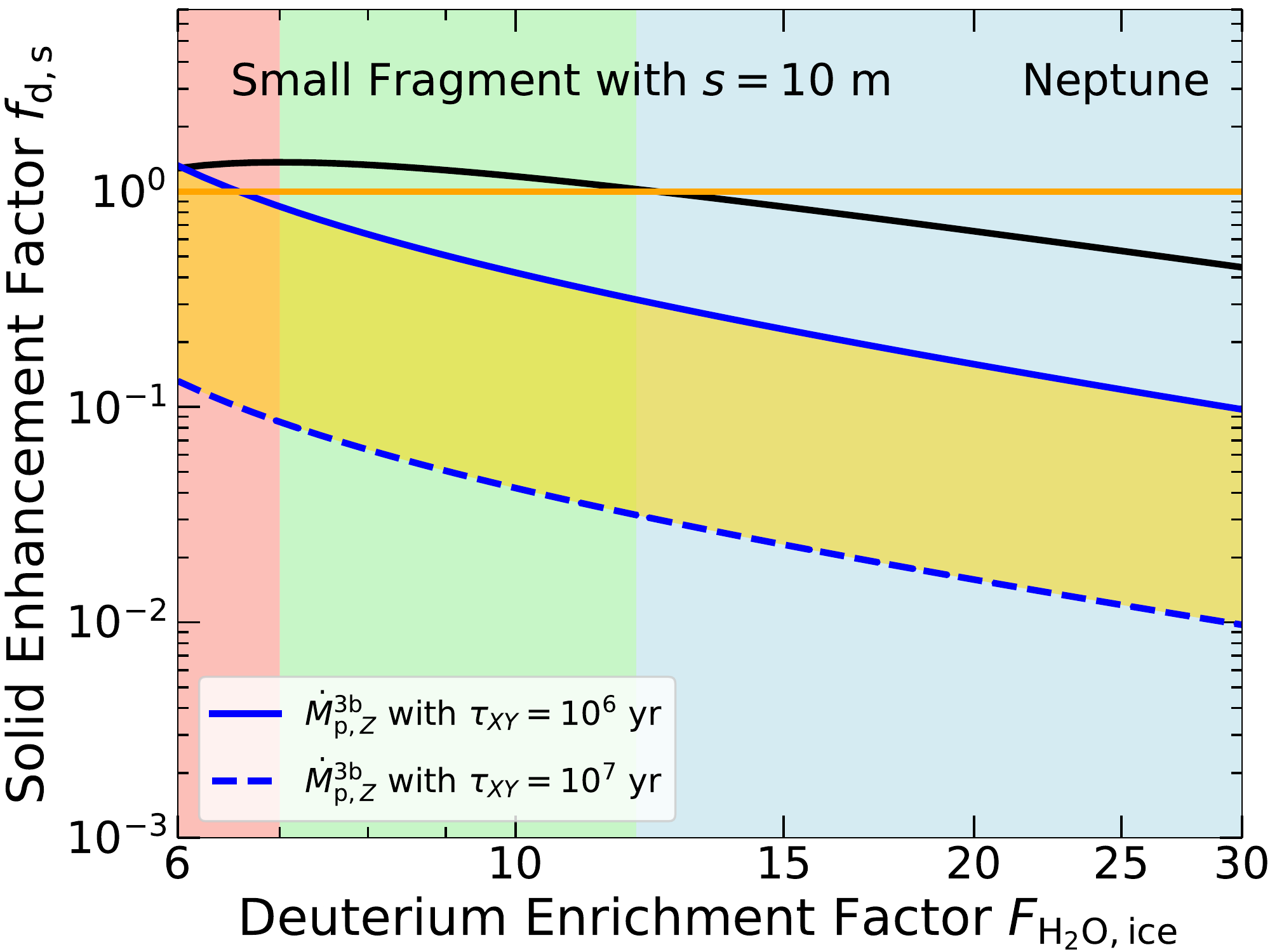}
\includegraphics[width=8.3cm]{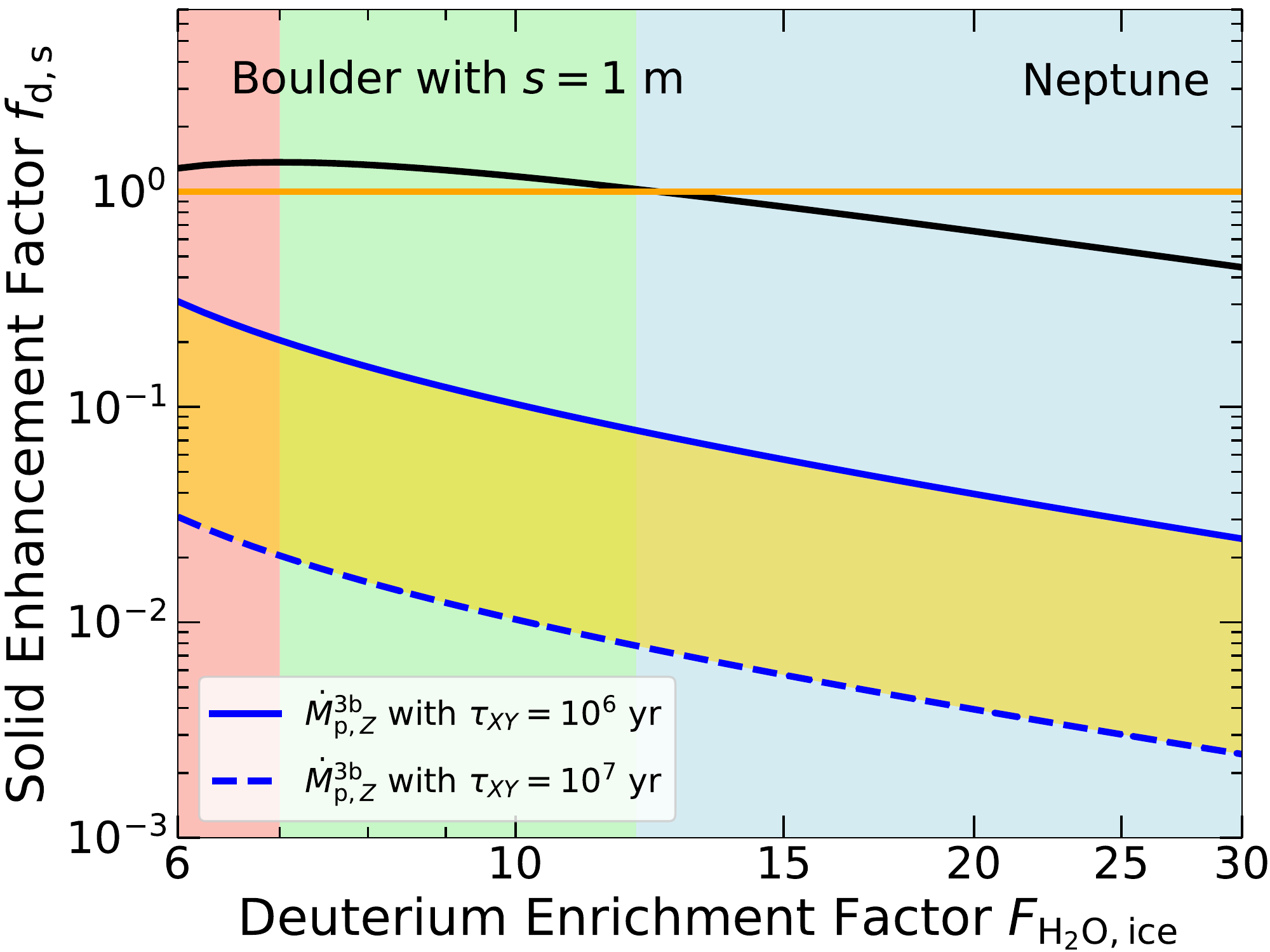}
\includegraphics[width=8.3cm]{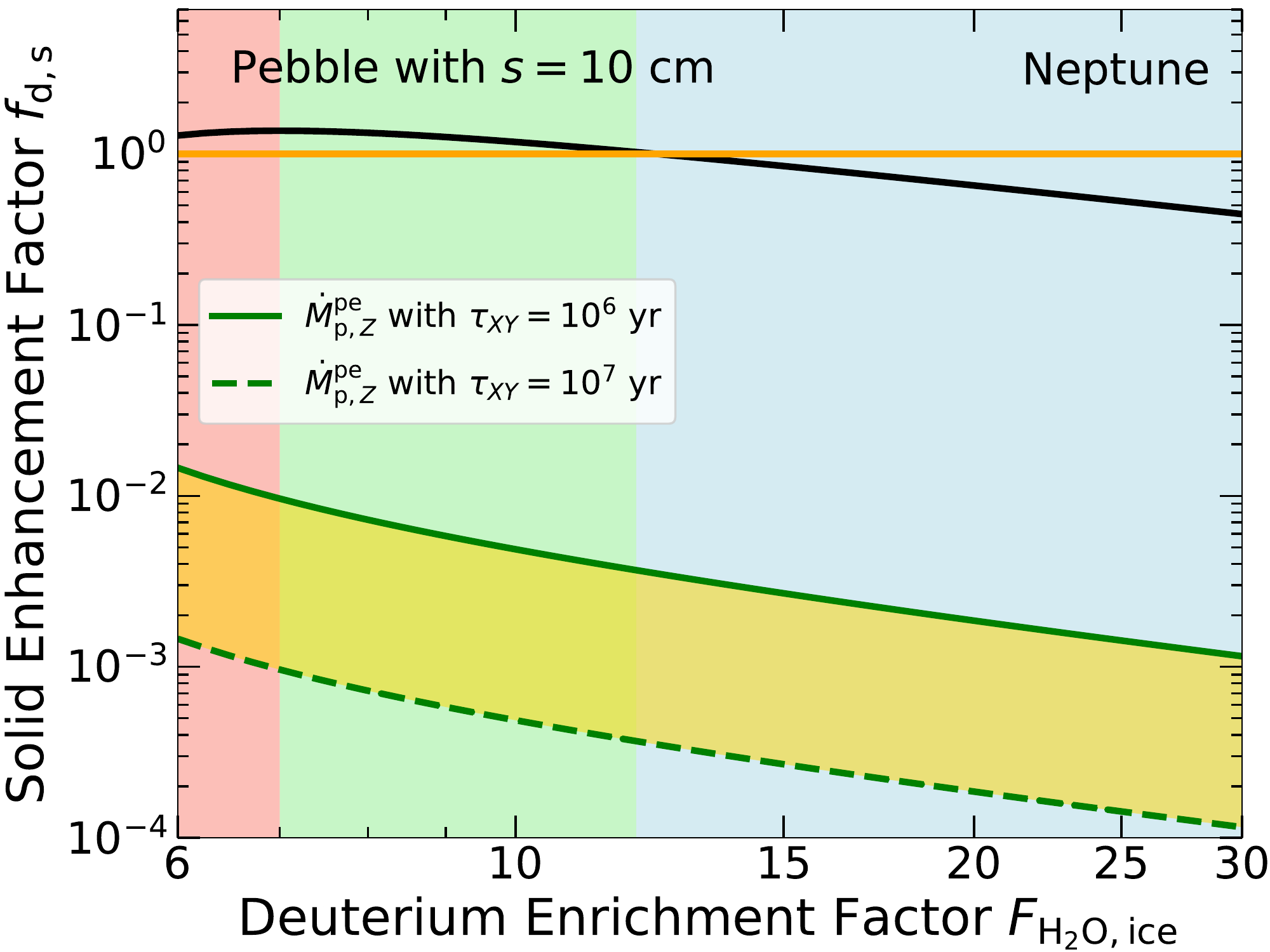}
\caption{The enhancement factor of the solid surface density ($f_{\rm d, s}$) required to reproduce the estimated values of $M_{Z, \rm e}$ for Neptune.
As done in Figure \ref{fig10}, $M_{Z, \rm c}$ is computed as a function of $F_{{\rm H}_2 {\rm O,ice}}$ (also see Figure \ref{fig1}).
Also, it is set that $r_{\rm p}= 10$ au.
The adopted accretion mode and model parameter are denoted as the legend;
the yellow shaded region corresponds to the possible values of $f_{\rm grain}$ (see Figure \ref{fig9}).
On all the panels, the black line represents $f_{\rm d,s}^{\rm crit}$, 
and as the eye's guide, the MMSN case (that is, $f_{\rm d, s}=1$) is shown by the orange horizontal line (as in Figure \ref{fig10}).
It is clear that significant enhancement of the solid surface density is required to reproduce the value of $M_{Z, \rm e}$ by the accretion of large-sized bodies.
The accretion of small-sized solids likely plays the primary role for deuterium enrichment of Uranus and Neptune.}
\label{fig11}
\end{center}
\end{minipage}
\end{figure*}

Figure \ref{fig11} summarizes the results.
We have confirmed that the results for both the cases of Uranus and Neptune are qualitatively similar,
and hence show the Neptune case only (see Section \ref{sec:app_UN_param} for the Uranus case).
As examples, we choose certain radii of solids in each regime.
All the cases are briefly discussed below; 
more discussions are provided in Section \ref{sec:res1}.

{\it Planetesimals.}
If planetesimals would be the main agent to produce the current value of the D/H ratio of Uranus and Neptune,
then the solid enhancement factor ($f_{\rm d, s}$) should be at least one order of magnitude higher than the MMSN model.
It is interesting that for the case of $\dot{M}_{{\rm p}, Z}^{\rm disp}$, $f_{\rm d, s}$ becomes higher for longer $\tau_{XY}$ (i.e., higher $f_{\rm grain}$),
while for the case of $\dot{M}_{{\rm p}, Z}^{\rm 3b}$, the trend is opposite.
This difference comes from that $\dot{M}_{{\rm p}, Z}^{\rm disp}$ is also a function of $f_{\rm grain}$ via $\eta_{\rm gap}$ (see equations (\ref{eq:Mdot_Z_disp}) and (\ref{eq:eta})).
Thus, massive disks are required to reproduce the current value of the D/H ratio of Uranus and Neptune by the in-situ accretion of planetesimals.

{\it Comets and large fragments.}
In contrast to the planetesimal case,
the accretion of comets and large fragments can significantly contribute to the deuterium enrichment of Uranus and Neptune 
without the enhancement of the solid surface density, when $F_{{\rm H}_2 {\rm O,ice}} \ga 12$.
It is therefore likely that these solids play the major role for the D/H ratio of Uranus and Neptune.

{\it Medium fragments.}
When the radius of accreted solids resides in the regime of medium fragments,
the corresponding accretion rate becomes too high to achieve concurrent gas accretion, when $F_{{\rm H}_2 {\rm O,ice}}  \ga 12$.
Therefore, the effect of the size distribution needs to be taken into account to reduce the solid surface density.

{\it Small fragments.}
As with the case of medium fragments,
the accretion rate of small fragments is high enough to prevent concurrent gas accretion.
This suggests that small fragments can contribute to deuterium enrichment of Uranus and Neptune considerably,
under certain conditions (that is, their surface density should be decreased by about one order of magnitude).

{\it Boulders.}
The accretion of boulders leads to even higher rates,
so that further reduction of $f_{\rm d,s}$ is needed.
Thus, boulders can also play an important role to better understand the current values of the D/H ratio of Uranus and Neptune.

{\it Pebbles.}
As expected, pebble accretion achieves the highest accretion rate.
This puts a constraint that $f_{\rm d,s} \la 10^{-3}$.
Consequently, pebbles are one important agent to better reproduce the D/H ratio of Uranus and Neptune.

\subsection{Parameter study} \label{sec:app_UN_param}

We now conduct a parameter study about the position of (proto)Uranus and (proto)Neptune.

\begin{figure*}
\begin{minipage}{17cm}
\begin{center}
\includegraphics[width=8.3cm]{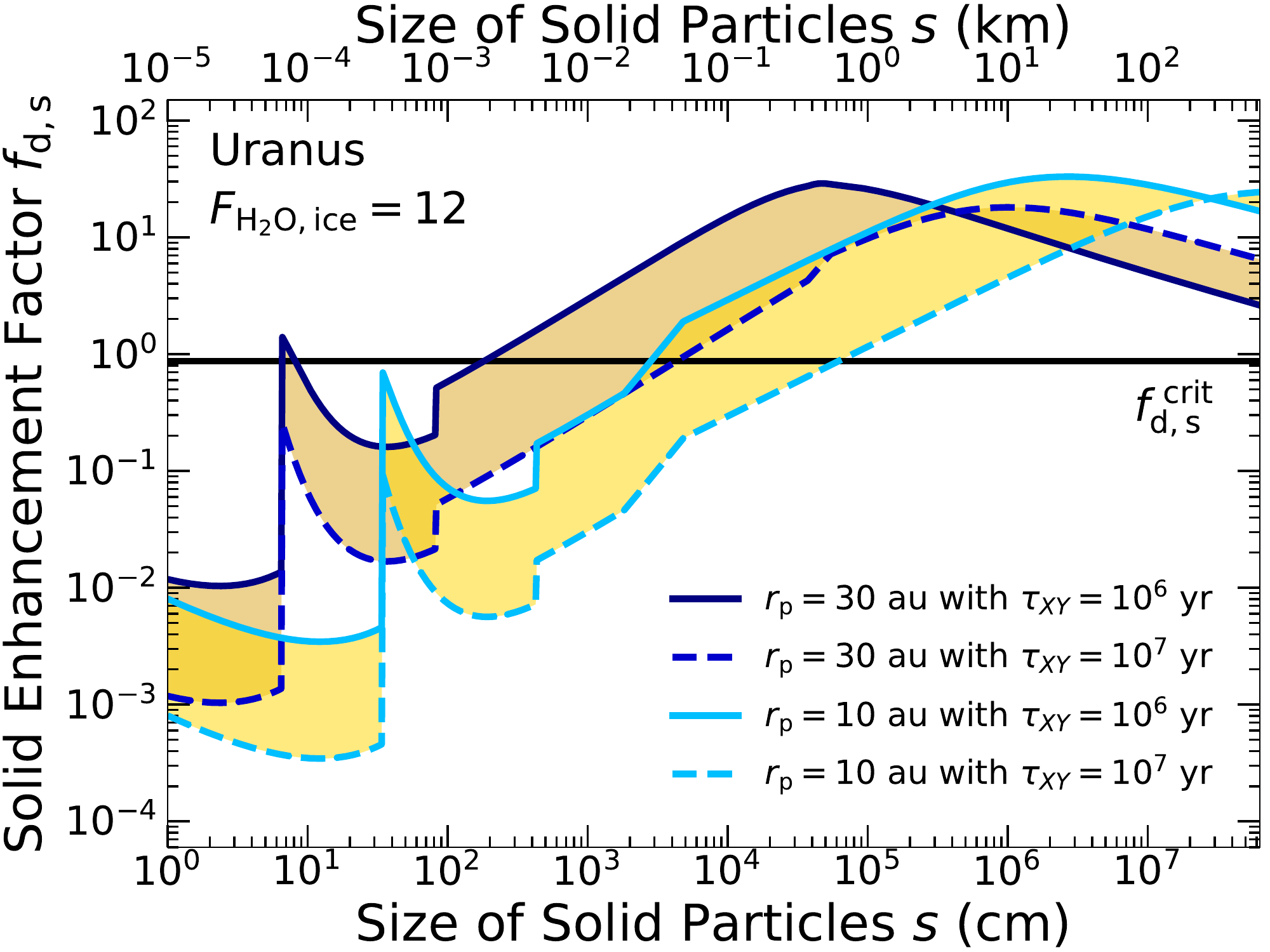}
\includegraphics[width=8.3cm]{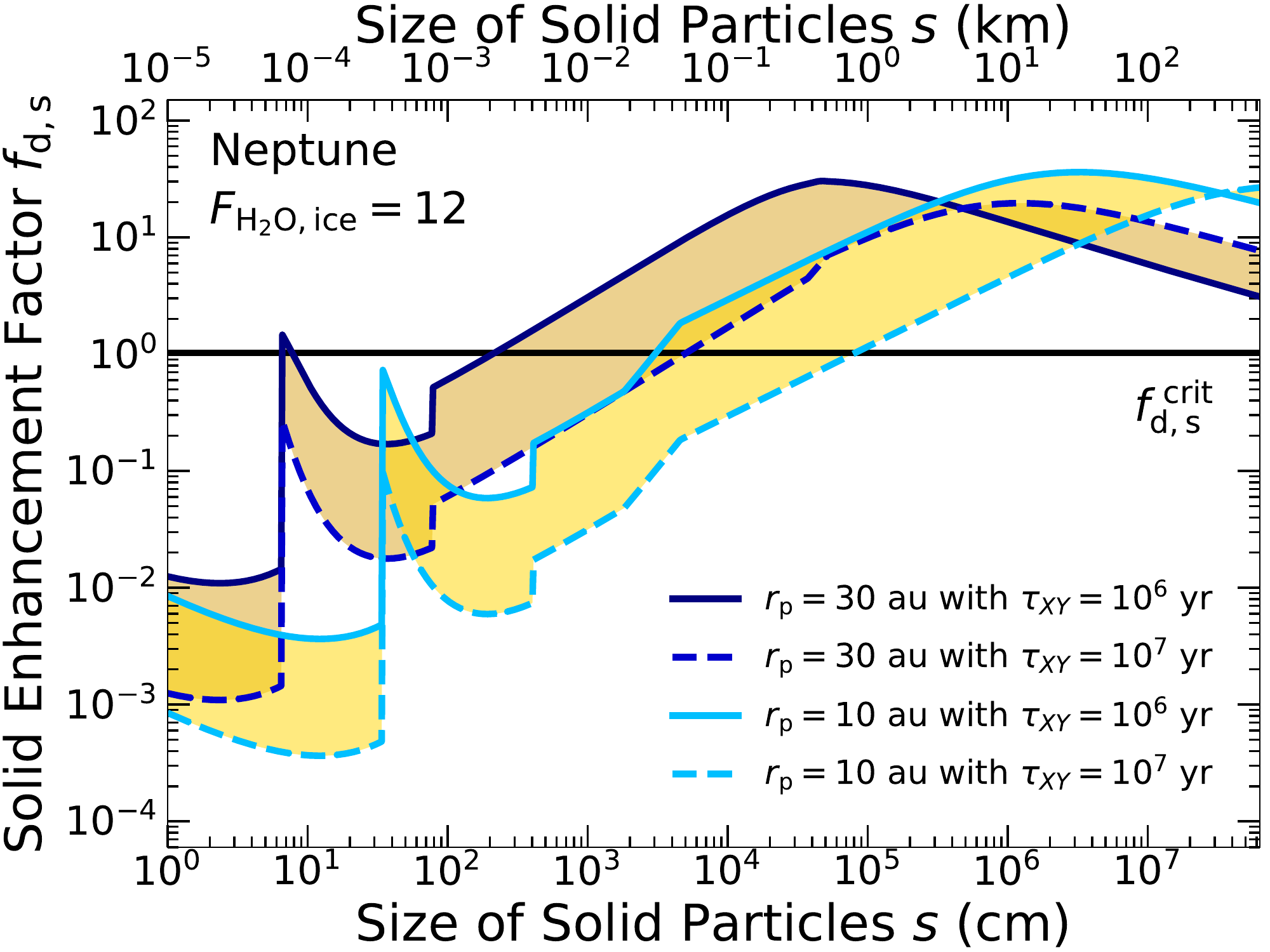}
\includegraphics[width=8.3cm]{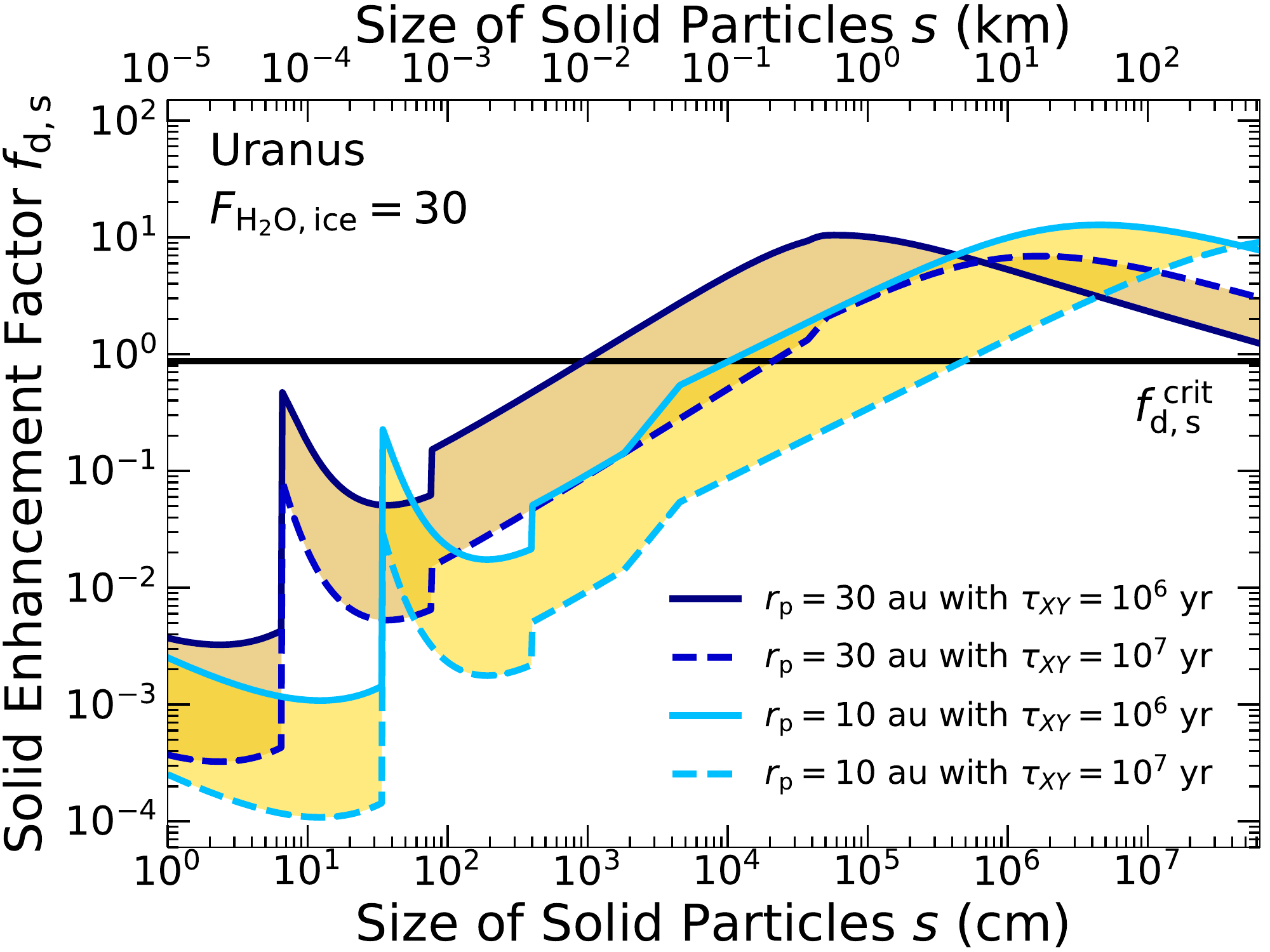}
\includegraphics[width=8.3cm]{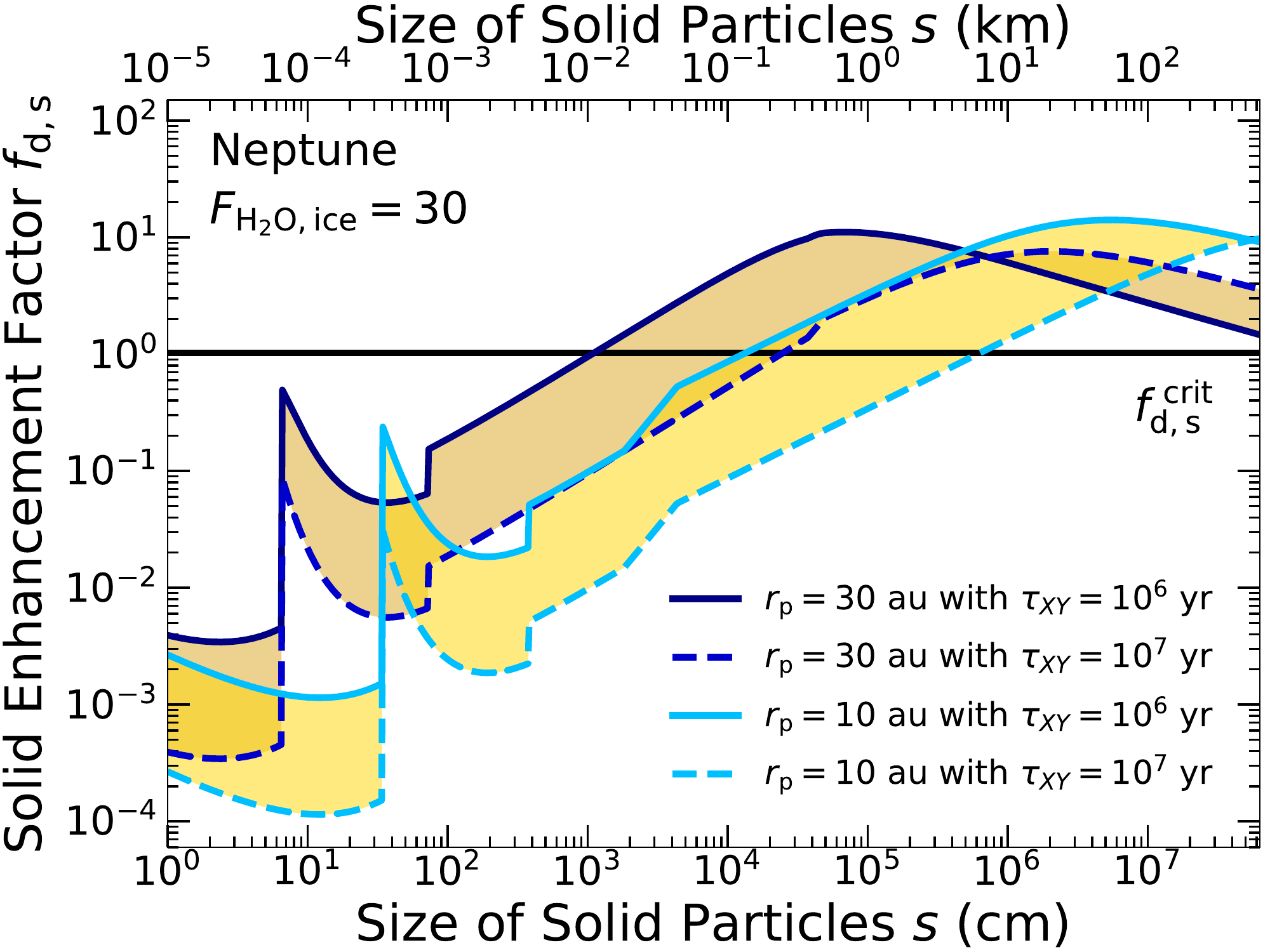}
\caption{
The enhancement factor of the solid surface density ($f_{\rm d, s}$) required to reproduce the estimated values of $M_{Z, \rm e}$ as a function of the size of accreted solids
for Uranus and Neptune on the left and right panels, respectively.
The two values are considered for $r_{\rm p}$ and $F_{{\rm H}_2 {\rm O,ice}}$: $r_{\rm p} = 10$ and 30 au and $F_{{\rm H}_2 {\rm O,ice}}=12$ and 30.
For comparison purpose, $f_{\rm d,s}^{\rm crit}$ is denoted by the black solid line on all the plots.
Plausible values of $f_{\rm d,s}$ are shown by the shaded region.
For small-sized solids, $f_{\rm d,s}$ becomes smaller than unity, and hence reduction in the solid surface density is needed for concurrent gas accretion.
For large-sized solids, $f_{\rm d,s}$ becomes larger than unity, and the solid surface density needs to be enhanced.
}
\label{fig12}
\end{center}
\end{minipage}
\end{figure*}

Figure \ref{fig12} summarizes the results.
In the plots, plausible values of $f_{\rm d,s}$ are shown as a function of $s$ for given values of $r_{\rm p}$ and $F_{{\rm H}_2 {\rm O,ice}}$.
We use the highest value of $\dot{M}_{{\rm p}, Z}$ (i.e., $\dot{M}_{{\rm p}, Z}^{\rm disp}$ vs $\dot{M}_{{\rm p}, Z}^{\rm 3b}$ vs $\dot{M}_{{\rm p}, Z}^{\rm pe}$) to compute $f_{\rm d,s}$.
We first confirm that the results for the Uranus and Neptune cases are almost identical;
in our problem setup, the difference between these two cases is only the planet mass, which is not significant.
Hence, the resulting behaviors of $f_{\rm d,s}$ do not differ very much.
Our results show that the variation of $r_{\rm p}$ shifts the plausible range of $f_{\rm d,s}$ toward smaller solid sizes.
This is simply because at larger $r_{\rm p}$, the gas surface density becomes smaller, 
which in turn increases the resulting Stokes number for given sizes of solids (see Section \ref{sec:chara_radius}, also see Table \ref{table1}).
We also find that the transitions of accretion modes become evident by noticeable changes in $f_{\rm d,s}$;
the transition from the pebble accretion mode to the drag-enhanced three-body one occurs at the sharp jump in $f_{\rm d,s}$,
and that from the drag-enhanced three-body mode to the two-body one is marked by the intersection between the solid and dashed lines.

In summary, the core accretion scenario can successfully reproduce the gas mass of both Uranus and Neptune within the gas disk lifetime,
under the assumption of the presence of planetary cores.
For solid accretion,
the resulting accretion rate becomes a function of solid size;
pebble accretion becomes effective for solids with size of $\sim 10$ cm or less.
The accretion rate is so high that the solid mass of both Uranus and Neptune can be reproduced 
even if the solid surface density is reduced by a few orders of magnitude relative to the MMSN model.
In fact, such reduction is needed to achieve concurrent gas accretion.
Drag-enhanced three body accretion becomes important for solids with the size range from $\sim 10$ cm to $\sim 10^2$ km.
The resulting accretion rate tends to be high for small-sized solids and low for large-sized solids.
The solid mass of Uranus and Neptune therefore can be accreted within the disk lifetime,
by adjusting the solid surface density by a couple of orders of magnitude relative to the MMSN model.
For the canonical planetesimal accretion, the solid mass ($M_{Z, \rm e}$) in their envelope at the formation stage can be reproduced
if the solid surface density is enhanced by at least a factor of few relative to the MMSN model.

It can be concluded that accretion of solids with radius of $\sim 1$ m to $\sim 10$ km likely played the major role for the deuterium enrichment of Uranus and Neptune,
if the MMSN model represents the solar nebula well.

\section{Discussion} \label{sec:disc}

We here discuss the applicability of our calculations to other atomic elements, the effect of dust opacity in planetary atmospheres,
and caveats that should be taken into account for our models and results.

\subsection{Other elements} \label{sec:disc_trac}

We have so far focused on the D/H ratio of Uranus and Neptune 
to constrain the solid mass that is accumulated in planetary envelopes via concurrent accretion of gas and solids after core formation (Section \ref{sec:d/h_2}).
This is just our choice.
In fact, similar equations (e.g., equations (\ref{eq:D/H_ratio}) -  (\ref{eq:Mz_Mgas})) can be derived for other elements (e.g., C via CH$_4$),
if their abundances both in the atmosphere of Uranus/Neptune and asteroids/comets are inferred.
Better constraints can be obtained when all the available elements are combined.

Thus, it would be fair to say that the D/H ratio is not a unique tracer of formation mechanisms of Uranus and Neptune.
Instead, the D/H ratio is one of metallicity indicators that can constrain their formation mechanisms.

\subsection{Effect of dust opacity in planetary atmospheres} \label{sec:disc_dust}

We have used the gas disk lifetime to constrain the value of $f_{\rm grain}$ in the above calculations (Section \ref{sec:app_UN_gas}).
The resulting values should be viewed as only an upper bound.
This is because a more reliable lower limit should come from either core formation timescales or the dust opacity in planetary envelopes;
if planetary cores form quite late (i.e., $\ga 10^6$ yrs) and/or dust evolution in planetary envelopes is very efficient,
then $f_{\rm grain}$ takes a smaller value.
Previous studies indeed show that $f_{\rm grain}$ can be $\sim 10^{-2}$ or even lower \citep[e.g.,][]{2010Icar..209..616M,2014A&A...566A.141M,2014ApJ...789L..18O}.
In order to examine this effect on the required solid surface density ($f_{\rm d,s}$), we here conduct another calculation.
 
\begin{figure}
\begin{center}
\includegraphics[width=8.3cm]{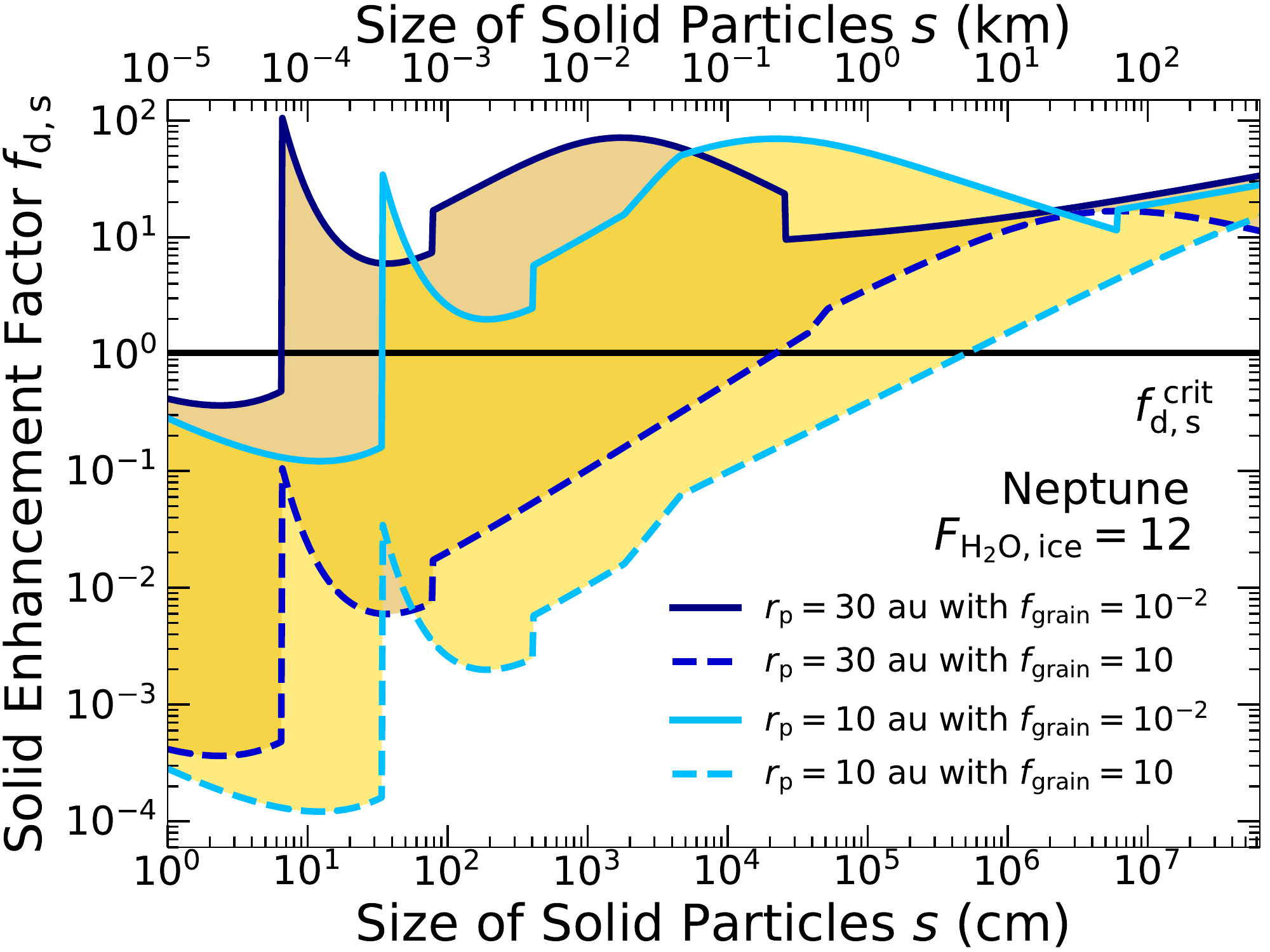}
\caption{
The enhancement factor of the solid surface density ($f_{\rm d, s}$) required to reproduce the estimated values of $M_{Z, \rm e}$ as a function of the size of accreted solids
for Neptune with $F_{{\rm H}_2 {\rm O,ice}}=12$ (as in Figure \ref{fig12}).
The two values are considered for $r_{\rm p}$ and $f_{\rm grain}$: $r_{\rm p} = 10$ and 30 au and $f_{\rm grain}=10^{-2}$ and 10.
Lower values of $f_{\rm grain}$ increase $f_{\rm d,s}$ and higher values of $f_{\rm grain}$ decreases $f_{\rm d,s}$.
}
\label{fig13}
\end{center}
\end{figure}

Figure \ref{fig13} shows how $f_{\rm d,s}$ behaves as a function of solid size for given values of $r_{\rm p}$ and $F_{{\rm H}_2 {\rm O,ice}}$ (i.e., $M_{Z, \rm c}$). 
In contrast to Figure \ref{fig12}, the value of $f_{\rm grain}$ is fixed.
Since the difference of results between Uranus and Neptune is minor and the variation of $F_{{\rm H}_2 {\rm O,ice}}$ shifts results only for the vertical direction (Figure \ref{fig12}),
we consider only the Neptune case with $F_{{\rm H}_2 {\rm O,ice}}=12$.
Our results indicate that the range of $f_{\rm d,s}$ expands for both directions;
lower values of $f_{\rm d,s}$ become possible when $f_{\rm grain}=10$.
This arises because the gas accretion timescale is an increasing function of $f_{\rm grain}$ (equation (\ref{eq:delta_t_XY})).
As a result, the solid mass of Uranus and Neptune can be reproduced even with slow solid accretion rates (i.e., low metallicity disks).
It should be noted that the resulting gas accretion timescale may become longer than the typical gas disk lifetime.

Higher values of $f_{\rm d,s}$ are required when $f_{\rm grain}=10^{-2}$.
This is also the direct outcome of concurrent gas and solid accretion, which is assumed in this calculation.
The gas accretion timescale is shortened for the case that $f_{\rm grain}=10^{-2}$.
In order to complete solid accretion within the shortened timescale, solid surface densities need to be enhanced more.
For this case, formation of Uranus and Neptune can be completed well before gas disks dissipate.
This in turn suggests that some mechanisms of stopping both gas and solid accretion are demanded for such a configuration.

One may notice sudden drops in $f_{\rm d,s}$ at $s \simeq 0.3$ km and $s \simeq 60$ km for the cases that $r_{\rm p}=30$ au and $r_{\rm p}=10$ au, respectively.
These features are not shown in Figure \ref{fig12}.
This is related to gap formation in the solid surface density and 
the resulting change in the solid accretion mode from $\dot{M}_{{\rm p}, Z}^{\rm disp}$ to $\dot{M}_{{\rm p}, Z}^{\rm shear}$ (equations (\ref{eq:Mdot_Z_disp}) and (\ref{eq:Mdot_Z_shear}), respectively).
As shown in equation (\ref{eq:eta_gap}), gaps tend to close readily (that is, $\eta_{\rm gap} >1$) when $f_{\rm grain}$ is smaller and $s$ is larger;
equivalently, gap formation is prevented when gas accretion occurs quickly and large-sized solids whose eccentricity damping is slow are accreted.
Accordingly, the drops are realized for the case that $f_{\rm grain}=10^{-2}$.

In summary, dust opacity in planetary atmospheres provides important effects on the solid surface density required 
to reproduce the current value of the D/H ratio of Uranus and Neptune.
More self-consistent calculations are necessary to explore this effect more carefully.

\subsection{Caveats} \label{sec:disc_cav}

We here list up caveats involved with this work.

One most important caveat is that our envelope model is very simplified.
We have confirmed that the envelope structure and adopted assumptions are reasonable within our problem setup (Appendix \ref{app_env}).
However, it would be ideal to self-consistently compute the atmospheric metallicity, the resulting gas accretion rate, and the accompanying accretion luminosity,
by taking into account the complicated interplay among solid accretion, the resulting metal enrichment of the atmosphere, the subsequent metal sedimentation,
and the eventually accelerated/slowed down gas accretion. 
In fact, detailed calculations show that the value of $f_{\rm grain}$ can be much smaller and larger than that constrained from the gas accretion timescale 
\citep[Figure \ref{fig9}, e.g.,][]{2010Icar..209..616M,2014ApJ...789L..18O,2020ApJ...900...96A,2021A&A...653A.103B}.
This complexity is beyond the scope of this work and remains to be the future work.

Another caveat is that the presence of planetary cores is assumed in this work.
There are two leading hypotheses to form cores in the literature:
runaway and oligarchic growth \citep[by planetesimal collisions, e.g., ][]{2000Icar..143...15K} and pebble accretion \citep[onto existing relatively massive bodies, e.g.,][]{2017AREPS..45..359J}.
As described in Section \ref{sec:disc_dust}, the core formation timescale may provide a better lower limit on $f_{\rm grain}$.
However, the timescale is also a function of the solid surface density and the position of protocores.
More self-consistent calculations will address how the core formation timescale affects the value of the D/H ratio of Uranus and Neptune.

Finally, this work assumes that both gas and solid accretion take place in-situ and concurrently.
As described in Section \ref{sec:d/h_3}, (proto)Uranus and (proto)Neptune may have underwent radial migration during the process of forming.
This migration may trigger additional solid accretion as migrating planets can encounter solids through the radial movement.
Thus, our results should be viewed as a most conservative one, since they provide a lower limit for the solid accretion rate.

\section{Summary \& Conclusions} \label{sec:conc}

We have explored solid accretion onto Neptune-mass planets, taking into account a wide range of the solid radius.
We have determined what radius of solids can play the primary role for reproducing the current values of the D/H ratio of Uranus and Neptune,
as a function of the solid surface density.
Our efforts have been conducted with the aim of quantifying 
how the known values of the D/H ratio of the solar system bodies can be used to constrain the accretion mechanisms of (proto)Uranus and (proto)Neptune.

Our investigation has begun with characterization of the solid mass distributions within Uranus and Neptune at their formation stage (see Section \ref{sec:d/h}).
These distributions can be constrained by the current values of the D/H ratio of these planets,
based on the simple mass conservation argument
and the assumption that the D/H ratio traces the solid accretion history.
Our results have shown that the optimal values of the core mass ($M_{Z, \rm c}$) and the solid mass in the envelope ($M_{Z, \rm c}$) are determined
as a function of the deuterium enrichment factor ($F_{{\rm H}_2 {\rm O,ice}}$) of accreted solids (Figure \ref{fig1}).
We have used these values to constrain the solid accretion mode that likely took place for the formation of Uranus and Neptune.

We have adopted the MMSN model to investigate the gas and solid accretion onto Neptune-mass planets in detail (see Section \ref{sec:gas_solid}).
We have also introduced the drag force arising from the nebular gas acting on accreted solids.
The definition of the drag force has allowed the computation of a number of characteristic radii of solids. 
We have found that our problem setup involves seven radius ranges of the solids (see Table \ref{table1});
in each radius range, the dominant accretion mode becomes different.

The detailed exploration of solid accretion onto Neptune-mass planets has been conducted in Sections \ref{sec:in-situ} and \ref{sec:res1}.
To proceed, we have assumed that planetary cores already form and the cores undergo both gas and solid accretion simultaneously.
Our results have shown that as the radius of the accreted solids decreases from a few 100 km to smaller values,
the accretion mode switches from the classical, two-body planetesimal accretion to the drag-enhanced three-body one;
for protoplanets with the core mass of $10 M_{\oplus}$, this transition occurs at the solid radius of $\sim 1.1 \times 10^2$ km.
A wide range of the solid radius is covered by the drag-enhanced three-body accretion,
and the corresponding accretion rate increases with decreasing the solid radius.
We have also found that the resulting values become high enough that concurrent gas accretion can be stopped
if all the solids reside in a certain radius.
This indicates that the full determination of the size distribution and abundance of solids with all the radii is crucial
to reliably specify the dominant accretion mode for the formation of Uranus and Neptune.
We have finally explored the so-called pebble accretion.
Our results confirm that this accretion mode is the most efficient and the size distribution is the key parameter to quantify its actual efficiency.

We have eventually applied these solid accretion modes to (proto)Uranus and (proto)Neptune 
to specify which radius ranges of solids are important for reproducing the current values of the D/H ratio of Uranus and Neptune in Section \ref{sec:app_UN} (see Figure \ref{fig11}).
Our calculations have shown that if the MMSN model is adopted,
solids with radius of $\sim 1$ m to $\sim 10$ km are likely to be responsible for deuterium enrichment of Uranus and Neptune due to modest accretion rates (Figure \ref{fig12});
for small-sized ($\la 1$ m) solids, the resulting accretion rate is too high to achieve concurrent gas accretion.
This issue can be avoided by reducing the solid surface density possibly due to the size distribution effect.
For large-sized ($\ga 10$ km) solids, enhancement in the solid surface density is required to reproduce the solid mass ($M_{Z, \rm e}$) in the envelope.
We therefore conclude that small- to medium-sized solids tend to be the main agent to account for the D/H ratio of Uranus and Neptune.
A tighter constrains can be obtained if the full size distribution of solids is taken into account, which is the target of our second paper.

We finally discuss the limitations of this work.
These include how our calculations can be applied to other atomic elements,
how dust opacity in planetary atmospheres can affect our results,
and caveats coming from the assumptions and idealization adopted in this work.
The most crucial simplification are the envelope model and the gas accretion formula.

In conclusion, the D/H ratio of solar system bodies is the important quantity for quantitatively tracing how (proto)Uranus and (proto)Neptune accreted solids from the solar nebula.
The direct application of the D/H ratio to exoplanetary systems might be possible for some systems 
after JWST becomes fully online \citep{2019ApJ...882L..29M}.

\begin{acknowledgments}

The author thanks two anonymous referee for useful comments, which significantly improved the quality of our manuscript,
and Paul Goldsmith, Dariusz Lis, Youngmin Seo, Robert West, and Karen Willacy for stimulating discussions.
This research was carried out at the Jet Propulsion Laboratory, California Institute of Technology, 
under a contract with the National Aeronautics and Space Administration (80NM0018D0004) and funded through the internal Research and Technology Development program.
Y.H. is supported by JPL/Caltech.
  
\end{acknowledgments}

\appendix

\section{Planetary envelope structures} \label{app_env}

We here describe briefly how planetary envelope structures are computed in this work and examine how our adopted assumptions are reasonable (Section \ref{sec:in-situ_1}).

The basic equations that govern an envelope structure of a protoplanet are
\begin{eqnarray}
\frac{dP_{\rm p}}{dR}     & = & - \frac{G M_r \rho_{XY}}{R^2} \\ \nonumber
\frac{dM_r}{dR} & = & 4 \pi R^2 \rho_{XY} \\  \nonumber
\frac{dT_{\rm p}}{dR} & = & \left\{ \begin{tabular}{@{}l@{}}                                     
                                                   $- \frac{3 \kappa_{\rm grain} L}{64 \pi \sigma_{\rm SB}} \frac{\rho_{XY}}{R^2 T^3_{\rm p}}$ if $- \frac{3 \kappa_{\rm grain} L}{64 \pi \sigma_{\rm SB}} \frac{\rho_{XY}}{R^2 T^3_{\rm p}} \frac{P_{\rm p}}{T_{\rm p}} \frac{dR}{dP_{\rm p}} < 1 - \frac{1}{\Gamma_2}$ (the radiative case) \\  \nonumber
                                                    $ \left( 1 - \frac{1}{\Gamma_2} \right) \frac{T_{\rm p}}{P_{\rm p}} \frac{dP_{\rm p}}{dR}$  (the convective case),   \nonumber
                                    \end{tabular} \right.
\end{eqnarray} 
where $P_{\rm p}$ is the gas pressure of the envelope, $M_{ r}$ is the mass of the protoplanet integrated from the core center to a certain radius ($R$),
$T_{\rm p}$ is the temperature of the envelope, and $\Gamma_2$ is the second adiabatic exponent.
Other quantities are defined in Section \ref{sec:solid_radius}.

Assuming that $M_r=M_{Z, \rm c}$ and $L$ is constant and adopting the ideal gas law, the above equations are simplified into
\begin{eqnarray}
\frac{dP_{\rm p}}{dR}     & = & - \frac{G M_{Z, \rm c} \rho_{XY}}{R^2} \\ \nonumber
P_{\rm p}                       & = &    \frac{k_{\rm B}}{m_{\rm g}} \rho_{XY} T_{\rm p}  \\ \nonumber
\frac{dT_{\rm p}}{dR} & = & \left\{ \begin{tabular}{@{}l@{}}                                     
                                                   $- \frac{3 \kappa_{\rm grain} L}{64 \pi \sigma_{\rm SB}} \frac{\rho_{XY}}{R^2 T^3_{\rm p}}$ if $\frac{3 \kappa_{\rm grain} L}{64 \pi \sigma_{\rm SB}} \frac{P_{\rm p}}{GM_{Z, \rm c} T^4_{\rm p}}  < 1 - \frac{1}{\Gamma_2}$ (the radiative case) \\    \nonumber
                                                    $-  \left( 1 - \frac{1}{\Gamma_2} \right) \frac{T_{\rm p}}{P_{\rm p}}  \frac{G M_{Z, \rm c} \rho_{XY}}{R^2}$  (the convective case). \nonumber
                                    \end{tabular} \right.
\end{eqnarray} 
We normalize the physical quantities ($x = \gamma R/ R_{\rm B}$, $p= P_{\rm p}/P_{\rm d}$, $\theta = T_{\rm p}/T_{\rm d}$, $\sigma=\rho_{XY}/\rho_{\rm d,g}$)
and adopt two additional assumptions: 
the opacity is constant in the radiative zone, and the second adiabatic exponent is constant in the convective zone
(equivalently, $\Gamma_2 \simeq \gamma$ in the ideal approximation).
Then the above simplified equations read
\begin{eqnarray}
\label{eq:dpdx}
\frac{dp}{dx}     & = & - \gamma  \frac{\sigma}{x^2} \\
\label{eq:p}
p                       & = &   \sigma \theta  \\
\label{eq:dthetadx}
\frac{d \theta}{dx} & = & \left\{ \begin{tabular}{@{}l@{}}                                     
                                                   $-  \gamma  \frac{W_0}{4} \frac{\sigma}{x^2 \theta^3}$ if $ \frac{W_0}{4} \frac{p}{\theta^4}  < 1 - \frac{1}{\gamma}$ (the radiative case) \\   
                                                    $-  \left( \gamma - 1 \right) \frac{\sigma \theta }{x^2 p} $  (the convective case),
                                    \end{tabular} \right. 
\end{eqnarray} 
where $ W_{\rm neb} = W_0/4$ (also see equation (\ref{eq:W_neb}) or (\ref{eq:W_neb2})).

Previous studies show that an analytical solution is obtained from equations (\ref{eq:dpdx}) - (\ref{eq:dthetadx}) \citep{2003A&A...410..711I}.
In a radiative zone, the temperature structure ($\theta$) is written as
\begin{equation}
\label{eq:theta}
\theta^4 = 1 + W_0 (p-1).
\end{equation}
This equation leads to \citep[see][for derivation]{2003A&A...410..711I}
\begin{equation}
\label{eq:1_x_rad}
\frac{1}{x} = 1 + \frac{1}{\gamma} [ 4 ( \theta -1 ) + f(\theta; w_0)],
\end{equation}
where 
\begin{equation}
w_0 = | 1- W_0 |^{1/4}
\end{equation}
and 
\begin{equation}
f(\theta; w_0) =   w_0 \left\{ \ln \left[ \left( \frac{\theta - w_0}{\theta + w_0} \right) \left( \frac{1 + w_0}{1 - w_0} \right) \right] - 2 \left[ \arctan{ \left( \frac{\theta}{w_0}  \right) }  - \arctan{\left( \frac{1}{w_0} \right) } \right]   \right\}.                                               
\end{equation}
The condition that $W_0 < 1$ is always satisfied in our case (see equation (\ref{eq:W_neb2})).
This solution can be further simplified, by considering two limits: $W_0 p \ll 1$ and $W_0 p \gg 1$ \citep{2012ApJ...747..115O}.
The former corresponds to the nearly isothermal regime, and the latter is for the pressure dominated regime.
Equations (\ref{eq:rho_XY}) are the resulting expressions.

In a convective zone, the famous adiabatic relation is retrieved:
\begin{equation}
\label{eq:p_conv}
p = \sigma^{\gamma}.
\end{equation}
Then, the density structure is given as
\begin{equation}
\label{eq:1_x_con}
\frac{1}{x} = \frac{1}{x_{\rm tr}} + \frac{1}{\gamma-1} \frac{ p_{\rm tr} }{ \sigma_{\rm tr} }\left[ \left( \frac{ \sigma }{ \sigma_{\rm tr} } \right)^{\gamma -1} -1 \right],
\end{equation}
where $x_{\rm tr} = \gamma R_{\rm tr} / R_{\rm B}$.
From the above equation, one can recover the adiabatic density profile derived in previous studies (e.g., equation (21a) in \citet{2014ApJ...786...21P}, also see \citet{2006ApJ...648..666R,2015ApJ...811...41L}),
by replacing $R_{\rm tr}$ with the radiative-convective boundary ($R_{\rm RCB}$):
\begin{equation}
\rho _{XY} = \rho_{XY} (R_{\rm RCB}) \left( 1+ \frac{R_{\rm B}'}{R} -\frac{R_{\rm B}'}{R_{\rm RCB}}   \right)^{1/(\gamma-1)},
\end{equation}
where $R_{\rm B}' = (\gamma-1)/ \gamma (R_{\rm B}T_{\rm d}/ T_{\rm p} (R_{\rm RCB}))$.
Note that in our formulation, $R_{\rm RCB}$ is defined as
\begin{equation}
\left. \frac{W_{\rm neb} p}{\theta^4} \right|_{R= R_{\rm RCB}} = 1 - \frac{1}{\gamma} = \frac{2}{7}.
\end{equation}
Since the isothermal assumption ($\theta \approx 1$) is acceptable at $R \simeq R_{\rm RCB}$, $W_{\rm neb} p \simeq 2/7$.
This condition is comparable to that used to define $R_{\rm tr}$, that is, $W_{\rm neb} p = 1/4$ \citep{2012ApJ...747..115O}.
Therefore, $R_{\rm tr} \simeq R_{\rm RCB}$ in our calculations.

Thus, the envelope structure is governed by equations (\ref{eq:p}), (\ref{eq:theta}), and (\ref{eq:1_x_rad}) for radiative zones, 
and by equations (\ref{eq:p}), (\ref{eq:p_conv}), and (\ref{eq:1_x_con}) for convective zones.

Using the above formulation, we here demonstrate the followings:
the radiative solution leads to self-consistent results within our framework;
and the increase in the mean molecular weight by a factor of $\la 2$ changes the envelope gas density only by a factor of few.
These two are used as a basis for the two assumptions adopted in this work  (Section \ref{sec:in-situ_1}).

\begin{figure*}
\begin{minipage}{17cm}
\begin{center}
\includegraphics[width=8.3cm]{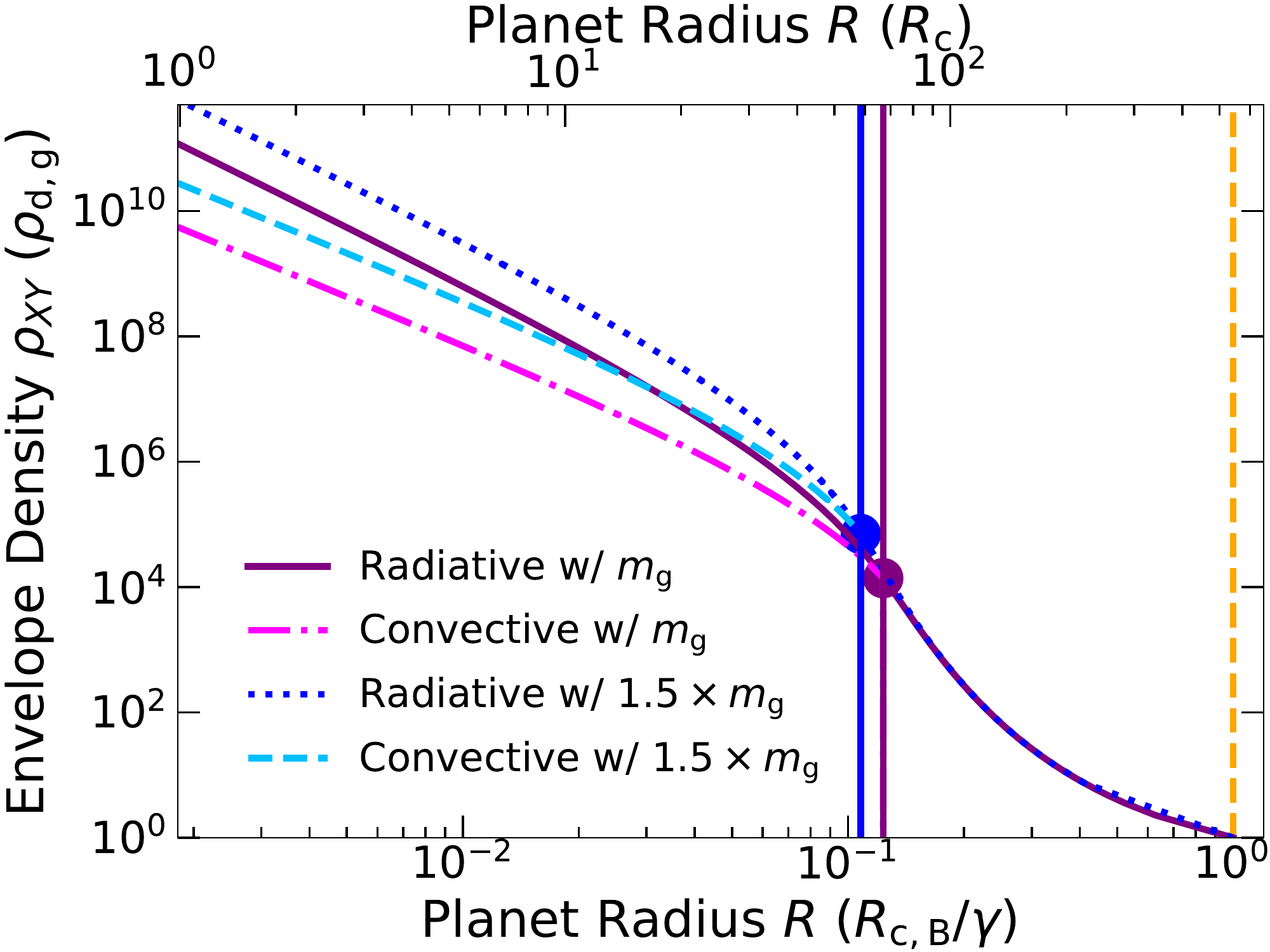}
\includegraphics[width=8.3cm]{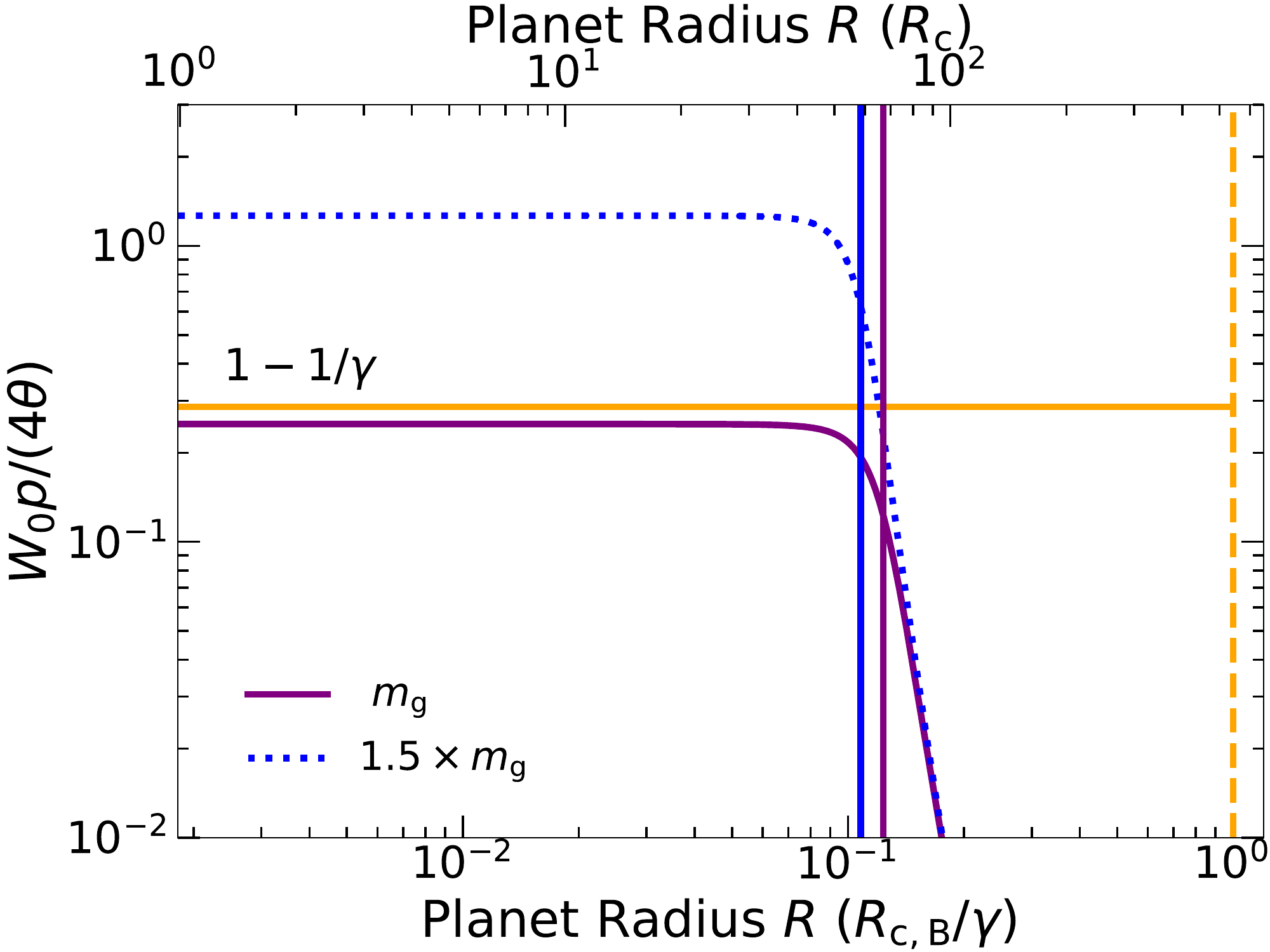}
\caption{
The envelope density profile as a function of the distance from the core center and the resulting behavior of the quantity, $W_0 p / ( 4 \theta^4 )$,
on the left and right panels, respectively.
On the left, four solutions are plotted: the radiative one with the mean molecular weight of $m_{\rm g}$ (the purple solid line), 
the convective one with the mean molecular weight of $m_{\rm g}$ (the magenta dash-dotted line),
the radiative one with the mean molecular weight of $1.5 \times m_{\rm g}$ (the blue dotted line),
and the convective one with the mean molecular weight of $1.5 \times m_{\rm g}$ (the light blue dashed line).
As in Figure \ref{fig3}, the transitions from the nearly isothermal regime to the pressure dominated one are denoted by the corresponding vertical solid lines and the dots.
The radiative solution with the standard case (i.e., $m_{\rm g}$) becomes comparable to the convective solution with the enhanced case (i.e., $1.5 \times m_{\rm g}$).
On the right, the quantity, $ W_0 p / ( 4 \theta^4 )$, is plotted to determine when the radiative solution becomes valid.
The standard case (i.e., $m_{\rm g}$) leads to the fully radiative envelope.}
\label{fig14}
\end{center}
\end{minipage}
\end{figure*}

We first explore how the envelope gas density changes as a function of the mean molecular weight.
Figure \ref{fig14} shows the resulting structure of a planetary envelope.
As examples, we consider two cases: the standard case with $m_{\rm g}$ and the enhanced case with $1.5 \times m_{\rm g}$.
Our results (the left panel) show that the radiative solution exceeds the convective solution and that the gas density is an increasing function of $m_{\rm g}$.
It is interesting that the radiative solution for the standard case eventually becomes comparable to the convective solution for the enhanced case.
On the right panel, we determine when the radiative solution becomes valid (equation (\ref{eq:dthetadx})).
We find that for the standard case, the condition that $ W_0 p / ( 4 \theta^4 )  < 1 - 1/\gamma$ is always met due to no enhancement of the mean molecular weight, 
and hence the planetary envelope becomes fully radiative;
for the enhanced case, the condition is not satisfied in the pressure dominated regime, and the envelope becomes convective in the interior region.

\begin{figure}
\begin{center}
\includegraphics[width=8.3cm]{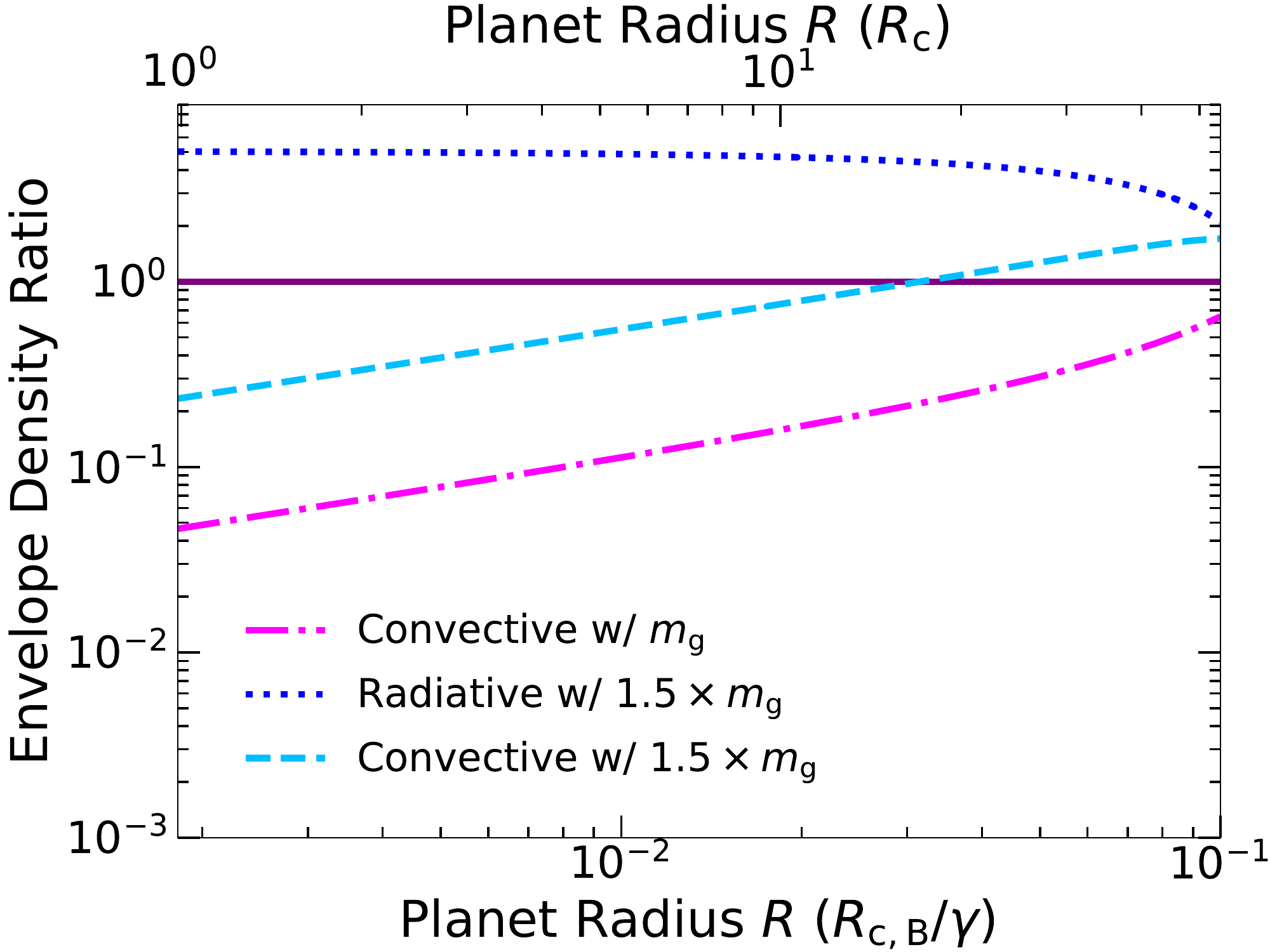}
\caption{
The envelope gas density ratio is shown as a function of the distance from the core center.
Three solutions (the convective solution for the standard case, and the radiative and convective solutions for the enhanced case) are
normalized by the radiative solution for the standard case.
The difference between the radiative solution for the standard case (which is adopted in this work) and the convective one for the enhanced case 
(which is the most realistic in our problem setup) is only about a factor of a few (the light blue dashed line).
}
\label{fig15}
\end{center}
\end{figure}

We then compute the envelope gas density difference for these four cases.
Figure \ref{fig15} summarizes the results. 
Three gas densities (the convective solution for the standard case, and the radiative and convective solutions for the enhanced case) are
normalized by that of the radiative solution for the standard case.
As expected from Figure \ref{fig14}, the density difference becomes the highest for the convective solution with the standard case,
and the lowest is for the convective solution with the enhanced case.

In summary, it can be concluded that adopting the radiative solution for the standard case is reasonable,
because it leads to self-consistent results within our framework
and the increase in the mean molecular weight by a factor of $\la 2$ changes the envelope gas density only by a factor of few.

\bibliographystyle{aasjournal}
\bibliography{adsbibliography}    

\begin{thebibliography}{}
\expandafter\ifx\csname natexlab\endcsname\relax\def\natexlab#1{#1}\fi
\providecommand{\url}[1]{\href{#1}{#1}}
\providecommand{\dodoi}[1]{doi:~\href{http://doi.org/#1}{\nolinkurl{#1}}}
\providecommand{\doeprint}[1]{\href{http://ascl.net/#1}{\nolinkurl{http://ascl.net/#1}}}
\providecommand{\doarXiv}[1]{\href{https://arxiv.org/abs/#1}{\nolinkurl{https://arxiv.org/abs/#1}}}

\bibitem[{{Adachi} {et~al.}(1976){Adachi}, {Hayashi}, \&
  {Nakazawa}}]{1976PThPh..56.1756A}
{Adachi}, I., {Hayashi}, C., \& {Nakazawa}, K. 1976, Progress of Theoretical
  Physics, 56, 1756, \dodoi{10.1143/PTP.56.1756}

\bibitem[{{Alexander} {et~al.}(2012){Alexander}, {Bowden}, {Fogel}, {Howard},
  {Herd}, \& {Nittler}}]{2012Sci...337..721A}
{Alexander}, C.~M.~O., {Bowden}, R., {Fogel}, M.~L., {et~al.} 2012, Science,
  337, 721, \dodoi{10.1126/science.1223474}

\bibitem[{{Ali-Dib} \& {Thompson}(2020)}]{2020ApJ...900...96A}
{Ali-Dib}, M., \& {Thompson}, C. 2020, \apj, 900, 96,
  \dodoi{10.3847/1538-4357/aba521}

\bibitem[{{Altwegg} {et~al.}(2019){Altwegg}, {Balsiger}, \&
  {Fuselier}}]{2019ARA&A..57..113A}
{Altwegg}, K., {Balsiger}, H., \& {Fuselier}, S.~A. 2019, \araa, 57, 113,
  \dodoi{10.1146/annurev-astro-091918-104409}

\bibitem[{{Altwegg} {et~al.}(2015){Altwegg}, {Balsiger}, {Bar-Nun},
  {Berthelier}, {Bieler}, {Bochsler}, {Briois}, {Calmonte}, {Combi}, {De
  Keyser}, {Eberhardt}, {Fiethe}, {Fuselier}, {Gasc}, {Gombosi}, {Hansen},
  {H{\"a}ssig}, {J{\"a}ckel}, {Kopp}, {Korth}, {LeRoy}, {Mall}, {Marty},
  {Mousis}, {Neefs}, {Owen}, {R{\`e}me}, {Rubin}, {S{\'e}mon}, {Tzou}, {Waite},
  \& {Wurz}}]{2015Sci...347A.387A}
{Altwegg}, K., {Balsiger}, H., {Bar-Nun}, A., {et~al.} 2015, Science, 347,
  1261952, \dodoi{10.1126/science.1261952}

\bibitem[{{Asplund} {et~al.}(2009){Asplund}, {Grevesse}, {Sauval}, \&
  {Scott}}]{2009ARA&A..47..481A}
{Asplund}, M., {Grevesse}, N., {Sauval}, A.~J., \& {Scott}, P. 2009, \araa, 47,
  481, \dodoi{10.1146/annurev.astro.46.060407.145222}

\bibitem[{{Balsiger} {et~al.}(1995){Balsiger}, {Altwegg}, \&
  {Geiss}}]{1995JGR...100.5827B}
{Balsiger}, H., {Altwegg}, K., \& {Geiss}, J. 1995, \jgr, 100, 5827,
  \dodoi{10.1029/94JA02936}

\bibitem[{{Benz} \& {Asphaug}(1999)}]{1999Icar..142....5B}
{Benz}, W., \& {Asphaug}, E. 1999, \icarus, 142, 5,
  \dodoi{10.1006/icar.1999.6204}

\bibitem[{{Bockel{\'e}e-Morvan} {et~al.}(2012){Bockel{\'e}e-Morvan}, {Biver},
  {Swinyard}, {de Val-Borro}, {Crovisier}, {Hartogh}, {Lis}, {Moreno},
  {Szutowicz}, {Lellouch}, {Emprechtinger}, {Blake}, {Courtin}, {Jarchow},
  {Kidger}, {K{\"u}ppers}, {Rengel}, {Davis}, {Fulton}, {Naylor}, {Sidher}, \&
  {Walker}}]{2012A&A...544L..15B}
{Bockel{\'e}e-Morvan}, D., {Biver}, N., {Swinyard}, B., {et~al.} 2012, \aap,
  544, L15, \dodoi{10.1051/0004-6361/201219744}

\bibitem[{{Bodenheimer} \& {Pollack}(1986)}]{1986Icar...67..391B}
{Bodenheimer}, P., \& {Pollack}, J.~B. 1986, \icarus, 67, 391,
  \dodoi{10.1016/0019-1035(86)90122-3}

\bibitem[{{Bolton} {et~al.}(2017){Bolton}, {Adriani}, {Adumitroaie}, {Allison},
  {Anderson}, {Atreya}, {Bloxham}, {Brown}, {Connerney}, {DeJong}, {Folkner},
  {Gautier}, {Grassi}, {Gulkis}, {Guillot}, {Hansen}, {Hubbard}, {Iess},
  {Ingersoll}, {Janssen}, {Jorgensen}, {Kaspi}, {Levin}, {Li}, {Lunine},
  {Miguel}, {Mura}, {Orton}, {Owen}, {Ravine}, {Smith}, {Steffes}, {Stone},
  {Stevenson}, {Thorne}, {Waite}, {Durante}, {Ebert}, {Greathouse}, {Hue},
  {Parisi}, {Szalay}, \& {Wilson}}]{2017Sci...356..821B}
{Bolton}, S.~J., {Adriani}, A., {Adumitroaie}, V., {et~al.} 2017, Science, 356,
  821, \dodoi{10.1126/science.aal2108}

\bibitem[{{Brouwers} {et~al.}(2021){Brouwers}, {Ormel}, {Bonsor}, \&
  {Vazan}}]{2021A&A...653A.103B}
{Brouwers}, M.~G., {Ormel}, C.~W., {Bonsor}, A., \& {Vazan}, A. 2021, \aap,
  653, A103, \dodoi{10.1051/0004-6361/202140476}

\bibitem[{{Ceccarelli} {et~al.}(2014){Ceccarelli}, {Caselli},
  {Bockel{\'e}e-Morvan}, {Mousis}, {Pizzarello}, {Robert}, \&
  {Semenov}}]{2014prpl.conf..859C}
{Ceccarelli}, C., {Caselli}, P., {Bockel{\'e}e-Morvan}, D., {et~al.} 2014, in
  Protostars and Planets VI, ed. H.~{Beuther}, R.~S. {Klessen}, C.~P.
  {Dullemond}, \& T.~{Henning}, 859,
  \dodoi{10.2458/azu\_uapress\_9780816531240-ch037}

\bibitem[{{Dauphas} \& {Pourmand}(2011)}]{2011Natur.473..489D}
{Dauphas}, N., \& {Pourmand}, A. 2011, \nat, 473, 489,
  \dodoi{10.1038/nature10077}

\bibitem[{{Feuchtgruber} {et~al.}(2013){Feuchtgruber}, {Lellouch}, {Orton}, {de
  Graauw}, {Vandenbussche}, {Swinyard}, {Moreno}, {Jarchow}, {Billebaud},
  {Cavali{\'e}}, {Sidher}, \& {Hartogh}}]{2013A&A...551A.126F}
{Feuchtgruber}, H., {Lellouch}, E., {Orton}, G., {et~al.} 2013, \aap, 551,
  A126, \dodoi{10.1051/0004-6361/201220857}

\bibitem[{{Frelikh} \& {Murray-Clay}(2017)}]{2017AJ....154...98F}
{Frelikh}, R., \& {Murray-Clay}, R.~A. 2017, \aj, 154, 98,
  \dodoi{10.3847/1538-3881/aa81c7}

\bibitem[{{Geiss} \& {Gloeckler}(1998)}]{1998SSRv...84..239G}
{Geiss}, J., \& {Gloeckler}, G. 1998, \ssr, 84, 239

\bibitem[{{Ginzburg} \& {Chiang}(2020)}]{2020MNRAS.498..680G}
{Ginzburg}, S., \& {Chiang}, E. 2020, \mnras, 498, 680,
  \dodoi{10.1093/mnras/staa2500}

\bibitem[{{Goldreich} {et~al.}(2004){Goldreich}, {Lithwick}, \&
  {Sari}}]{2004ARA&A..42..549G}
{Goldreich}, P., {Lithwick}, Y., \& {Sari}, R. 2004, \araa, 42, 549,
  \dodoi{10.1146/annurev.astro.42.053102.134004}

\bibitem[{{Hasegawa} {et~al.}(2018){Hasegawa}, {Bryden}, {Ikoma}, {Vasisht}, \&
  {Swain}}]{2018ApJ...865...32H}
{Hasegawa}, Y., {Bryden}, G., {Ikoma}, M., {Vasisht}, G., \& {Swain}, M. 2018,
  \apj, 865, 32, \dodoi{10.3847/1538-4357/aad912}

\bibitem[{{Hasegawa} {et~al.}(2019){Hasegawa}, {Hansen}, \&
  {Vasisht}}]{2019ApJ...876L..32H}
{Hasegawa}, Y., {Hansen}, B. M.~S., \& {Vasisht}, G. 2019, \apjl, 876, L32,
  \dodoi{10.3847/2041-8213/ab1b5a}

\bibitem[{{Hayashi}(1981)}]{1981PThPS..70...35H}
{Hayashi}, C. 1981, Progress of Theoretical Physics Supplement, 70, 35,
  \dodoi{10.1143/PTPS.70.35}

\bibitem[{{Helled} {et~al.}(2011){Helled}, {Anderson}, {Podolak}, \&
  {Schubert}}]{2011ApJ...726...15H}
{Helled}, R., {Anderson}, J.~D., {Podolak}, M., \& {Schubert}, G. 2011, \apj,
  726, 15, \dodoi{10.1088/0004-637X/726/1/15}

\bibitem[{{Helled} {et~al.}(2020){Helled}, {Nettelmann}, \&
  {Guillot}}]{2020SSRv..216...38H}
{Helled}, R., {Nettelmann}, N., \& {Guillot}, T. 2020, \ssr, 216, 38,
  \dodoi{10.1007/s11214-020-00660-3}

\bibitem[{{Helled} {et~al.}(2014){Helled}, {Bodenheimer}, {Podolak}, {Boley},
  {Meru}, {Nayakshin}, {Fortney}, {Mayer}, {Alibert}, \&
  {Boss}}]{2014prpl.conf..643H}
{Helled}, R., {Bodenheimer}, P., {Podolak}, M., {et~al.} 2014, in Protostars
  and Planets VI, ed. H.~{Beuther}, R.~S. {Klessen}, C.~P. {Dullemond}, \&
  T.~{Henning}, 643, \dodoi{10.2458/azu\_uapress\_9780816531240-ch028}

\bibitem[{{Ikoma} {et~al.}(2000){Ikoma}, {Nakazawa}, \&
  {Emori}}]{2000ApJ...537.1013I}
{Ikoma}, M., {Nakazawa}, K., \& {Emori}, H. 2000, \apj, 537, 1013,
  \dodoi{10.1086/309050}

\bibitem[{{Inaba} \& {Ikoma}(2003)}]{2003A&A...410..711I}
{Inaba}, S., \& {Ikoma}, M. 2003, \aap, 410, 711,
  \dodoi{10.1051/0004-6361:20031248}

\bibitem[{{Johansen} \& {Lambrechts}(2017)}]{2017AREPS..45..359J}
{Johansen}, A., \& {Lambrechts}, M. 2017, Annual Review of Earth and Planetary
  Sciences, 45, 359, \dodoi{10.1146/annurev-earth-063016-020226}

\bibitem[{{J{\o}rgensen} {et~al.}(2020){J{\o}rgensen}, {Belloche}, \&
  {Garrod}}]{2020ARA&A..58..727J}
{J{\o}rgensen}, J.~K., {Belloche}, A., \& {Garrod}, R.~T. 2020, \araa, 58, 727,
  \dodoi{10.1146/annurev-astro-032620-021927}

\bibitem[{{Kavelaars} {et~al.}(2011){Kavelaars}, {Mousis}, {Petit}, \&
  {Weaver}}]{2011ApJ...734L..30K}
{Kavelaars}, J.~J., {Mousis}, O., {Petit}, J.-M., \& {Weaver}, H.~A. 2011,
  \apjl, 734, L30, \dodoi{10.1088/2041-8205/734/2/L30}

\bibitem[{{Kokubo} \& {Ida}(2000)}]{2000Icar..143...15K}
{Kokubo}, E., \& {Ida}, S. 2000, \icarus, 143, 15,
  \dodoi{10.1006/icar.1999.6237}

\bibitem[{{Kreidberg} {et~al.}(2014){Kreidberg}, {Bean}, {D{\'e}sert}, {Line},
  {Fortney}, {Madhusudhan}, {Stevenson}, {Showman}, {Charbonneau},
  {McCullough}, {Seager}, {Burrows}, {Henry}, {Williamson}, {Kataria}, \&
  {Homeier}}]{2014ApJ...793L..27K}
{Kreidberg}, L., {Bean}, J.~L., {D{\'e}sert}, J.-M., {et~al.} 2014, \apjl, 793,
  L27, \dodoi{10.1088/2041-8205/793/2/L27}

\bibitem[{{Lambrechts} \& {Johansen}(2012)}]{2012A&A...544A..32L}
{Lambrechts}, M., \& {Johansen}, A. 2012, \aap, 544, A32,
  \dodoi{10.1051/0004-6361/201219127}

\bibitem[{{Lammer} {et~al.}(2021){Lammer}, {Brasser}, {Johansen}, {Scherf}, \&
  {Leitzinger}}]{2021SSRv..217....7L}
{Lammer}, H., {Brasser}, R., {Johansen}, A., {Scherf}, M., \& {Leitzinger}, M.
  2021, \ssr, 217, 7, \dodoi{10.1007/s11214-020-00778-4}

\bibitem[{{Lecluse} {et~al.}(1996){Lecluse}, {Robert}, {Gautier}, \&
  {Guiraud}}]{1996P&SS...44.1579L}
{Lecluse}, C., {Robert}, F., {Gautier}, D., \& {Guiraud}, M. 1996, \planss, 44,
  1579, \dodoi{10.1016/S0032-0633(96)00070-0}

\bibitem[{{Lee} \& {Chiang}(2015)}]{2015ApJ...811...41L}
{Lee}, E.~J., \& {Chiang}, E. 2015, \apj, 811, 41,
  \dodoi{10.1088/0004-637X/811/1/41}

\bibitem[{{Lodders}(2003)}]{2003ApJ...591.1220L}
{Lodders}, K. 2003, \apj, 591, 1220, \dodoi{10.1086/375492}

\bibitem[{{Madhusudhan} {et~al.}(2014){Madhusudhan}, {Amin}, \&
  {Kennedy}}]{2014ApJ...794L..12M}
{Madhusudhan}, N., {Amin}, M.~A., \& {Kennedy}, G.~M. 2014, \apjl, 794, L12,
  \dodoi{10.1088/2041-8205/794/1/L12}

\bibitem[{{Mahaffy} {et~al.}(1998){Mahaffy}, {Donahue}, {Atreya}, {Owen}, \&
  {Niemann}}]{1998SSRv...84..251M}
{Mahaffy}, P.~R., {Donahue}, T.~M., {Atreya}, S.~K., {Owen}, T.~C., \&
  {Niemann}, H.~B. 1998, \ssr, 84, 251

\bibitem[{{Mandt} {et~al.}(2020){Mandt}, {Mousis}, {Lunine}, {Marty}, {Smith},
  {Luspay-Kuti}, \& {Aguichine}}]{2020SSRv..216...99M}
{Mandt}, K.~E., {Mousis}, O., {Lunine}, J., {et~al.} 2020, \ssr, 216, 99,
  \dodoi{10.1007/s11214-020-00723-5}

\bibitem[{{Marboeuf} {et~al.}(2018){Marboeuf}, {Thiabaud}, {Alibert}, \&
  {Benz}}]{2018MNRAS.475.2355M}
{Marboeuf}, U., {Thiabaud}, A., {Alibert}, Y., \& {Benz}, W. 2018, \mnras, 475,
  2355, \dodoi{10.1093/mnras/stx3315}

\bibitem[{{Marty}(2012)}]{2012E&PSL.313...56M}
{Marty}, B. 2012, Earth and Planetary Science Letters, 313, 56,
  \dodoi{10.1016/j.epsl.2011.10.040}

\bibitem[{{Mayor} {et~al.}(2011){Mayor}, {Marmier}, {Lovis}, {Udry},
  {S{\'e}gransan}, {Pepe}, {Benz}, {Bertaux}, {Bouchy}, {Dumusque}, {Lo Curto},
  {Mordasini}, {Queloz}, \& {Santos}}]{2011arXiv1109.2497M}
{Mayor}, M., {Marmier}, M., {Lovis}, C., {et~al.} 2011, arXiv e-prints,
  arXiv:1109.2497.
\newblock \doarXiv{1109.2497}

\bibitem[{{Mizuno}(1980)}]{1980PThPh..64..544M}
{Mizuno}, H. 1980, Progress of Theoretical Physics, 64, 544,
  \dodoi{10.1143/PTP.64.544}

\bibitem[{{Mordasini} {et~al.}(2014){Mordasini}, {Klahr}, {Alibert}, {Miller},
  \& {Henning}}]{2014A&A...566A.141M}
{Mordasini}, C., {Klahr}, H., {Alibert}, Y., {Miller}, N., \& {Henning}, T.
  2014, \aap, 566, A141, \dodoi{10.1051/0004-6361/201321479}

\bibitem[{{Mordasini} {et~al.}(2016){Mordasini}, {van Boekel}, {Molli{\`e}re},
  {Henning}, \& {Benneke}}]{2016ApJ...832...41M}
{Mordasini}, C., {van Boekel}, R., {Molli{\`e}re}, P., {Henning}, T., \&
  {Benneke}, B. 2016, \apj, 832, 41, \dodoi{10.3847/0004-637X/832/1/41}

\bibitem[{{Morley} {et~al.}(2019){Morley}, {Skemer}, {Miles}, {Line}, {Lopez},
  {Brogi}, {Freedman}, \& {Marley}}]{2019ApJ...882L..29M}
{Morley}, C.~V., {Skemer}, A.~J., {Miles}, B.~E., {et~al.} 2019, \apjl, 882,
  L29, \dodoi{10.3847/2041-8213/ab3c65}

\bibitem[{{Movshovitz} {et~al.}(2010){Movshovitz}, {Bodenheimer}, {Podolak}, \&
  {Lissauer}}]{2010Icar..209..616M}
{Movshovitz}, N., {Bodenheimer}, P., {Podolak}, M., \& {Lissauer}, J.~J. 2010,
  \icarus, 209, 616, \dodoi{10.1016/j.icarus.2010.06.009}

\bibitem[{{Mumma} \& {Charnley}(2011)}]{2011ARA&A..49..471M}
{Mumma}, M.~J., \& {Charnley}, S.~B. 2011, \araa, 49, 471,
  \dodoi{10.1146/annurev-astro-081309-130811}

\bibitem[{{Nakagawa} {et~al.}(1986){Nakagawa}, {Sekiya}, \&
  {Hayashi}}]{1986Icar...67..375N}
{Nakagawa}, Y., {Sekiya}, M., \& {Hayashi}, C. 1986, \icarus, 67, 375,
  \dodoi{10.1016/0019-1035(86)90121-1}

\bibitem[{{Nettelmann} {et~al.}(2013){Nettelmann}, {Helled}, {Fortney}, \&
  {Redmer}}]{2013P&SS...77..143N}
{Nettelmann}, N., {Helled}, R., {Fortney}, J.~J., \& {Redmer}, R. 2013,
  \planss, 77, 143, \dodoi{10.1016/j.pss.2012.06.019}

\bibitem[{{{\"O}berg} {et~al.}(2011){{\"O}berg}, {Murray-Clay}, \&
  {Bergin}}]{2011ApJ...743L..16O}
{{\"O}berg}, K.~I., {Murray-Clay}, R., \& {Bergin}, E.~A. 2011, \apjl, 743,
  L16, \dodoi{10.1088/2041-8205/743/1/L16}

\bibitem[{{Ormel}(2014)}]{2014ApJ...789L..18O}
{Ormel}, C.~W. 2014, \apjl, 789, L18, \dodoi{10.1088/2041-8205/789/1/L18}

\bibitem[{{Ormel} \& {Klahr}(2010)}]{2010A&A...520A..43O}
{Ormel}, C.~W., \& {Klahr}, H.~H. 2010, \aap, 520, A43,
  \dodoi{10.1051/0004-6361/201014903}

\bibitem[{{Ormel} \& {Kobayashi}(2012)}]{2012ApJ...747..115O}
{Ormel}, C.~W., \& {Kobayashi}, H. 2012, \apj, 747, 115,
  \dodoi{10.1088/0004-637X/747/2/115}

\bibitem[{{Piso} \& {Youdin}(2014)}]{2014ApJ...786...21P}
{Piso}, A.-M.~A., \& {Youdin}, A.~N. 2014, \apj, 786, 21,
  \dodoi{10.1088/0004-637X/786/1/21}

\bibitem[{{Podolak} {et~al.}(1995){Podolak}, {Weizman}, \&
  {Marley}}]{1995P&SS...43.1517P}
{Podolak}, M., {Weizman}, A., \& {Marley}, M. 1995, \planss, 43, 1517,
  \dodoi{10.1016/0032-0633(95)00061-5}

\bibitem[{{Pollack} {et~al.}(1996){Pollack}, {Hubickyj}, {Bodenheimer},
  {Lissauer}, {Podolak}, \& {Greenzweig}}]{1996Icar..124...62P}
{Pollack}, J.~B., {Hubickyj}, O., {Bodenheimer}, P., {et~al.} 1996, \icarus,
  124, 62, \dodoi{10.1006/icar.1996.0190}

\bibitem[{{Rafikov}(2004)}]{2004AJ....128.1348R}
{Rafikov}, R.~R. 2004, \aj, 128, 1348, \dodoi{10.1086/423216}

\bibitem[{{Rafikov}(2006)}]{2006ApJ...648..666R}
---. 2006, \apj, 648, 666, \dodoi{10.1086/505695}

\bibitem[{{Saito} {et~al.}(2006){Saito}, {Miyamoto}, {Nakamura}, {Ishiguro},
  {Michikami}, {Nakamura}, {Demura}, {Sasaki}, {Hirata}, {Honda}, {Yamamoto},
  {Yokota}, {Fuse}, {Yoshida}, {Tholen}, {Gaskell}, {Hashimoto}, {Kubota},
  {Higuchi}, {Nakamura}, {Smith}, {Hiraoka}, {Honda}, {Kobayashi}, {Furuya},
  {Matsumoto}, {Nemoto}, {Yukishita}, {Kitazato}, {Dermawan}, {Sogame},
  {Terazono}, {Shinohara}, \& {Akiyama}}]{2006Sci...312.1341S}
{Saito}, J., {Miyamoto}, H., {Nakamura}, R., {et~al.} 2006, Science, 312, 1341,
  \dodoi{10.1126/science.1125722}

\bibitem[{{Sato} {et~al.}(2016){Sato}, {Okuzumi}, \&
  {Ida}}]{2016A&A...589A..15S}
{Sato}, T., {Okuzumi}, S., \& {Ida}, S. 2016, \aap, 589, A15,
  \dodoi{10.1051/0004-6361/201527069}

\bibitem[{{Schwieterman} {et~al.}(2018){Schwieterman}, {Kiang}, {Parenteau},
  {Harman}, {DasSarma}, {Fisher}, {Arney}, {Hartnett}, {Reinhard}, {Olson},
  {Meadows}, {Cockell}, {Walker}, {Grenfell}, {Hegde}, {Rugheimer}, {Hu}, \&
  {Lyons}}]{2018AsBio..18..663S}
{Schwieterman}, E.~W., {Kiang}, N.~Y., {Parenteau}, M.~N., {et~al.} 2018,
  Astrobiology, 18, 663, \dodoi{10.1089/ast.2017.1729}

\bibitem[{{Shiraishi} \& {Ida}(2008)}]{2008ApJ...684.1416S}
{Shiraishi}, M., \& {Ida}, S. 2008, \apj, 684, 1416, \dodoi{10.1086/590226}

\bibitem[{{Stern} {et~al.}(2015){Stern}, {Bagenal}, {Ennico}, {Gladstone},
  {Grundy}, {McKinnon}, {Moore}, {Olkin}, {Spencer}, {Weaver}, {Young},
  {Andert}, {Andrews}, {Banks}, {Bauer}, {Bauman}, {Barnouin}, {Bedini},
  {Beisser}, {Beyer}, {Bhaskaran}, {Binzel}, {Birath}, {Bird}, {Bogan},
  {Bowman}, {Bray}, {Brozovic}, {Bryan}, {Buckley}, {Buie}, {Buratti},
  {Bushman}, {Calloway}, {Carcich}, {Cheng}, {Conard}, {Conrad}, {Cook},
  {Cruikshank}, {Custodio}, {Dalle Ore}, {Deboy}, {Dischner}, {Dumont},
  {Earle}, {Elliott}, {Ercol}, {Ernst}, {Finley}, {Flanigan}, {Fountain},
  {Freeze}, {Greathouse}, {Green}, {Guo}, {Hahn}, {Hamilton}, {Hamilton},
  {Hanley}, {Harch}, {Hart}, {Hersman}, {Hill}, {Hill}, {Hinson}, {Holdridge},
  {Horanyi}, {Howard}, {Howett}, {Jackman}, {Jacobson}, {Jennings}, {Kammer},
  {Kang}, {Kaufmann}, {Kollmann}, {Krimigis}, {Kusnierkiewicz}, {Lauer}, {Lee},
  {Lindstrom}, {Linscott}, {Lisse}, {Lunsford}, {Mallder}, {Martin}, {McComas},
  {McNutt}, {Mehoke}, {Mehoke}, {Melin}, {Mutchler}, {Nelson}, {Nimmo},
  {Nunez}, {Ocampo}, {Owen}, {Paetzold}, {Page}, {Parker}, {Parker},
  {Pelletier}, {Peterson}, {Pinkine}, {Piquette}, {Porter}, {Protopapa},
  {Redfern}, {Reitsema}, {Reuter}, {Roberts}, {Robbins}, {Rogers}, {Rose},
  {Runyon}, {Retherford}, {Ryschkewitsch}, {Schenk}, {Schindhelm}, {Sepan},
  {Showalter}, {Singer}, {Soluri}, {Stanbridge}, {Steffl}, {Strobel}, {Stryk},
  {Summers}, {Szalay}, {Tapley}, {Taylor}, {Taylor}, {Throop}, {Tsang},
  {Tyler}, {Umurhan}, {Verbiscer}, {Versteeg}, {Vincent}, {Webbert}, {Weidner},
  {Weigle}, {White}, {Whittenburg}, {Williams}, {Williams}, {Williams},
  {Woods}, {Zangari}, \& {Zirnstein}}]{2015Sci...350.1815S}
{Stern}, S.~A., {Bagenal}, F., {Ennico}, K., {et~al.} 2015, Science, 350,
  aad1815, \dodoi{10.1126/science.aad1815}

\bibitem[{{Stevenson}(1982)}]{1982P&SS...30..755S}
{Stevenson}, D.~J. 1982, \planss, 30, 755, \dodoi{10.1016/0032-0633(82)90108-8}

\bibitem[{{Swain} {et~al.}(2021){Swain}, {Estrela}, {Roudier}, {Sotin},
  {Rimmer}, {Valio}, {West}, {Pearson}, {Huber-Feely}, \&
  {Zellem}}]{2021AJ....161..213S}
{Swain}, M.~R., {Estrela}, R., {Roudier}, G.~M., {et~al.} 2021, \aj, 161, 213,
  \dodoi{10.3847/1538-3881/abe879}

\bibitem[{{Tajima} \& {Nakagawa}(1997)}]{1997Icar..126..282T}
{Tajima}, N., \& {Nakagawa}, Y. 1997, \icarus, 126, 282,
  \dodoi{10.1006/icar.1996.5651}

\bibitem[{{Tanigawa} \& {Tanaka}(2016)}]{2016ApJ...823...48T}
{Tanigawa}, T., \& {Tanaka}, H. 2016, \apj, 823, 48,
  \dodoi{10.3847/0004-637X/823/1/48}

\bibitem[{{Tanigawa} \& {Watanabe}(2002)}]{2002ApJ...580..506T}
{Tanigawa}, T., \& {Watanabe}, S.-i. 2002, \apj, 580, 506,
  \dodoi{10.1086/343069}

\bibitem[{{Thommes} {et~al.}(1999){Thommes}, {Duncan}, \&
  {Levison}}]{1999Natur.402..635T}
{Thommes}, E.~W., {Duncan}, M.~J., \& {Levison}, H.~F. 1999, \nat, 402, 635,
  \dodoi{10.1038/45185}

\bibitem[{{Tsiganis} {et~al.}(2005){Tsiganis}, {Gomes}, {Morbidelli}, \&
  {Levison}}]{2005Natur.435..459T}
{Tsiganis}, K., {Gomes}, R., {Morbidelli}, A., \& {Levison}, H.~F. 2005, \nat,
  435, 459, \dodoi{10.1038/nature03539}

\bibitem[{{Venturini} \& {Helled}(2020)}]{2020A&A...634A..31V}
{Venturini}, J., \& {Helled}, R. 2020, \aap, 634, A31,
  \dodoi{10.1051/0004-6361/201936591}

\bibitem[{{Wahl} {et~al.}(2017){Wahl}, {Hubbard}, {Militzer}, {Guillot},
  {Miguel}, {Movshovitz}, {Kaspi}, {Helled}, {Reese}, {Galanti}, {Levin},
  {Connerney}, \& {Bolton}}]{2017GeoRL..44.4649W}
{Wahl}, S.~M., {Hubbard}, W.~B., {Militzer}, B., {et~al.} 2017, \grl, 44, 4649,
  \dodoi{10.1002/2017GL073160}

\bibitem[{{Walsh} {et~al.}(2011){Walsh}, {Morbidelli}, {Raymond}, {O'Brien}, \&
  {Mandell}}]{2011Natur.475..206W}
{Walsh}, K.~J., {Morbidelli}, A., {Raymond}, S.~N., {O'Brien}, D.~P., \&
  {Mandell}, A.~M. 2011, \nat, 475, 206, \dodoi{10.1038/nature10201}

\bibitem[{{Weidenschilling}(1977)}]{1977MNRAS.180...57W}
{Weidenschilling}, S.~J. 1977, \mnras, 180, 57, \dodoi{10.1093/mnras/180.2.57}

\bibitem[{{Winn} \& {Fabrycky}(2015)}]{2015ARA&A..53..409W}
{Winn}, J.~N., \& {Fabrycky}, D.~C. 2015, \araa, 53, 409,
  \dodoi{10.1146/annurev-astro-082214-122246}

\bibitem[{{Zhou} \& {Lin}(2007)}]{2007ApJ...666..447Z}
{Zhou}, J.-L., \& {Lin}, D. N.~C. 2007, \apj, 666, 447, \dodoi{10.1086/520043}

\end{thebibliography}



\end{document}